\documentclass[10pt,a4]{article}

\usepackage{Lbook}
\usepackage{a4wide}
\usepackage{amsmath}
\usepackage{graphicx}
\usepackage{url}
\usepackage{xcolor}
\usepackage{xfrac}
\usepackage{multirow}

\newcommand{\floor}[1]{\left\lfloor #1 \right\rfloor}

\newcommand{\donotshow}[1]{}

\newcommand{\apostrophe}{'}

\newcommand{\ignore}[1]{}

\newcommand{\lref}[1]{(\ref{#1})}

\providecommand{\CC}{C\raisebox{.08ex}{\hbox{\tt ++}}}

\newcommand {\abs}[1] {| #1 |}

\newlength{\mysetspacing}
\providecommand{\sset}[1]{\{\hspace{\mysetspacing} #1 \hspace{\mysetspacing}\}}

\newcommand{\mbegin}{\{\ \ }
\newcommand{\mend}{\}}

\newlength{\mleftindent}
\newlength{\mindent}
\newlength{\mboxwidth}
\newcommand{\mincrement}{\addtolength{\mboxwidth}{-\mindent}}
\newcommand{\mdecrement}{\addtolength{\mboxwidth}{\mindent}}

\newlength{\preprogramskip}
\newlength{\postprogramskip}
\newlength{\mexpwidth}
\newlength{\mexpindent}
\newcommand{\indentafterkeyword}{\hspace*{0.5em}}

\newcommand{\mslifelse}[3]  
{\setlength{\mexpwidth}{\mboxwidth}%
\settowidth{\mexpindent}{{\bf if\indentafterkeyword}}%
\addtolength{\mexpwidth}{-\mexpindent}%
{\bf if\indentafterkeyword}\parbox[t]{\mexpwidth}{#1}\\
\mincrement \mbegin \parbox[t]{\mboxwidth}{#2 \mend} \mdecrement \\
{\bf else} \\
\mincrement \mbegin \parbox[t]{\mboxwidth}{#3}\\
\mend \mdecrement
}

{\vspace{-0.3em}\begin{itemize}
\setlength{\itemsep}{0.2\itemsep}%
\setlength{\parskip}{0.2\parskip}%
\setlength{\topsep}{0.0\topsep}%
}
{\end{itemize}}

{\begin{itemize}
\setlength{\itemsep}{0.2\itemsep}%
\setlength{\parskip}{0.2\parskip}%
\setlength{\topsep}{0.0\topsep}%
}
{\end{itemize}}

{\begin{itemize}
\setlength{\itemsep}{1.5\itemsep}%
\setlength{\parskip}{1.5\parskip}%
\setlength{\topsep}{0.0\topsep}%
}
{\end{itemize}}

{\begin{itemize}
\setlength{\itemsep}{1.5\itemsep}%
\setlength{\parskip}{1.5\parskip}%
\par \smallskip
}
{\end{itemize}}

{\begin{itemize}
\setlength{\itemsep}{1.5\itemsep}%
\setlength{\parskip}{1.5\parskip}%
\par \medskip
}
{\end{itemize}}

{\vspace{-0.3em}\begin{description}
\setlength{\itemsep}{0.2\itemsep}%
\setlength{\parskip}{0.2\parskip}%
\setlength{\topsep}{0.0\topsep}%
}
{\end{description}}

{\begin{description}
\setlength{\itemsep}{0.2\itemsep}%
\setlength{\parskip}{0.2\parskip}%
\setlength{\topsep}{0.0\topsep}%
}
{\end{description}}

{\vspace{-0.3em}\begin{enumerate}
\setlength{\itemsep}{0.2\itemsep}%
\setlength{\parskip}{0.2\parskip}%
\setlength{\topsep}{0.0\topsep}%
}
{\end{enumerate}}

{\begin{enumerate}
\setlength{\itemsep}{0.2\itemsep}%
\setlength{\parskip}{0.2\parskip}%
\setlength{\topsep}{0.0\topsep}%
}
{\end{enumerate}}

\newlength{\proofpostskipamount}\newlength{\proofpreskipamount}
\newenvironment{proof}%
               {\par\vspace{\proofpreskipamount}\noindent{\bf Proof:}\hspace{0.5em}}
               {\nopagebreak%
                \strut\nopagebreak%
                \hspace{\fill}\qed\par\vspace{\proofpostskipamount}\noindent}

               {\par\vspace{0.5ex}\noindent{\bf Proof #1:}\hspace{0.5em}}%
               {\nopagebreak%
                \strut\nopagebreak%
                \hspace{\fill}\qed\par\medskip\noindent}

\newtheorem{lemma}{Lemma}

\providecommand{\qed}{\rule[-0.2ex]{0.3em}{1.4ex}}

\newlength{\mydefwidth}
\newlength{\mytextwidth}

\newcommand{\myurl}[1]{{\footnotesize \url{#1}}}

\begin{document}

\title{Gabow's Cardinality Matching Algorithm in General Graphs\\Implementation and Experiments\footnote{Part of the work was done as an internship project of the first two authors at MPI for Informatics.}}
\author{Matin Ansaripour\footnote{Department of Computer Science, EPFL, Lausanne, Switzerland} and Alireza Danaei\footnote{Department of Computer Enginering, Sharif University, Tehran, Iran} and Kurt Mehlhorn\footnote{MPI for Informatics, Saarbr\"ucken, Germany}}

\maketitle

\begin{abstract} It is known since 1975 (\cite{HK75}) that maximum cardinality matchings in bipartite graphs with $n$ nodes and $m$ edges can be computed in time $O(\sqrt{n} m)$. Asymptotically faster algorithms were found in the last decade and maximum cardinality bipartite matchings can now be computed in near-linear time~\cite{NearlyLinearTimeBipartiteMatching, AlmostLinearTimeMaxFlow,AlmostLinearTimeMinCostFlow}. For general graphs, the problem seems harder. Algorithms with running time $O(\sqrt{n} m)$ were given in~\cite{MV80,Vazirani94,Vazirani12,Vazirani20,Vazirani23,Goldberg-Karzanov,GT91,Gabow:GeneralMatching}. Mattingly and Ritchey~\cite{Mattingly-Ritchey} and Huang and Stein~\cite{Huang-Stein} discuss implementations of the Micali-Vazirani Algorithm. We describe an implementation of Gabow's algorithm~\cite{Gabow:GeneralMatching} in C++ based on LEDA~\cite{LEDAsystem,LEDAbook} and report on running time experiments. On worst-case graphs, the asymptotic improvement pays off dramatically. On random graphs, there is no improvement with respect to algorithms that have a worst-case running time of $O(n m)$. The performance seems to be near-linear. The implementation is available open-source. 

\end{abstract}


\tableofcontents

\section{Introduction} The maximum matching problem is one of the basic problems in graph theory and graph algorithms. Given an undirected graph, the goal is to find a matching, i.e., a set of edges no two of which share an endpoint, of maximum cardinality. Edmonds gave a polynomial time algorithm as early as 1965~\cite{Edmonds:matching}. The running time of the algorithm was improved over time, culminating in the $O(nm\alpha(n))$ algorithm of Gabow~\cite{Gabow:edmonds} and the $O(nm)$ algorithm of Gabow and Tarjan~\cite{Gabow-Tarjan:union-find}. An implementation of the former algorithm is available in LEDA~\cite{LEDAsystem,LEDAbook}. Kececioglu and Pecqueur~\cite{Kececioglu:matching} give heuristic improvements that often lead to considerably smaller running times.  Algorithms with running time $O(\sqrt{n} m)$ were given in~\cite{MV80,Vazirani94,Vazirani12,Vazirani20,Vazirani23,Goldberg-Karzanov,GT91,Gabow:GeneralMatching}. Mattingly and Ritchey~\cite{Mattingly-Ritchey} and Huang and Stein~\cite{Huang-Stein} discuss implementations of the Micali-Vazirani Algorithm. We describe an implementation of Gabow's algorithm~\cite{Gabow:GeneralMatching} in C++ based on LEDA~\cite{LEDAsystem,LEDAbook} and report on running time experiments. On worst-case graphs, the asymptotic improvement pays off dramatically. On random graphs, there is no improvement with respect to algorithms that have a worst-case running time of $O(n m)$. The performance seems to be near-linear.

In Section~\ref{Gabow} we describe our implementation of Gabow\apostrophe s algorithm. We follow his description closely~\cite{Gabow:GeneralMatching}. 

in Section~\ref{experiments} we define a worst-case family of sparse graphs and report on various experiments. On the family, Gabow\apostrophe s new algorithm has a running time $\Theta(n^{\sfrac{3}{2}})$ and Gabow\apostrophe s older algorithm has a running time of $\Theta(n^2)$. We also give a family, where the respective running times are $O(n)$ and $\Omega(n^{\sfrac{3}{2}})$, respectively. We make sure that these running times do not only hold for a particular representation of the graph, i.e., a particular numbering of the vertices and edges, but for most representations. 

In the appendices we include listing of LEDA\apostrophe s matching algorithm, of LEDA\apostrophe s matching algorithm with the Kececioglu-Pecqueur Heuristics, and of our program for the running time experiments.

We present the program as a literate programming document~\cite{Knuth-Levy} using the adaption of the Ramsey\apostrophe s noweb-system developed for LEDA~\cite{LEDAbook}. In particular, the program is defined as a sequence of chunks.

\nwfilename{./GabowImplementation.nw}\nwbegincode{1}\moddef{chunk}\endmoddef\nwstartdeflinemarkup\nwenddeflinemarkup\nwcodepenalty=\Lhighpen
some \\CC text that may contain \LA{}subchuncks\RA{}\nwcodepenalty=\Lhighpen
\vspace{\Lemptyline}\nwendcode{}\nwbegindocs{2}In our case, the top-level chunck is \ensuremath{\mathit{Gabow}\nspacedot.h}. The corresponding file is obtained in a process similar to macro-expansion. Subchunks are replaced by their definitions until all subchuncks are expanded. 

\section{Gabow's Matching Algorithm}\label{Gabow}
We define a C++ class \ensuremath{\mathit{G\nspaceunderscore\_card\nspaceunderscore\_matching}} that realizes Gabow's algorithm~\cite{Gabow:GeneralMatching} for maximum matchings in general graphs. It runs in time $O(\sqrt{n} m)$ and generalizes the Hopcroft and Karp algorithm from bipartite to general graphs.\footnote{As customary, we use $n$ and $m$ for the number of vertices and edges, respectively.} It follows the same principle. It works in iterations and in each iteration augments a maximal number of edge-disjoint shortest augmenting paths (\emph{sap}s) to the current matching. Each iteration works in linear time (in our implementation in time $O(m \alpha(n))$) and the number of iterations is $O(\sqrt{n})$.
Each iteration consists of two phases. Let $M$ be the current matching and let $G$ be the input graph.

\paragraph{First Phase:} In the first phase, the length of the \ensuremath{\mathit{sap}}s is determined and, if an augmenting path exists, an auxiliary graph $H$ is constructed. The graph $H$ has the (amazing) property that augmenting paths in $H$ are in one-to-one correspondence to \ensuremath{\mathit{sap}}s in $G$.
The first phase is an adaption of Edmonds' weighted matching algorithm; this algorithm constructs a maximum weight matching by iteratively constructing maximum weight matchings of cardinalities 0, 1, 2, \ldots. Each matching is obtained from the previous one by augmenting a maximum weight augmenting path, where the weight of a path $P$ is defined as
\begin{equation} w(P) = w(P \setminus M) - w(P \cap M), \label{weight of path} \end{equation}
i.e., the increase in weight obtained by augmenting $P$ to $M$. We want $P$ to be a \ensuremath{\mathit{sap}} and therefore define $w(e) = 2$ for $e \in M$ and $w(e) = 0$ for $e \not\in M$. Then a maximum weight augmenting path is a \ensuremath{\mathit{sap}}; notice that the weight of an augmenting path is the negative of 2 times the number of edges contained in the current matching. In principle, any positive weight could be used for the edges in $M$. The choice of two guarantees that the algorithm has to deal with integers only. Edmonds' algorithm is based on the linear programming formulation of maximum weighted matching and makes extensive use of the dual linear program. The dual linear program has a variable for each vertex of $G$ and for each set of vertices of odd cardinality three or more. The latter are non-negative, the former are unconstrained. The \emph{reduced weight} $\hat{w}(e)$ of an edge $e = xy$ is defined as
\begin{equation}  \hat{w}(e) = d(x) + d(y) + \sum_{x,y \in B} z(B) - w(e), \label{reduced weight} \end{equation}
where $d$ denotes dual values of vertices and $z$ denotes dual values of odd sets (of cardinality three or more). The sum is over all odd sets containing $x$ and $y$. \emph{Reduced weights are always non-negative and an edge is called tight if its reduced weight is zero}.
When the search for an augmenting path starts, $d(v) = 1$ for all $v$ and $z(B) = 0$ for all odd sets $B$. Then matching edges are tight and non-matching edges are non-tight. The search for an augmenting path grows search structures (usually called trees) rooted at free nodes. The trees are grown concurrently and the endpoints of a matching edge either both belong to a tree or none belongs to a tree. The trees are initialized with the free nodes. A tree node is \ensuremath{\mathit{EVEN}} if there is an even length path in the search structure connecting it to its root and \ensuremath{\mathit{ODD}} otherwise. The incoming tree-edges of even nodes are matching edges and are non-matching for odd nodes. Trees are grown by the addition of \emph{tight} edges incident to even nodes. So assume that $x$ is an even node and let $xy$ be a tight edge incident to $x$. If $y$ does not belong to any search structure yet (then $y$ is matched), we add $y$ and its mate to the tree containing $x$. $y$ becomes odd and \ensuremath{\mathit{mate}[y]} becomes even. When $y$ already belongs to a search structure and is \ensuremath{\mathit{ODD}}, we do nothing as we have simply discovered another odd length path to $y$. If $y$ is \ensuremath{\mathit{EVEN}} and belongs to a different tree, we have discovered an augmenting path. If $y$ is \ensuremath{\mathit{EVEN}} and belongs to the same tree, we have discovered a so-called blossom. Let $b$ be the lowest common ancestor of $x$ and $y$. Then all odd nodes on the paths from $b$ to $x$ and $y$ become even. For example, an odd node $z$ between $b$ and $y$ can now be reached by going from the root to $x$ (even length), then to $y$ (one step) and then from $y$ towards $z$ (odd length). The edge $xy$ is called the \emph{bridge} of the blossom and the blossom consists of all nodes on the paths from $b$ to $x$ and $y$, respectively. A blossom contains an odd number of nodes. We contract all nodes of the blossom into a single node. In this way, the search structures stay trees. Note that blossoms can be nested.

In what order do we add edges to the search structures? We maintain the invariant that all edges in the search structures are tight and that all roots have the same dual value. We grow the trees by adding non-matching tight edges, i.e., edges $xy$ with $\hat{w}(e) = 0$. Then necessarily, $d(x) + d(y) = 0$ as $w(e) = 0$. In particular, if $y$ does not belong to a search structure yet, $d(x) = -1$ and $d(y) = +1$. Suppose now that we cannot further grow trees, i.e., $\hat{w}(e) > 0$ for any edge $xy$ with at least one even endpoint. Then we perform a dual update. We decrease $d(v)$ by one for every even vertex, increase $d(v)$ by one for every odd vertex and increase $z(B)$ by two for every maximal blossom, i.e., any blossom not contained in any other blossom. Then $z$-values are always even, and the $d$-values of all vertices in the search structures have the same parity. Also, all edges in the search structures stay tight: if one endpoint is even and one is odd, this is obvious, if both endpoints are even, the endpoints are contained in the same maximal blossom, and again the claim is obvious. Consider now an edge $xy$ not belonging to any search structure. If no endpoint belongs to the search structures, $d(x) = d(y) = 1$ and the reduced weight does not change. If at least one endpoint is odd, the reduced weight does not decrease. If one endpoint is even, and the other endpoint does not belong to a search structure, the reduced weight decreases by one and hence stays non-negative. If both endpoints are even and do not belong to the same blossom, the reduced weight decreases by two and hence stays non-negative since the $d$-values of the endpoints have the same parity. We have now shown that reduced weights stay non-negative, that edges in the search structure are tight, and that all roots have the same dual value. 
After $\Delta$ dual updates, we have $d(f) = 1 - \Delta$ for any free node $f$.



\newcommand{\sap}{\emph{sap}}

The auxiliary graph $H$ is a contraction of $G$. All maximal positive blossoms are contracted and the tight edges between them are kept. In this way, the augmenting paths in $H$ correspond to \ensuremath{\mathit{sap}}s in $G$. Since a \ensuremath{\mathit{sap}} must use the matching edge incident to the base of a maximal blossom (except if the base is free)\footnote{See Lemma~\ref{structure of saps}.}, a \ensuremath{\mathit{sap}} contracts to an augmenting path in $H$. Conversely, an augmenting path in $H$ is easily lifted to a augmenting path in $G$ by filling in the parts inside the contracted nodes. 

\paragraph{Phase Two:} We give only a very rough sketch of the second phase at this point and give more details in Section~\ref{phase2}. In the second phase a maximal number of disjoint augmenting paths in $H$ is constructed. The second phase uses depth-first search for growing the search structures. Consider the growth of a tree rooted at a free node $f$ and assume first that our graph is bipartite. Then there are two important facts about DFS. When we add an edge $xy$ where $x$ is even and $y$ is free, the tree path to $x$ together with the edge $xy$ forms an augmenting path. When we return from a recursive call for node $x$ without having found an augmenting path, there is none passing through $x$. In general graphs, we also have blossoms. Since, we are performing DFS on an undirected graph, blossoms are discovered when we explore a forward or backward edge. We have the choice and delay the formation of the blossom to the exploration of the forward edge. Consider a forward edge $xy$ and let $b$ be the lowest common ancestor\footnote{If $xy$ is a forward edge, shouldn't $x$ be the lowest common ancestor of $x$ and $y$? How can the lowest common ancestor be different from $x$? Recall that we are working on a graph in which some blossoms are already be shrunken. In particular, $x$ may be part of a blossom with base $b$. Then $b$ is the common ancestor of $x$ and $y$ in the original graph.} of $x$ and $y$. Then $x$ belongs to the blossom with base $b$ and the blossom step makes the odd nodes between $y$ and $b$ even. In order to maintain the crucial properties of DFS, it suffices to make the recursive calls in a particular order. We start with the odd node closest to $b$ and then work our way up to $y$. 

\subsection{The Overall Structure of the Program}

We define all the variables used in the program and introduce the two program chunks \emph{init} and \emph{solve}. The space requirement of the program is enormous. In particular, it uses more than 20 arrays indexed by nodes (in addition to the ones visible as \ensuremath{\mathit{node\nspaceunderscore\_array}}s, there are two hidden in the \ensuremath{\mathit{node\nspaceunderscore\_partition}}s base and dbase and one hidden in the auxiliary graph $H$). We give more information about the space usage in Section~\ref{instruction counts}. It would be interesting to reduce the space requirement. There is no need to read all the declarations in detail. 

The chunk \ensuremath{\mathit{init}} contains the functions for greedy initialization and for initialization with a known matching. The chunck \ensuremath{\mathit{solve}} is the workhorse of the program.

\nwenddocs{}\nwbegincode{3}\moddef{Gabow.h}\endmoddef\nwstartdeflinemarkup\nwenddeflinemarkup\nwcodepenalty=\Lhighpen
\vspace{\Lemptyline}\LA{}priority\_queue.h\RA{}\nwcodepenalty=\Lhighpen
\vspace{\Lemptyline}enum LABEL \{EVEN, ODD, UNLABELED\};  // also declared in  _mc_matching.cpp\nwcodepenalty=\Lhighpen
\vspace{\Lemptyline}class G_card_matching\{\nwcodepenalty=\Llowpen
private:\nwcodepenalty=\Llowpen
  const graph& G;\nwcodepenalty=\Llowpen
  list<edge> M; // matching in G\nwcodepenalty=\Llowpen
  node_array<node> mate; // mates in M\nwcodepenalty=\Llowpen
  node_partition base;   // dbase and base are almost the same\nwcodepenalty=\Llowpen
  node_partition dbase; // the maximal positive blossoms in G\nwcodepenalty=\Llowpen
  node_array<node> rep;  // rep[v] is the base of the maximal positive blossom containing v\nwcodepenalty=\Llowpen
  node_array<node> parent; // parent[v] is the parent of v in the alternating tree\nwcodepenalty=\Llowpen
  node_array<node> source_bridge;  // bridges close blossoms; a blossom consists of two paths\nwcodepenalty=\Llowpen
  node_array<node> target_bridge;  // x--z and y--z plus the edge xy; z is the base of the blossom.\nwcodepenalty=\Llowpen
  // the nodes on the path from x to z store x as source_bridge and y as target_bridge.\nwcodepenalty=\Llowpen
  node_array<double> path1;\nwcodepenalty=\Llowpen
  node_array<double> path2;\nwcodepenalty=\Llowpen
  edge_array<int> w;  // w[e] = 2 for e in M and w[e] = 0 otherwise\nwcodepenalty=\Llowpen
  node_array<LABEL> label;\nwcodepenalty=\Llowpen
  simple_queue PQ;\nwcodepenalty=\Llowpen
  list<node> P;\nwcodepenalty=\Llowpen
  list<node> T;\nwcodepenalty=\Llowpen
  node_array<int> bd;    // initialized to 1\nwcodepenalty=\Llowpen
  node_array<int> bDelta;  // correct value for free nodes\nwcodepenalty=\Llowpen
  graph H;   // H is obtained from G by contracting the maximal positive blossoms\nwcodepenalty=\Llowpen
  node_array<node> mateHG;\nwcodepenalty=\Llowpen
  edge_array<bool> is_edge_of_H;\nwcodepenalty=\Llowpen
  node_array<list<node> > contracted_into;\nwcodepenalty=\Llowpen
  // node_partition baseHG;\nwcodepenalty=\Llowpen
  node_array<LABEL> labelHG; \nwcodepenalty=\Llowpen
  node_array<edge> parentHG;    // only odd nodes use it. \nwcodepenalty=\Llowpen
  node_array<int> even_timeHG;   // timestep at which a node of H became even;\nwcodepenalty=\Llowpen
  node_array<edge> bridgeHG;\nwcodepenalty=\Llowpen
  node_array<int> dirHG;\nwcodepenalty=\Llowpen
  int size_of_M;\nwcodepenalty=\Llowpen
  int tG;\nwcodepenalty=\Llowpen
  int Delta;\nwcodepenalty=\Llowpen
  double strue;\nwcodepenalty=\Llowpen
  node_array<int> num;\nwcodepenalty=\Llowpen
  node_array<int> even_time;\nwcodepenalty=\Llowpen
  int even_count;\nwcodepenalty=\Llowpen
\vspace{\Lemptyline}  \LA{}helper functions\RA{}\nwcodepenalty=\Llowpen
\vspace{\Lemptyline}public:\nwcodepenalty=\Llowpen
G_card_matching(const graph& g): G(g), mate(node_array<node>(G,nil)),\nwcodepenalty=\Llowpen
base(node_partition(G)), dbase(node_partition(G)),\nwcodepenalty=\Llowpen
// no need to initialize rep, parent, source_bridge, target_bridgt, parentHG, bridgeHG, dirHG, mateHG\nwcodepenalty=\Llowpen
rep(node_array<node>(G)), parent(node_array<node>(G)),\nwcodepenalty=\Llowpen
source_bridge(node_array<node>(G)), target_bridge(node_array<node>(G)),\nwcodepenalty=\Llowpen
path1(node_array<double>(G)), path2(node_array<double>(G)),\nwcodepenalty=\Llowpen
w(edge_array<int>(G,0)), label(node_array<LABEL>(G,UNLABELED)),\nwcodepenalty=\Llowpen
PQ(simple_queue(G)), P(list<node>()), T(list<node>()), bd(node_array<int>(G)), bDelta(node_array<int>(G)),\nwcodepenalty=\Llowpen
mateHG(node_array<node>(G)),\nwcodepenalty=\Llowpen
is_edge_of_H(edge_array<bool>(G,false)), contracted_into(node_array<list<node> >(G)), \nwcodepenalty=\Llowpen
labelHG(node_array<LABEL>(G,UNLABELED)), \nwcodepenalty=\Llowpen
parentHG(node_array<edge>(G)), even_timeHG(node_array<int>(G,0)),\nwcodepenalty=\Llowpen
bridgeHG(node_array<edge>(G)), dirHG(node_array<int>(G)), size_of_M(0), tG(0), strue(0),\nwcodepenalty=\Llowpen
num(node_array<int>(G)), even_time(node_array<int>(G)), even_count(0)\nwcodepenalty=\Llowpen
\{ \nwcodepenalty=\Llowpen
\}\nwcodepenalty=\Llowpen
\vspace{\Lemptyline}\LA{}init: initializes with a greedy matching or a given matching\RA{}\nwcodepenalty=\Lhighpen
\vspace{\Lemptyline}\LA{}solve: computes matching and odd set cover\RA{}\nwcodepenalty=\Lhighpen
\vspace{\Lemptyline}\};\nwcodepenalty=\Llowpen
\vspace{\Lemptyline}\nwendcode{}\nwbegindocs{4}\paragraph{Initialization:} Initialization with a greedy matching is trivial. We iterate over all edges and add an edge to the matching if both endpoints are still free. We are careful and skip over self-loops. Alternatively, we can initialize with a matching \ensuremath{\mathit{M0}}.

\nwenddocs{}\nwbegincode{5}\moddef{init: initializes with a greedy matching or a given matching}\endmoddef\nwstartdeflinemarkup\nwenddeflinemarkup\nwcodepenalty=\Lhighpen
int init()\{\nwcodepenalty=\Lhighpen
  edge e;\nwcodepenalty=\Lhighpen
  forall_edges(e,G)\{\nwcodepenalty=\Llowpen
    node u = G.source(e); node v = G.target(e);\nwcodepenalty=\Llowpen
    if (u != v && mate[u] == nil && mate[v] == nil) \{\nwcodepenalty=\Llowpen
      mate[u] = v;\nwcodepenalty=\Llowpen
      mate[v] = u;\nwcodepenalty=\Llowpen
      size_of_M++;\nwcodepenalty=\Llowpen
    \}\nwcodepenalty=\Llowpen
  \}\nwcodepenalty=\Llowpen
  return size_of_M;\nwcodepenalty=\Llowpen
\};\nwcodepenalty=\Llowpen
\vspace{\Lemptyline}int init(const list<edge>& M0)\{\nwcodepenalty=\Llowpen
  edge e;  \nwcodepenalty=\Llowpen
  forall(e,M0)\{\nwcodepenalty=\Llowpen
    node u = G.source(e); node v = G.target(e);\nwcodepenalty=\Llowpen
    if (u != v && mate[u] == nil && mate[v] == nil) \{\nwcodepenalty=\Llowpen
      mate[u] = v;\nwcodepenalty=\Llowpen
      mate[v] = u;\nwcodepenalty=\Llowpen
      size_of_M++;\nwcodepenalty=\Llowpen
    \}\nwcodepenalty=\Llowpen
  \}\nwcodepenalty=\Lhighpen
  return G.number_of_nodes();   // there is no guarantee on the size of |M0|\nwcodepenalty=\Lhighpen
\};\nwcodepenalty=\Llowpen
\vspace{\Lemptyline}\nwendcode{}\nwbegindocs{6}\nwdocspar
\paragraph{The Computation of a Maximum Matching: Function \ensuremath{\mathit{solve}}:} We come to the work-horse, the function \ensuremath{\mathit{solve}}. As already stated, it works in iterations. Each iteration consists of two phases. In the first phase, the length of a \ensuremath{\mathit{sap}} and an auxiliary graph $H$ containing all $\sap$s is determined. If the auxiliary graph is empty, a maximum matching has been found, we construct an odd-set cover as a witness of optimality and return.  If the auxiliary graph is non-empty, we start the second phase, determine a maximal set of edge-disjoint $\sap$s, lift them to $G$, and augment the current matching. Then the next iteration begins.

The size of a maximum matching is at most twice the size of a greedy matching and never more than $n/2$. The number of iterations till completion is at most the difference between the size of the maximum matching and the size of the current matching. This is true even if future iterations find only a single augmenting path each. Since finding a single augmenting path is simpler and we have already implemented an efficient search for a single augmenting path in LEDA's general matching algorithm, we switch to this algorithm once the distance to the maximum matching is about the number of already executed iterations.

\nwenddocs{}\nwbegincode{7}\moddef{solve: computes matching and odd set cover}\endmoddef\nwstartdeflinemarkup\nwenddeflinemarkup\nwcodepenalty=\Lhighpen
\vspace{\Lemptyline}list<edge> solve(node_array<int>& OSC, int& number_of_iterations,\nwcodepenalty=\Lhighpen
                 double heur_factor = 1, bool heur = true)\{\nwcodepenalty=\Lhighpen
  int max_size_of_M = min(G.number_of_nodes()/2, 2*init());\nwcodepenalty=\Llowpen
  // the size of a greedy matching is at least half of the maximum matching.\nwcodepenalty=\Llowpen
  number_of_iterations = 0;\nwcodepenalty=\Llowpen
  node v; edge e;\nwcodepenalty=\Llowpen
  int count = 1; T = G.all_nodes();\nwcodepenalty=\Llowpen
  forall_nodes(v,G) num[v] = count++;\nwcodepenalty=\Llowpen
  while ( true )\{ \nwcodepenalty=\Llowpen
    \LA{}setting up w\RA{}\nwcodepenalty=\Llowpen
    number_of_iterations++;\nwcodepenalty=\Llowpen
    if( heur && (number_of_iterations > 0.5 * heur_factor * (max_size_of_M - size_of_M)) )\{\nwcodepenalty=\Llowpen
      \LA{}finish the computation by means of LEDA's matching algorithm\RA{}\nwcodepenalty=\Llowpen
      break; \nwcodepenalty=\Llowpen
    \}\nwcodepenalty=\Llowpen
    else\{\nwcodepenalty=\Llowpen
      if( phase_1() ) // returns true if there is an augmenting path\nwcodepenalty=\Llowpen
        phase_2();\nwcodepenalty=\Llowpen
      else break; \nwcodepenalty=\Llowpen
    \}\nwcodepenalty=\Llowpen
  \}\nwcodepenalty=\Llowpen
  \LA{}compute matching and odd-set-cover\RA{}\nwcodepenalty=\Lhighpen
  return M;\nwcodepenalty=\Lhighpen
\};\nwcodepenalty=\Llowpen
\vspace{\Lemptyline}\nwendcode{}\nwbegindocs{8}\nwdocspar
\subsection{Computation of the Matching and the Odd-Set-Cover}

When no augmenting path exists, the matching is maximum. The matching is encoded in \ensuremath{\mathit{mate}}. We iterate over all edges of $G$ and whenever the endpoints are mated, we add the edge to the output $M$. Since only one edge in a bundle of parallel edges should go into $M$, we unmate the endpoints. 

Once the maximum matching is determined, we also compute an odd-set-cover to witness its optimality. Recall that an odd-set-cover is a labeling of the nodes with non-negative integers such that
\begin{itemize}
\item the equality $\abs{M} = n_1 + \sum_{i \ge 2} \floor{n_i/2}$
  holds, where for $i \ge 0$, $n_i$ is the number of nodes labeled $i$, and
\item for every edge of $G$ either at least one endpoint is labeled one or both endpoints are labeled $i$ for some $i \ge 2$.
\end{itemize}
It is easy to see that $\abs{M}$ is at most the sum on the right-hand side for any matching $M$. Let $M_1$ be the edges in $M$ that are incident to a node labeled one and, for $i \ge 2$, let $M_i$ be the edges in $M$ that have both endpoints labeled $i$. Then $M = M_1 \cup \bigcup_{i \ge 2} M_i$ by the third property, $\abs{M_1} \le n_1$ by the second property, and $\abs{M_i} \le \floor{n_i/2}$ by the definition of $M_i$ for $i \ge 2$ and the fact that $\abs{M_i}$ is an integer. 

The odd-set-cover is readily computed. All nodes in a non-trivial blossom receive the same number and nodes in trivial blossoms receive the number zero if even and the number one if odd. Unlabeled nodes are matched and come in pairs. If there is no unlabeled node, we are done. If there are exactly two unlabeled nodes, we label one of them one and the other one zero. If there are more than two unlabeled nodes, we label one of the then one, and all others with two. We then start labeling the blossoms at three.

\nwenddocs{}\nwbegincode{9}\moddef{compute matching and odd-set-cover}\endmoddef\nwstartdeflinemarkup\nwenddeflinemarkup\nwcodepenalty=\Lhighpen
\vspace{\Lemptyline}M.clear(); // make empty list\nwcodepenalty=\Lhighpen
forall_edges(e,G)\{\nwcodepenalty=\Lhighpen
  node u = G.source(e); node v = G.target(e);\nwcodepenalty=\Llowpen
  if (mate[u] == v)\{ \nwcodepenalty=\Llowpen
    M.append(e);\nwcodepenalty=\Llowpen
    mate[u] = mate[v] = nil; // only one edge from a bundle of parallel edges\nwcodepenalty=\Llowpen
  \}\nwcodepenalty=\Llowpen
  \}\nwcodepenalty=\Llowpen
\vspace{\Lemptyline}OSC.init(G,-1);\nwcodepenalty=\Llowpen
\vspace{\Lemptyline}int number_of_unlabeled = 0;\nwcodepenalty=\Llowpen
node arb_u_node = nil;\nwcodepenalty=\Llowpen
\vspace{\Lemptyline}forall_nodes(v,G) \nwcodepenalty=\Llowpen
  if ( label[v] == UNLABELED )\{\nwcodepenalty=\Llowpen
    number_of_unlabeled++;\nwcodepenalty=\Llowpen
    arb_u_node = v;\nwcodepenalty=\Llowpen
  \}\nwcodepenalty=\Llowpen
\vspace{\Lemptyline}int L = 0;\nwcodepenalty=\Llowpen
if ( number_of_unlabeled > 0 )\{\nwcodepenalty=\Llowpen
  OSC[arb_u_node] = 1;\nwcodepenalty=\Llowpen
  if (number_of_unlabeled > 2) L = 2;\nwcodepenalty=\Llowpen
  forall_nodes(v,G) \nwcodepenalty=\Llowpen
    if ( label[v] == UNLABELED && v != arb_u_node ) OSC[v] = L;\nwcodepenalty=\Llowpen
\}\nwcodepenalty=\Llowpen
\vspace{\Lemptyline}int K = ( L == 0? 2 : 3);\nwcodepenalty=\Llowpen
\vspace{\Lemptyline}forall_nodes(v,G)\nwcodepenalty=\Llowpen
   if ( dbase(v) != v && OSC[dbase(v)] == -1 ) OSC[dbase(v)] = K++;\nwcodepenalty=\Llowpen
\vspace{\Lemptyline}forall_nodes(v,G)\nwcodepenalty=\Llowpen
  \{ if ( dbase(v) == v && OSC[v] == -1 )\nwcodepenalty=\Llowpen
    \{ if ( label[v] == EVEN ) OSC[v] = 0;\nwcodepenalty=\Llowpen
      if ( label[v] == ODD  ) OSC[v] = 1;\nwcodepenalty=\Llowpen
    \}\nwcodepenalty=\Lhighpen
    if ( dbase(v) != v ) OSC[v] = OSC[dbase(v)];\nwcodepenalty=\Lhighpen
  \}\nwcodepenalty=\Llowpen
\vspace{\Lemptyline}\nwendcode{}\nwbegindocs{10}\nwdocspar
\subsection{Phase One: The Search for a sap and the Construction of the Auxiliary Graph $H$}

We assume that the reader is familiar with the standard general matching algorithm as, for example, described in Section~7.7\ of~\cite{LEDAbook}. We give a brief review. The algorithm searches for an augmenting path by growing alternating trees rooted at free vertices. Vertices are either unlabeled, even, or odd. A vertex is \emph{unlabeled} if it does not belong to any tree, \emph{even} if there is an even length alternating tree path ending in it (there may also be an odd length alternating path ending in it), and \emph{odd} if all alternating tree paths ending in it have odd length. For any node one of the alternating paths ending in it is designated the \emph{canonical path} of the vertex. Roots are even and free and, at the beginning, all matched vertices are unlabeled. The canonical path of a root is the trivial path consisting only of the root. In each step of the search a non-matching non-trivial\footnote{Self-loops are not explored.} edge incident to an even vertex is explored, say $xy$ with $x$ even. We follow~\cite{Gabow:GeneralMatching} and use $xy$ for the edge to be explored; in~\cite{LEDAbook} we use $vw$. If there is a matching edge incident to $x$, it connects $x$ with its parent. 

If $y$ is unlabeled, we add $y$ and its mate to the search structure; $y$ becomes odd and its mate becomes even. The canonical paths are the canonical path of $x$ extended by $y$ and extended by $y$ and the mate of $y$, respectively. 

If $y$ is odd, we do nothing, as we have simply discovered another odd length path to $y$. The canonical path of $y$ is already defined. 

If $y$ is even and in a different tree, we have found an augmenting path. The path consists of the edge $xy$ plus the canonical paths from $x$ and $y$ to their respective roots.

If $y$ is even and in the same tree, we have discovered new even length alternating paths. Let $z$ be the lowest common ancestor of $x$ and $y$. $z$ is an even vertex. All odd vertices on the paths from $x$ and $y$ to $z$ become even. For example, consider a odd vertex on the path from $y$ to $z$. The even length path to it consists of the even length path ending in $x$ followed by the edge $xy$, followed by the odd length path from $y$ to the vertex (equivalently, the suffix of the canonical path of $y$ starting in the vertex). This path is also the canonical path of the vertex. The edge $xy$ is added to the search structure and all nodes on the paths from $x$ and $y$ to $z$ are contracted into $z$; one says that these nodes form a \emph{blossom} with $z$ as its \emph{base}. Contraction of blossoms ensures that the search structure stays a forest. Also, all unexplored edges connecting two vertices of the blossom become self-loops and will not be explored in the future. These edges do not become part of the search structure.

We use $B_x$ to denote the maximal blossom containing $x$. For uniformity of notation, a vertex is also considered a blossom. So, if $x$ was never part of a shrinking process, $B_x = \sset{x}$.

In the implementation, we use $\ensuremath{\mathit{label}[x]} \in \sset{\ensuremath{\mathit{EVEN}},\ensuremath{\mathit{ODD}},\ensuremath{\mathit{UNLABELED}}}$ for the vertex label. We use LEDA's node partition class \ensuremath{\mathit{node\nspaceunderscore\_partition}} to keep track of blossoms. A node partition keeps track of a partition of the vertices of a graph into disjoint blocks. Each block has a representative; in our case, the base of the blossom will be the representative of the block. \ensuremath{\mathit{node\nspaceunderscore\_partition}(G)\ \mathit{base}} creates a node-partition \ensuremath{\mathit{base}} for $G$ and makes each vertex a block of its own; the vertex is also the representative of the block. \ensuremath{\mathit{int}\ \mathit{base}\nspacedot.\mathit{same\nspaceunderscore\_block}(v,w)} returns a non-zero integer if $v$ and $w$ belong to the same block and $0$ otherwise, \ensuremath{\mathit{base}\nspacedot.\mathit{union\nspaceunderscore\_blocks}(v,w)} unites the blocks containing $v$ and $w$ (the canonical representative of the new block is the canonical representative of one of the blocks), \ensuremath{\mathit{node}\ \mathit{base}\nspacedot.\mathit{find}(v)} or even simpler \ensuremath{\mathit{node}\ \mathit{base}(v)} return the canonical representative node of the block containing \ensuremath{v}, \ensuremath{\mathit{base}\nspacedot.\mathit{make\nspaceunderscore\_rep}(v)} makes $v$ the canonical representative of the block containing $v$, and \ensuremath{\mathit{base}\nspacedot.\mathit{split}(G\nspacedot.\mathit{all\nspaceunderscore\_nodes}(\;))} restores the initial state. 

When a blossom is formed, we use union operations (either between each vertex of the blossom and the base of the blossom or between the endpoints of each edge forming the blossom). We do \textbf{not} change the label of the odd nodes in the blossom to even, we change the label only implicitly by always asking for the label of the base of a vertex instead of the label of the vertex itself, i.e, we write \ensuremath{\mathit{label}[\mathit{base}(v)]} instead of \ensuremath{\mathit{label}[v]}. For vertices in trivial blossoms, both calls give the same result, and for vertices in non-trivial blossoms, the former call returns \ensuremath{\mathit{EVEN}}. In this way, we save some updates and, more importantly, we still have the original label of each node. We will need the original label, when we construct augmenting paths in \ensuremath{\mathit{phase\nspaceunderscore\_2}}.

\subsubsection{Determining a $\sap$} The algorithm as outlined above finds some augmenting path if there is one. We need to find a shortest augmenting path, a $\sap$, and therefore need to explore edges in a judicious order. To this end, we organize the exploration of edges into $\Delta$-phases, where $\Delta = 0, 1, 2, \ldots$. In a $\Delta$-phase, a non-matching edge $xy$ incident to an even node $x$ is explored where either (see Lemma~\ref{characterization})
\begin{itemize}
\item $y$ is unlabeled, and the canonical path of $x$ has length $\Delta - 2$ (the canonical path to the mate of $y$ will then have length $\Delta)$ or
\item $y$ is even, and the sum of the lengths of the canonical paths to $x$ and $y$ is $2\Delta - 2$ (together with the edge $xy$, the length is $2\Delta - 1$).
\end{itemize}
In the former case, the search structure is grown by two edges, and in the latter case, the search structure is extended by one edge and either a blossom or an augmenting path has been found. We say that the edge $e = xy$ is explored at level $\Delta$. In the program, we refer to $e$ as a \emph{tight} edge. This concept is defined below. An augmenting path contains at most $n - 1$ edges. Thus $2 \Delta - 1 \le n - 1$ when an augmenting path is found. Thus we have $2\Delta \le n$ as the condition of the while-loop. The function returns true if an augmenting path exists, and false otherwise.

\nwenddocs{}\nwbegincode{11}\moddef{helper functions}\endmoddef\nwstartdeflinemarkup\nwenddeflinemarkup\nwcodepenalty=\Lhighpen
bool phase_1()\{\nwcodepenalty=\Lhighpen
  node v; edge e; // the generic vertex and edge\nwcodepenalty=\Lhighpen
  Delta = 0; int n = G.number_of_nodes();\nwcodepenalty=\Llowpen
  bool found_sap = false;\nwcodepenalty=\Llowpen
  \LA{}local declarations in phase\_1\RA{}\nwcodepenalty=\Llowpen
  while (2 * Delta <= n)\{ \nwcodepenalty=\Llowpen
    while (\LA{}there is a tight unexplored edge e at level Delta incident to an even node\RA{})\{\nwcodepenalty=\Llowpen
      node x = G.source(e), y = G.target(e); // one of the endpoints must be even, none odd\nwcodepenalty=\Llowpen
      if ( label[base(x)] != EVEN) swap(x,y);\nwcodepenalty=\Llowpen
      if (y == mate[x] || base(x) == base(y) || label[base(y)] == ODD) continue;\nwcodepenalty=\Llowpen
        // only non-matching and non-self-loops and y not odd\nwcodepenalty=\Llowpen
      if ( label[base(y)] == UNLABELED )\{\nwcodepenalty=\Llowpen
        \LA{}grow step\RA{}\nwcodepenalty=\Llowpen
      \}\nwcodepenalty=\Llowpen
      else\nwcodepenalty=\Llowpen
        if ( label[base(y)] == EVEN )\{// chunk sets found_sap if an augmenting path is found\nwcodepenalty=\Llowpen
        \LA{}blossom step or augmentation\RA{}\nwcodepenalty=\Llowpen
      \}\nwcodepenalty=\Llowpen
    \} // there is no further tight edge and no augmenting path at level Delta\nwcodepenalty=\Llowpen
    if (found_sap)\{\nwcodepenalty=\Llowpen
      \LA{}augmentation: construction of H\RA{}\nwcodepenalty=\Llowpen
      return true;\nwcodepenalty=\Llowpen
    \}\nwcodepenalty=\Llowpen
    \LA{}commit to unions\RA{}   // see section 3.4 Construction of H\nwcodepenalty=\Llowpen
    Delta++;\nwcodepenalty=\Llowpen
  \} // end Delta-Loop\nwcodepenalty=\Lhighpen
  return false; // no augmenting path found\nwcodepenalty=\Lhighpen
\};\nwcodepenalty=\Llowpen
\vspace{\Lemptyline}\nwendcode{}\nwbegindocs{12}The remainder of the section expands on the introduction; there will be some repetition.  Exploring edges
in a carefully chosen order is at the core of Edmonds' weighted matching algorithm~\cite{Edmonds:matching}. Since we have only edge weights zero and two, the full generality of Edmonds' algorithm is not needed.  It uses augmenting paths of maximum  weight, where the weight $w(P)$ of an alternating path is defined as the total weight of the edges in $P \setminus M$ minus the total weight of the edges in $P \cap M$,
\begin{equation}\label{weight of path 2}   w(P) = w(P \setminus M) - w(P \cap M), \end{equation}
i.e., the increase in weight of $M$ obtained by augmenting $P$. The weighted matching algorithm iteratively constructs maximum weight matchings of increasing cardinality.
We define
\[     w(e) = \begin{cases} 2 & \text{if $e \in M$} \\
    0 & \text{if $e \not\in M$.}
  \end{cases} \]
Then a maximum weight augmenting path is an augmenting path with a minimum number of matching edges, i.e., a $\sap$. Note that the weight of an augmenting path is the negative of twice the number of matching edges contained in it. 

The weighted matching algorithm originates from the formulation of the matching problem as a linear program. The variables of the dual play a crucial role. We have a dual value\footnote{Gabow uses $y$ for the vertex duals and somehow manages to never write $y(y)$ for the dual value of vertex $y$. We prefer to use $d$ for the duals of vertices.} $d(v)$ for each vertex $v$ and a non-negative dual value $z(B)$ for each non-trivial blossom. The duals \emph{dominate} an edge $e = uv$ if
\begin{equation}\label{reduced weight 2}    \hat{w}(e) = d(u) + d(v) + \sum_{u,v \in B} z(B) - w(e) \ge 0.\end{equation}
We call $\hat{w}(e)$ the \emph{reduced weight} of $e$. An edge $e$ is \emph{tight} if $\hat{w}(e) = 0$. We initialize $d(v)$ to one for all $v$. Then matching edges are tight and non-matching edges are non-tight. 

\emph{The search maintains the invariant that all edges are dominated and that all edges in the search structure and in $M$ are tight. It grows the trees by exploring a tight non-matching edge incident to an even vertex.} We will show in Lemma~\ref{equivalence} that the dual values of nodes reflect the length of canonical paths: If $v$ is a node in a tree with root $f$, then, for an even node, $d(v) - d(f)$ is the length of the canonical path to $v$, and, for an odd node, $-d(v) - d(f) + 1$ is the length of the canonical path to $v$. 

The trees are grown until either an augmenting path is found or there are no further tight non-matching edges incident to an even vertex. In the latter case, a dual update might make further edges tight. It could also be that the matching is maximum. We come back to this alternative below. We change the duals as follows:
\begin{align*}
  d(v) &= \begin{cases} d(v) - 1 & \text{if $v$ is even}\\
    d(v) + 1 & \text{if $v$ is odd.}
          \end{cases}\\
  z(B) &= z(B) + 2  \quad\text{for every maximal non-trivial blossom.}
\end{align*}
The effect of the update on the reduced weight $\hat{w}(e)$ of an edge $e = uv$ is as follows:
\begin{enumerate}
  \item Both endpoints already belong to the search structure: If both endpoints are odd, the reduced weight  increases, if one endpoint is even and one endpoint is odd and hence the edge is not contained in a blossom (recall that all nodes contained in a non-trivial blossom are even), the reduced weight does not change, if both endpoints are even and contained in the same maximal blossom, the reduced weight does not change, and if both endpoints are even and contained in different maximal blossoms, the reduced weight decreases by two. Lemma~\ref{properties of dual values} shows that reduced weights stay non-negative. Also, tight edges in the search structure stay tight, because such edges either connect an even and an odd node or two even nodes in the same maximal blossom. 

\item No endpoint already belongs to the search structure: The reduced weight does not change.

\item Exactly one endpoint belongs to the search structure: If this endpoint is odd, the reduced weight goes up. If this endpoint is even, the reduced weight decreases by one.
\end{enumerate}

\begin{figure}[t]
  \includegraphics[width=\textwidth,trim=60 355 60 100,clip=]{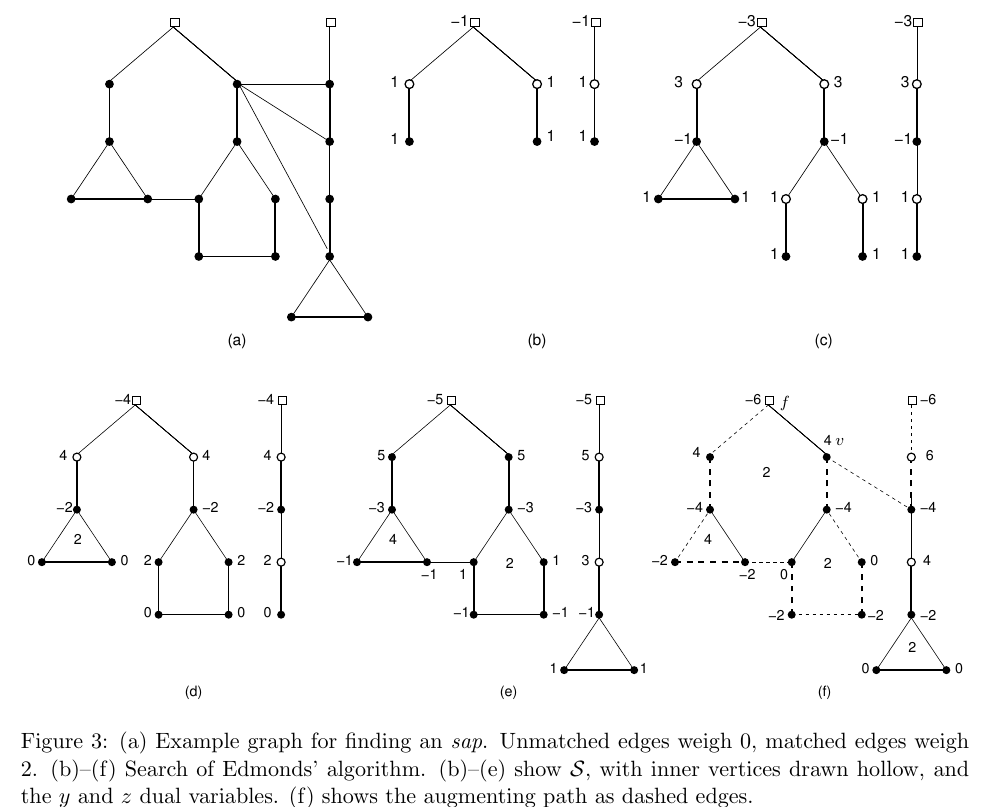}
  \caption{Reprinted from~\cite{Gabow:GeneralMatching}. Note that between (b) and (c), there are two updates of the vertex duals. The z-values of blossoms are shown inside the blossom. The augmenting path has length 13. It consists of 7 non-matching edges and 6 matching edges. The total weight of the matching edges in the path is 12. }
  \label{while-loop in first phase}
\end{figure}

Figure~\ref{while-loop in first phase} illustrates the tree growing process and the dual updates. Next, we state basic properties of dual values and then give some more intuition about the dual values.

\begin{lemma}\label{properties of dual values} $z$-values are always even and the $d$-values of all vertices in the search structure have the same parity. When a tight edge $xy$ is explored from $x$, we have $d(y) = -d(x)$. In a grow step, we have $d(x) = -1$. All free vertices have the same dual value $1 - \Delta$.  The reduced weight of an edge having both endpoints in the search structure is even. \end{lemma}
\begin{proof} The first claim is obvious as $z$-values are always increased by two. The second claim is true initially as all $d$-values are initialized to one. When $d$-values are changed, the $d$-values of even vertices are incremented and the $d$-values of odd nodes are decremented and the claim stays true. When a tight edge $e = xy$ is explored, $x$ and $y$ do not belong to the same blossom and $e$ is unmatched. Thus $d(x) + d(y) = 0$. In a grow step, $d(y) = 1$ and hence $d(x) = -1$. The $d$-value of the mate of $y$ is also one.

  The initial dual value of a free vertex in $+1$ and free vertices are even. Every dual update decrements the value.

  If an edge has both endpoints in the search structures, the parity of the dual values of the endpoints is the same. Thus the reduced weight is even.\end{proof}

\begin{lemma}\label{equivalence} Let $v$ be any vertex in the search structure and let $f$ be the root of the tree containing $v$. If $v$ is even, $d(v) - d(f)$ is the length of the canonical path from $f$ to $v$, if $v$ is odd, $-d(v) - d(f) + 1$ is the length of the canonical path from $f$ to $v$.\end{lemma}
\begin{proof} Note first, that dual updates do not change the values of $d(v) - d(f)$ and $-d(v) - d(f) - 1$. So, we only have to prove the claim for the point in time, when $v$ got is current label. If $v$ is the root of a search structure, $v$ is even, $d(v) - d(f) = 0$, and the length of the path is zero.

  If $v$ is odd, then $v$ was added in a grow step, say from node $x$. We have $d(v) + d(x) = 0$, the length of the path to $x$ is $d(x) - d(f)$, and the length of the path to $v$ is one more. Thus the length is $d(x) - d(f) + 1 = -d(v) - d(f) + 1$. If $v$ is even and was added in a grow step, its mate $m$ is odd and added in the same grow step, and $d(v) = d(m) = 1$. The path to $v$ is one longer than the path to $m$. Thus the length of the path is $1  - d(m) - d(f) +1  = d(v) - d(f)$.

Finally, if $v$ became even after the addition of the bridge $xy$, $d(x) + d(y) = 0$, and the length of the path is the length of the path from $f$ to $x$ plus 1 plus the length of the path from $y$ to $v$. Thus the length is $(d(x) - d(f)) + 1 + (d(y) - d(f)) - (-d(v) - d(f) + 1) = d(v) - d(f)$. \end{proof}

\begin{lemma}\label{characterization} Let $x$ be an even node and $xy$ a tight edge incident to $x$ in a $\Delta$-phase. If $y$ is unlabeled, the canonical path to $x$ has length $\Delta - 2$, if $y$ is even, the sum of the length of the canonical paths to $x$ and $y$ is $2\Delta - 2$. 
\end{lemma}
\begin{proof} The initial $d$-value of a free node is 1. The value is decreased by one in each dual update. Thus $d(f) = 1 - \Delta$ in a $\Delta$-phase.
  
In a grow step $x$ along edge $xy$, we have $d(y) = 1$, $d(x) + d(y) = 0$ and the path to $x$ has length $d(x) - d(f) = -1 - 1 + \Delta = \Delta - 2$.

In a blossom step or augmentation via the edge $xy$, we have $d(x) + d(y) = 0$, the paths to $x$ and $y$ have length $d(x) - d(f)$ and $d(y) - d(f')$ respectively (here $f = f'$ in a blossom step and $f \not= f'$ in a augmentation), and $d(f) = d(f') = 1 - \Delta$. Thus the sum of the length of the canonical paths of $x$ and $y$ are $d(x) - d(f) + d(y) - d(f')= 2\Delta - 2$.
\end{proof}

\begin{figure}
  \begin{center}
    \includegraphics[width = 0.4\textwidth,trim=0 150 100 250,clip=]{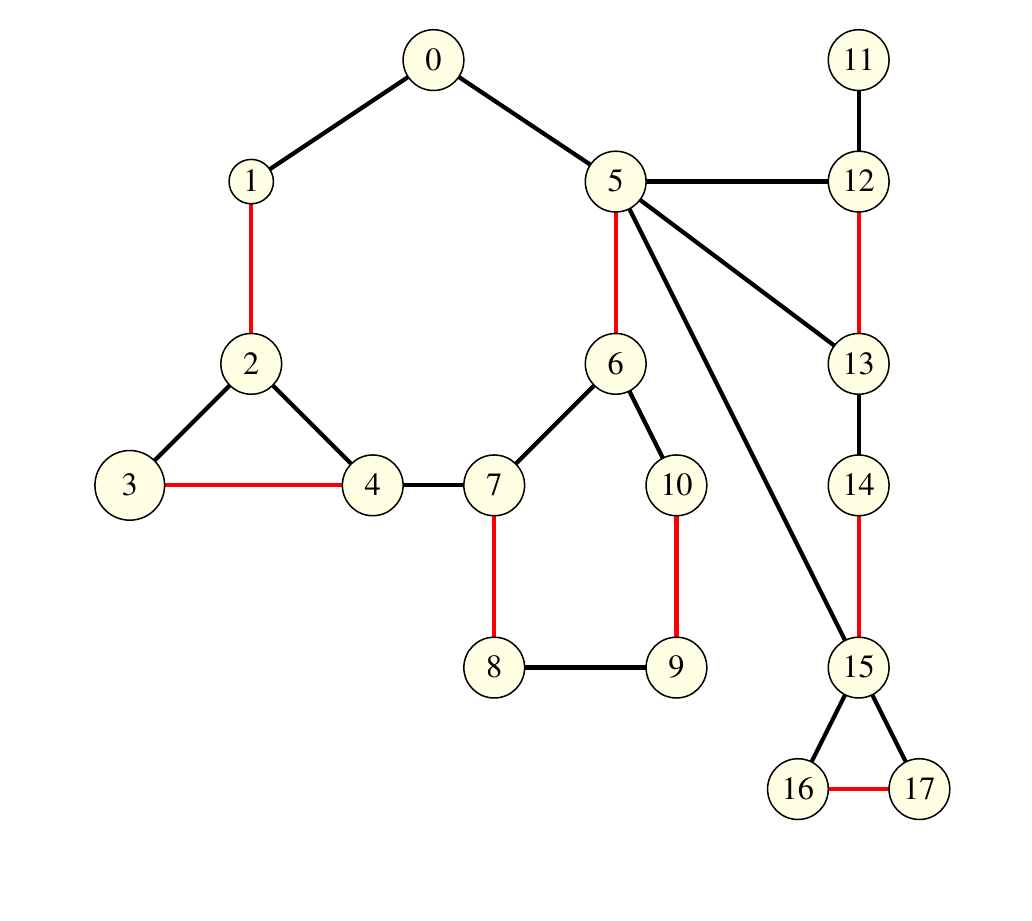}\qquad\includegraphics[width = 0.4\textwidth,trim=100 100 100 200,clip=]{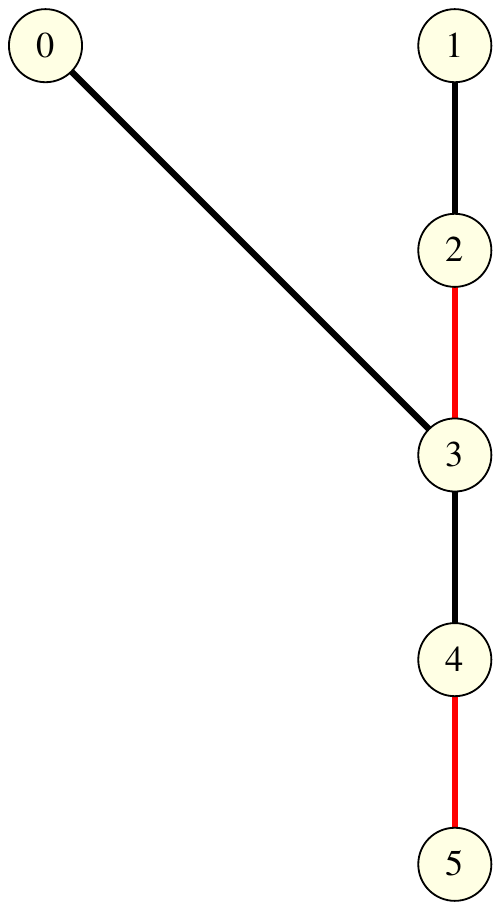}
    \end{center}
  \caption{The original and the contracted graph for the graph in Figure~\ref{while-loop in first phase}. The original graph is shown on the left with matching edges in red. The contracted graph is shown on the right. Nodes 0 to 10 of $G$ are contracted into node 0 of H and nodes 15, 16, and 17 are contracted into node 5. The edges $(5,12)$ and $(5,15)$ are not tight and hence are not added to $H$.}\label{Graph H}
\end{figure}

Call a blossom \emph{positive} if it contained (not necessarily proper) in a blossom with positive $z$-value. A blossom is positive iff it was formed before the last dual adjustment. For a blossom $B$ and an path $P$, $\gamma(B,P)$ is the part of $P$ using edges of $B$.  

\begin{lemma}[Lemma 3.1 and Corollary 3.2 in~\cite{Gabow:GeneralMatching}]\mbox{}\label{structure of saps}
  \begin{itemize}
  \item At all times during the execution of the while-loop in \ensuremath{\mathit{phase\nspaceunderscore\_1}}: $w(P) \le 2d(f)$ for any augmenting path $P$ and any free node $f$.
  \item An augmenting path has maximum weight iff all its edges are tight and for every blossom $B$, $\gamma(B,P)$ is an even-length alternating path (and hence passes through the base of the blossom). 
  \end{itemize}
\end{lemma}
\begin{proof} Substituting~\lref{reduced weight} into~\lref{weight of path} for all edges of the path yields
\begin{equation}\label{upper bound on w(P)}, 2 w(P) \le \sum_{uv \in P \setminus M} (d(u) + d(v) + \sum_{u,v \in B} z(B)) - \sum_{uv \in P \cap M} (d(u) + d(v) + \sum_{u,v \in B} z(B))\end{equation}
Now observe that the $d$-values of all internal vertices of $P$ cancel and that the contribution of a blossom $B$ is zero if $P$ uses an even number of edges of the blossom and is negative if it uses an odd number of edges of the blossom.

For the second item, observe that the path determined by the first phase satisfies~\lref{upper bound on w(P)} with equality as all edges of the path are tight.\footnote{An alternative argument uses Lemma~\ref{characterization}. Let $L$ be the length of the augmenting path $P$ found by the algorithm. Then $1 - w(P) = L = 1 + 2\Delta - 2 = 1 - 2d(f)$.}   Hence any other sap must also satisfy it with equality.  \end{proof}

\begin{lemma} If the matching is not maximum, the while-loop terminates with a $\sap$ in a $\Delta$-phase where $2\Delta \le n$.  \end{lemma}
\begin{proof} Let $P$ be a $\sap$. Then $\abs{P}$ is 1 plus twice the number of matching edges in $P$ or $1 - w(P)$. So the dual values of free vertices cannot go below $-w(P)/2$ and hence the growth of the search structure must come to a halt. If the matching is not maximum, an augmenting path exists. Hence the search stops having found an $\sap$.

The length of a $\sap$ is at most $n - 1$. When a $\sap$ is found in a $\Delta$-phase, the $\sap$ has length $2\Delta - 1$. Thus $2\Delta \le n$.  \end{proof}

\subsubsection{From One to All $\sap$s} At this point, we have determined one $\sap$ but our goal is to construct a graph $H$ containing all $\sap$s. The growth of the search structure is non-deterministic in the sense that the choice of which tight edge to explore is arbitrary. However, the state just after the last dual update is unique\footnote{The maximal positive blossoms are precisely the non-trivial biconnected components of the tight edges with at least one endpoint even.}. We therefore (implicitly) revert to this state and form $H$ from $G$ by contracting every maximal positive blossom\footnote{In other words, the vertices of $H$ are the maximal positive and the trivial blossoms.} and keeping only the tight edges that join distinct non-odd vertices.

\begin{lemma}\label{H and G}[Corollary 3.3 in~\cite{Gabow:GeneralMatching}] A set of edges $P$ forms an augmenting path in $H$ iff it is the image of a $\sap$ $Q$ in $G$. \end{lemma}
\begin{proof}(Sketch) Consider any $\sap$ $Q$ in $G$ and any maximal positive blossom $B$ with $\gamma(P,B) \not= \emptyset$. By Lemma~\ref{structure of saps}, $\gamma(P,B)$ is an even-length alternating path connecting the base of $B$ to some vertex of $B$. So contraction of all such $\gamma(P,.)$ yields an augmenting path in $H$. Note that no edge of $Q$ can have both endpoints odd. 

  Conversely, consider an augmenting path $P$ in $H$ and any vertex of $H$ on $P$. If it represents a trivial blossom, the incident edges in $H$ correspond to edges in $G$. If it represents a maximal positive blossom $B$, one of the incident edge in $H$ is non-matching and the other (if it exists) is matching. Let $v$ be the endpoint in $B$ of the preimage of the non-matching edge. Then there is a unique even-length alternating path connecting $v$ to the base of $B$. The base of $B$ is free in $G$ if no matching edge is incident to the contraction of $B$ in $G$. If there is a matching edge incident to the contraction of $B$ in $P$, its preimage must be the matching edge incident to the base of $B$ in $G$. Thus we can lift $P$ to a path $Q$ in $G$. The path is a $\sap$ by Lemma~\ref{structure of saps}. \end{proof}

Note that the second part of the proof of Lemma~\ref{H and G} shows how to lift augmenting paths from $H$ to $G$.

\subsubsection{Details of the While-Loop in \ensuremath{\mathit{phase\nspaceunderscore\_1}}}

\paragraph{Dual Values and Tight Edges:} The dual values of vertices are maintained implicitly\footnote{Note a dual update might change many dual values and that there might be $\Omega(n)$ dual updates in phase one. So maintaining dual values explicitly would result in running time $\Omega(n^2)$ for \ensuremath{\mathit{phase\nspaceunderscore\_1}}.}. We have a counter $\Delta$ (\ensuremath{\mathit{Delta}} in the program) that counts the number of dual adjustments. For each vertex $v$, we keep two values \ensuremath{\mathit{bd}[v]} (\emph{base dual}) and \ensuremath{\mathit{bDelta}[v]} (\emph{base Delta})  in which we store the values of $d[v]$ and $\Delta$ at the moment of time, when $v$ received its current label; \ensuremath{\mathit{bd}[v]} is one for unlabeled vertices. Then
\[    d[v] = \begin{cases}  \ensuremath{\mathit{bd}[v]} - (\Delta - \ensuremath{\mathit{bDelta}[v]})  &  \text{if $v$ is even} \\
    \ensuremath{\mathit{bd}[v]} + (\Delta - \ensuremath{\mathit{bDelta}[v]}) & \text{if $v$ is odd} \\
    1  & \text{if $v$ is unlabeled.} \end{cases} \]
When $v$ becomes labeled for the first time, we set $\ensuremath{\mathit{bd}[v]}$ to one and $\ensuremath{\mathit{bDelta}[v]}$ to the current value of $\Delta$, and when $v$ changes its label from odd to even, we set $\ensuremath{\mathit{bd}[v]}$ to $1 + (\Delta - \ensuremath{\mathit{bDelta}[v]})$ and \ensuremath{\mathit{bDelta}[v]} to the current value of $\Delta$. In order to determine whether a node is even or odd, we ask whether its base is even or odd. Note that an odd node is its own base, bases are even always, and a node is even iff is base is even.

\nwenddocs{}\nwbegincode{13}\moddef{helper functions}\plusendmoddef\nwstartdeflinemarkup\nwenddeflinemarkup\nwcodepenalty=\Lhighpen
int d(const node& v)\nwcodepenalty=\Lhighpen
\{\nwcodepenalty=\Lhighpen
  if (label[base(v)] == UNLABELED) return 1;\nwcodepenalty=\Llowpen
  if (label[base(v)] == EVEN)      return bd[v] - (Delta - bDelta[v]);\nwcodepenalty=\Lhighpen
  return bd[v] + (Delta - bDelta[v]);\nwcodepenalty=\Lhighpen
\}\nwcodepenalty=\Llowpen
\vspace{\Lemptyline}\nwendcode{}\nwbegindocs{14}There is no need to store the dual values of blossoms. Note that in the algorithm (not in the proof!), we are only interested in the reduced weight of edges running between distinct maximal blossoms. For them the contribution of the blossom duals to the reduced weight is zero.

So consider a non-matching edge $e = xy$ with even endpoint $x$, and $x$ and $y$ in distinct maximal blossoms.  A dual update decreases $\hat{w}(e)$ by one if $y$ is unlabeled, leaves $\hat{w}(e)$ unchanged if $y$ is odd, and decreases $\hat{w}(e)$ by two if $y$ is even. So, if we (re)compute the reduced weight of $e$ whenever the label of one of its endpoints changes, we know how many further dual updates are needed until the edge becomes tight. Namely $d(x) + d(y)$ if $y$ is unlabeled and $(d(x) + d(y))/2$ if $y$ is even. Note that in the latter case, $d(x)$ and $d(y)$ have the same parity and hence $d(x) + d(y)$ is even.

\paragraph{The Priority Queue:} So a simple priority queue suffices: For each integer $k$, we maintain the set of edges that will become tight when $\Delta$ reaches $k$ in a linear list, and we store the lists in an array. What size of array is needed? A free vertex $f$ is even from the beginning of \ensuremath{\mathit{phase\nspaceunderscore\_1}} and hence $d(f) = 1 - \Delta$. Assume an augmenting $P$ exists. It can contain at most $n - 1$ edges and hence
\[  n - 1 \ge \abs{P} = -w(P) + 1 \ge -2d(f) + 1.\]
Thus $d(f) \ge -(n - 2)/2 = -n/2 + 1$ and hence $\Delta \le n/2$. So an array with entries $0$ to $\floor{n/2}$ suffices. When $\Delta$ exceeds $n/2$, there is no augmenting path. We are now ready for the definition of the priority queue.

\nwenddocs{}\nwbegincode{15}\moddef{priority\_queue.h}\endmoddef\nwstartdeflinemarkup\nwenddeflinemarkup\nwcodepenalty=\Lhighpen
\vspace{\Lemptyline}class simple_queue\{\nwcodepenalty=\Lhighpen
private:\nwcodepenalty=\Lhighpen
  int D = 0;;   // Q[0] to Q[D-1] are already empty\nwcodepenalty=\Llowpen
  int max; // maximum value + 1\nwcodepenalty=\Llowpen
  array<list<edge> > Q;\nwcodepenalty=\Llowpen
public:\nwcodepenalty=\Llowpen
simple_queue(const graph& G) :\nwcodepenalty=\Llowpen
    max(G.number_of_nodes()/2 + 1), Q(array<list<edge> >(max)) \{\nwcodepenalty=\Llowpen
\};\nwcodepenalty=\Llowpen
\vspace{\Lemptyline}void init()\{\nwcodepenalty=\Llowpen
  for (int i = 0; i < max; i++)\nwcodepenalty=\Llowpen
    Q[i].clear();\nwcodepenalty=\Llowpen
  D = 0;\nwcodepenalty=\Llowpen
\}\nwcodepenalty=\Llowpen
\vspace{\Lemptyline}void insert(edge e, int d)\{// insert e into queue Q[d]\nwcodepenalty=\Llowpen
  if (d >= max) return;\nwcodepenalty=\Llowpen
  Q[d].append(e);\nwcodepenalty=\Llowpen
\};\nwcodepenalty=\Llowpen
\vspace{\Lemptyline}edge delete_at_Delta(const int& Delta)\{\nwcodepenalty=\Llowpen
  if (Delta > D) D = Delta;\nwcodepenalty=\Llowpen
  if (Delta >= max || Q[Delta].empty()) return nil;\nwcodepenalty=\Llowpen
  return Q[Delta].pop();\nwcodepenalty=\Lhighpen
\};\nwcodepenalty=\Lhighpen
\};\nwcodepenalty=\Llowpen
\vspace{\Lemptyline}\nwendcode{}\nwbegindocs{16}The selection of a tight unexplored edge $e$ incident to an even node is simple. We ask the priority queue to give us one.

\nwenddocs{}\nwbegincode{17}\moddef{there is a tight unexplored edge e at level Delta incident to an even node}\endmoddef\nwstartdeflinemarkup\nwenddeflinemarkup\nwcodepenalty=\Lhighpen
(e = PQ.delete_at_Delta(Delta)) != nil\nwcodepenalty=\Lhighpen
\vspace{\Lemptyline}\nwendcode{}\nwbegindocs{18}\nwdocspar
\paragraph{Reinitialization:} At the beginning of \ensuremath{\mathit{phase\nspaceunderscore\_1}}, we reinitialize $\Delta$, the priority queue \ensuremath{\mathit{PQ}}, the node partitions \ensuremath{\mathit{base}} and \ensuremath{\mathit{dbase}}, and the node labels \ensuremath{\mathit{label}}. It suffices to reinitialize the labels and the node partitions for the nodes that were added to the search structure in the previous iteration. We collected these nodes in the node list $T$. The free nodes for the next iteration are the nodes in $T$ that are not matched. For them we put the incident edges into the priority queue. Matched nodes can be deleted from $T$. 
There is no need to reinitialize \ensuremath{\mathit{bd}}, \ensuremath{\mathit{bDelta}}, \ensuremath{\mathit{parent}}, \ensuremath{\mathit{source\nspaceunderscore\_bridge}}, and \ensuremath{\mathit{target\nspaceunderscore\_bridge}}.

\nwenddocs{}\nwbegincode{19}\moddef{local declarations in phase\_1}\endmoddef\nwstartdeflinemarkup\nwenddeflinemarkup\nwcodepenalty=\Lhighpen
\vspace{\Lemptyline}PQ.init();\nwcodepenalty=\Lhighpen
\vspace{\Lemptyline}// reinitialize the part of the node-partitions used in the previous iteration. \nwcodepenalty=\Lhighpen
base.split(T);\nwcodepenalty=\Llowpen
dbase.split(T);\nwcodepenalty=\Llowpen
\vspace{\Lemptyline}forall(v,T) label[v] = ((mate[v] == nil)? EVEN : UNLABELED);\nwcodepenalty=\Llowpen
\vspace{\Lemptyline}list_item it;\nwcodepenalty=\Llowpen
forall_items(it,T)\nwcodepenalty=\Llowpen
  if ( mate[T[it]] == nil )\nwcodepenalty=\Llowpen
    forall_inout_edges(e,T[it]) scan_edge(e,T[it]);\nwcodepenalty=\Lhighpen
  else\nwcodepenalty=\Lhighpen
  T.del_item(it);\nwcodepenalty=\Llowpen
\nwendcode{}\nwbegindocs{20}\nwdocspar

\paragraph{Grow Step:}

The edge $e = xy$ goes from the even vertex $x$ to the unlabeled vertex $y$. We add $y$ and its mate, call it $z$, to the search structure; $y$ becomes odd and $z$ becomes even. We iterate over all non-matching edges incident to $z$ and insert the ones going to an unlabeled or an even node into the priority queue. We also add the nodes to $T$.

\nwenddocs{}\nwbegincode{21}\moddef{grow step}\endmoddef\nwstartdeflinemarkup\nwenddeflinemarkup\nwcodepenalty=\Lhighpen
node z = mate[y];\nwcodepenalty=\Lhighpen
bd[y] = bd[z] = 1;  \nwcodepenalty=\Lhighpen
bDelta[y] = bDelta[z] = Delta;\nwcodepenalty=\Llowpen
parent[z] = y; parent[y] = x;\nwcodepenalty=\Llowpen
label[y] = ODD; label[z] = EVEN;\nwcodepenalty=\Llowpen
T.push(y); T.push(z);\nwcodepenalty=\Lhighpen
forall_inout_edges(e,z)\nwcodepenalty=\Lhighpen
  scan_edge(e,z); \nwcodepenalty=\Llowpen
\vspace{\Lemptyline}\nwendcode{}\nwbegindocs{22}where \ensuremath{\mathit{scan\nspaceunderscore\_edge}} is defined in the following chunk. Note that $p$ or $p/2$ further dual updates are needed until $e$ becomes tight depending on whether $u$ is unlabeled or even. 

\nwenddocs{}\nwbegincode{23}\moddef{helper functions}\plusendmoddef\nwstartdeflinemarkup\nwenddeflinemarkup\nwcodepenalty=\Lhighpen
void scan_edge(const edge& e, const node& z)\{\nwcodepenalty=\Lhighpen
  node u = G.opposite(e,z);\nwcodepenalty=\Lhighpen
  if (mate[u] == z || label[base(u)] == ODD) return; \nwcodepenalty=\Llowpen
  int p = d(z) + d(u); \nwcodepenalty=\Llowpen
  if (label[u] == UNLABELED)\nwcodepenalty=\Llowpen
    PQ.insert(e,Delta + p);\nwcodepenalty=\Llowpen
  else\nwcodepenalty=\Llowpen
    PQ.insert(e,Delta + p/2);\nwcodepenalty=\Lhighpen
  return;\nwcodepenalty=\Lhighpen
\}\nwcodepenalty=\Llowpen
\vspace{\Lemptyline}\nwendcode{}\nwbegindocs{24}\paragraph{Blossom Step or Augmentation:}

The edge $e = xy$ connects two even nodes that are not contained in the same blossom. If they belong to the same tree, we have discovered a new blossom, if they belong to distinct trees, we have found an augmenting path. In order to decide whether $x$ and $y$ belong to the same tree or not, we walk down the paths from $x$ and $y$ in lock-step fashion until we find a node $z$ lying on both paths or we reach two roots.

In the former case, $z$ be the lowest common ancestor of $x$ and $y$. We contract all nodes on the paths from $x$ and $y$ back to $z$ into a blossom with base $z$. All odd nodes on both paths become even and their incident edges will be checked for addition to the priority queue.

There is nothing new here and our presentation follows the LEDA-book~\cite[page 407]{LEDAbook}. We use a counter \ensuremath{\mathit{strue}}. We increment it and then label all nodes on the paths from $x$ and $y$ towards the root with \ensuremath{\mathit{strue}}. We walk the path in lock-step fashion and stop once we encounter a node that is already labeled \ensuremath{\mathit{strue}}.

\nwenddocs{}\nwbegincode{25}\moddef{blossom step or augmentation}\endmoddef\nwstartdeflinemarkup\nwenddeflinemarkup\nwcodepenalty=\Lhighpen
strue++;\nwcodepenalty=\Lhighpen
node hx = base(x), hy = base(y);\nwcodepenalty=\Lhighpen
path1[hx] = path2[hy] = strue;\nwcodepenalty=\Llowpen
\vspace{\Lemptyline}while( (path1[hy] != strue && path2[hx] != strue) && (mate[hx] != nil || mate[hy] != nil))\{\nwcodepenalty=\Llowpen
// hy does not lie on the first path, hx does not lie on the second, and one is not free\nwcodepenalty=\Llowpen
  if (mate[hx] != nil)\{\nwcodepenalty=\Llowpen
    hx = base(parent[mate[hx]]);\nwcodepenalty=\Llowpen
    path1[hx] = strue;\nwcodepenalty=\Llowpen
  \}\nwcodepenalty=\Llowpen
  if (mate[hy] != nil)\{ \nwcodepenalty=\Llowpen
    hy = base(parent[mate[hy]]);\nwcodepenalty=\Llowpen
    path2[hy] = strue;\nwcodepenalty=\Llowpen
  \}\nwcodepenalty=\Llowpen
\}\nwcodepenalty=\Llowpen
\vspace{\Lemptyline}if (path1[hy] == strue || path2[hx] == strue)\{\nwcodepenalty=\Llowpen
  \LA{}shrink blossom\RA{}\nwcodepenalty=\Llowpen
\}\nwcodepenalty=\Llowpen
else\{\nwcodepenalty=\Lhighpen
  found_sap = true;\nwcodepenalty=\Lhighpen
\}\nwcodepenalty=\Llowpen
\vspace{\Lemptyline}\nwendcode{}\nwbegindocs{26}We rephrase~\cite[page 407]{LEDAbook}. Let us see how we shrink a blossom. The base $b$ of the blossom is either \ensuremath{\mathit{hx}} or \ensuremath{\mathit{hy}}, depending on which of the two nodes also lies on the other path. We call \ensuremath{\mathit{shrink\nspaceunderscore\_path}(b,x,y,\nspacedot.\nspacedot.\nspacedot.)} to shrink the path from $x$ to $b$. The call also has the other end of the edge that closes the blossom as an argument.

\nwenddocs{}\nwbegincode{27}\moddef{shrink blossom}\endmoddef\nwstartdeflinemarkup\nwenddeflinemarkup\nwcodepenalty=\Lhighpen
\vspace{\Lemptyline}node b = (path1[hy] == strue) ? hy : hx; // base\nwcodepenalty=\Lhighpen
\vspace{\Lemptyline}shrink_path(b,x,y,dunions); \nwcodepenalty=\Lhighpen
shrink_path(b,y,x,dunions); \nwcodepenalty=\Llowpen
\vspace{\Lemptyline}\nwendcode{}\nwbegindocs{28}Again, we rephrase~\cite[page 407]{LEDAbook}. When an edge $xy$ closes a blossom, all odd nodes in the blossom also get an even length path to the root of their alternating tree. This path goes through the edge that closes the blossom. We call this edge the \emph{bridge} of the blossom. The odd nodes on the tree path from $x$ to $b$, use the bridge in the direction from $x$ to $y$ and the \ldots. We use the node arrays \ensuremath{\mathit{source\nspaceunderscore\_bridge}} and \ensuremath{\mathit{target\nspaceunderscore\_bridge}} to record the source and the target of the bridge for each odd node shrunken into the new blossom. The details of collapsing the tree path from $x$ to $b$ into $b$ are now simple. For each node of the path $v$ on the path, we perform \ensuremath{\mathit{union\nspaceunderscore\_blocks}(v,b)} to union the blocks containing $v$ and $b$, for each odd node, we set \ensuremath{\mathit{source\nspaceunderscore\_bridge}} to $x$ and \ensuremath{\mathit{target\nspaceunderscore\_bridge}} to $y$. After a union-operation the representative of the newly formed block is undefined. We want it to be $b$ and therefore call \ensuremath{\mathit{make\nspaceunderscore\_rep}(b)}.

Some additional steps are required. The previously odd nodes reset their \ensuremath{\mathit{bd}} and \ensuremath{\mathit{bDelta}} values. We also scan their incident edges for addition to the priority queue. We do \textbf{not} change the label of the odd nodes that become even. Recall that we make this change only implicitly by always asking for the label of the base of a node.

\nwenddocs{}\nwbegincode{29}\moddef{helper functions}\plusendmoddef\nwstartdeflinemarkup\nwenddeflinemarkup\nwcodepenalty=\Lhighpen
\vspace{\Lemptyline}void shrink_path(node b, node x, node y, list<node>& dunions)\{\nwcodepenalty=\Lhighpen
  node v = base(x);\nwcodepenalty=\Lhighpen
  while (v != b)\{\nwcodepenalty=\Llowpen
    base.union_blocks(v,b); dunions.append(v); dunions.append(b);\nwcodepenalty=\Llowpen
    v = mate[v]; \nwcodepenalty=\Llowpen
    base.union_blocks(v,b);  dunions.append(v); dunions.append(b);\nwcodepenalty=\Llowpen
    base.make_rep(b); \nwcodepenalty=\Llowpen
    source_bridge[v] = x; target_bridge[v] = y;\nwcodepenalty=\Llowpen
    // label[v] = EVEN; we don't have to do this because we always ask for the label of the base of v.\nwcodepenalty=\Llowpen
    bd[v] = bd[v] + (Delta - bDelta[v]); bDelta[v] = Delta;\nwcodepenalty=\Llowpen
    edge e; \nwcodepenalty=\Llowpen
    forall_inout_edges(e,v)\nwcodepenalty=\Llowpen
      scan_edge(e,v);\nwcodepenalty=\Llowpen
    v = base(parent[v]);\nwcodepenalty=\Llowpen
  \}\nwcodepenalty=\Lhighpen
  dunions.append(b); dunions.append(b);  // signal to dbase that this round of unions is over. \nwcodepenalty=\Lhighpen
\}\nwcodepenalty=\Llowpen
\nwendcode{}\nwbegindocs{30}\nwdocspar

\subsubsection{Construction of $H$}\label{construction of H} The vertices of $H$ are the trivial blossoms and the contractions of the maximal positive blossoms of $G$ and the edges of $H$ correspond to the tight edges of $G$ connecting distinct vertices of $H$ (Section 3.3 in~\cite{Gabow:GeneralMatching}). The blocks of \ensuremath{\mathit{base}} are \textbf{not} the vertices of $H$ as we may have performed further unions in the $\Delta$-phase in which we found the $\sap$.

We therefore keep a second node-partition \ensuremath{\mathit{dbase}} (= \emph{delayed base}) and build in each iteration a list \ensuremath{\mathit{dunions}} of the union-operations performed. When an iteration ends with a dual update, we perform these unions on \ensuremath{\mathit{dbase}} and empty \ensuremath{\mathit{dunions}}. In this way, the blocks of \ensuremath{\mathit{dbase}} are the trivial and the maximal positive blossoms and it is a simple matter to construct $H$.

\nwenddocs{}\nwbegincode{31}\moddef{local declarations in phase\_1}\plusendmoddef\nwstartdeflinemarkup\nwenddeflinemarkup\nwcodepenalty=\Lhighpen
list<node> dunions;\nwcodepenalty=\Lhighpen
\nwendcode{}\nwbegindocs{32}\nwdocspar

\nwenddocs{}\nwbegincode{33}\moddef{commit to unions}\endmoddef\nwstartdeflinemarkup\nwenddeflinemarkup\nwcodepenalty=\Lhighpen
while (!dunions.empty())\{\nwcodepenalty=\Lhighpen
  node u = dunions.pop(); node v = dunions.pop();\nwcodepenalty=\Lhighpen
  if (u == v) dbase.make_rep(u);\nwcodepenalty=\Lhighpen
  else dbase.union_blocks(u,v);\nwcodepenalty=\Lhighpen
\}\nwcodepenalty=\Llowpen
\nwendcode{}\nwbegindocs{34}\nwdocspar
The edges of $H$ are given by the edges of $G$ that run between blocks of \ensuremath{\mathit{dbase}} and are tight. For each edge of $H$, we remember its preimage in $G$ in the edge-map \ensuremath{\mathit{edge\nspaceunderscore\_in\nspaceunderscore\_G}}. 

There is a small lacuna in Gabow's paper in the construction of $H$ in Section 5, where he writes: Then scan every edge $e \in E$ and add $e$ to $E(H)$ if it is tight and joins distinct $V(H)$ vertices, at least one being outer. We first implemented this and it does not seem to work. Consider a graph $G$ consisting of two paths of length three; in both paths, the middle edge belongs to the current matching and the two endpoints are free. We start by making the four free nodes even. The edges incident to the four endpoints are added to $Q[2]$. After two dual changes, the free nodes have dual values $-1$ and the edges incident to the free nodes are tight. The matching edges are also tight. We explore one of the edges incident to a free node and add its other endpoint and its mate to the search structure. Suppose that we scan the edge incident to the other endpoint of the path next. Then we have found an augmenting path. Now consider the other path of length three. Its endpoints are even and the other nodes of the path are unlabeled. So the matching edge is not added to $H$ according to the second description. This is incorrect as we will miss the second augmenting path of length three. Even worse, there are now mistakenly two augmenting paths of length one in $H$.

The correct definition of $H$ is as given in the first paragraph of this section. We contract the maximal positive blossoms and keep all tight edges connecting distinct nodes. Actually we never construct $H$
explicitly. Rather we re-use the base-nodes of blossoms as the nodes of $H$. 
We write \ensuremath{\mathit{vh}} and \ensuremath{\mathit{uh}} for nodes of $G$ when we refer to them as nodes of $H$. For a node \ensuremath{\mathit{vh}}, the list \ensuremath{\mathit{contracted\nspaceunderscore\_into}[\mathit{vh}]} contains all nodes of $G$ contracted into \ensuremath{\mathit{vh}}. An matching edge $(u,v)$ induces the matching edge $(\ensuremath{\mathit{dbase}(u)},\ensuremath{\mathit{dbase}(v)})$ in $H$. We use \ensuremath{\mathit{mateHG}} for mates in $H$.

\nwenddocs{}\nwbegincode{35}\moddef{augmentation: construction of H}\endmoddef\nwstartdeflinemarkup\nwenddeflinemarkup\nwcodepenalty=\Lhighpen
\vspace{\Lemptyline}node vh; node v; node u; edge e;\nwcodepenalty=\Lhighpen
\vspace{\Lemptyline}int number_of_nodes_of_H = 0;\nwcodepenalty=\Lhighpen
\vspace{\Lemptyline}forall(v,T)\{\nwcodepenalty=\Llowpen
  contracted_into[dbase(v)].append(v);\nwcodepenalty=\Llowpen
  mateHG[v] = nil;\nwcodepenalty=\Llowpen
\}\nwcodepenalty=\Llowpen
\vspace{\Lemptyline}forall(u,T)\nwcodepenalty=\Llowpen
  forall_inout_edges(e,u) is_edge_of_H[e] = false;\nwcodepenalty=\Llowpen
\vspace{\Lemptyline}forall(u,T)\{\nwcodepenalty=\Llowpen
   node uh = dbase(u);\nwcodepenalty=\Llowpen
   forall_out_edges(e,u)\{\nwcodepenalty=\Llowpen
     is_edge_of_H[e] = false;\nwcodepenalty=\Llowpen
     node v = G.target(e); node vh = dbase(v);\nwcodepenalty=\Llowpen
     if( uh != vh &&  d(u) + d(v) == w[e] )\{\nwcodepenalty=\Llowpen
       is_edge_of_H[e] = true;  //assert( in_T[u] == in_T[v] && ((in_T[u] && in_T[v]) || mate[u] == v));\nwcodepenalty=\Llowpen
       if(w[e] == 2)\{       // e is a matching edge\nwcodepenalty=\Llowpen
         mateHG[uh] = vh; mateHG[vh] = uh;\nwcodepenalty=\Llowpen
       \}\nwcodepenalty=\Llowpen
     \}\nwcodepenalty=\Lhighpen
  \}\nwcodepenalty=\Lhighpen
\}\nwcodepenalty=\Llowpen
\nwendcode{}\nwbegindocs{36}\nwdocspar
\subsection{Using LEDA's Matching Algorithm to Finish Off}

This section is brief as the second phase uses a similar strategy for finding augmenting path. We refer the reader to Section~\ref{Phase2} for a more detailed treatment. 

When there are only a few augmentations left, Gabow's algorithm is overkill and we switch to LEDA's matching algorithm. It is based on~\cite{GT91}. We refer to the LEDA book for an explanation of the program. We use the two most effective heuristic improvements suggested by Kececioglu and Pecqueur~\cite{Kececioglu:matching}. The first improvement is the use of depth-first search.  Then bridges are either backward- or forward edges. For a bridge $e = (v,w)$ either \ensuremath{\mathit{base}(v)} or \ensuremath{\mathit{base}(w)} is the base of the newly formed blossom, whoever is the ancestor of the other. We use DFS-numbers to distinguish the two possibilities and shrink blossoms only on forward-edges. The node with the smaller DFS-number is the base of the newly formed blossom.

Initially, all nodes are \ensuremath{\mathit{UNLABELED}}. We iterate over all nodes and grow a search structure at every free node \ensuremath{\mathit{v0}}. The call \ensuremath{\mathit{find\nspaceunderscore\_aug\nspaceunderscore\_path}(\mathit{v0},\mathit{v0})} uses DFS for finding an augmenting path starting at the second argument \ensuremath{\mathit{v0}}. The first argument is the current node of the DFS. Nodes are numbered by the time at which they become even. The call returns the other end of the augmenting path, return \ensuremath{\mathit{nil}} if the search is unsuccessful. If the call is successful, all non-matching edges are pushed onto a list $P$ as pairs of nodes; the procedure \ensuremath{\mathit{find\nspaceunderscore\_path}(\mathit{pw},\mathit{v0})} traces the even length path from \ensuremath{\mathit{pw}} to \ensuremath{\mathit{v0}}. We mate the pairs in $P$. We have also collected all nodes of the search structure in \ensuremath{T}. In an augmenting path is found, \ensuremath{T} is dismantled and all nodes in $T$ are returned to the pool of unlabeled nodes.

\nwenddocs{}\nwbegincode{37}\moddef{finish the computation by means of LEDA's matching algorithm}\endmoddef\nwstartdeflinemarkup\nwenddeflinemarkup\nwcodepenalty=\Lhighpen
\vspace{\Lemptyline}dbase.split(G.all_nodes());\nwcodepenalty=\Lhighpen
\vspace{\Lemptyline}node v0, v, w;\nwcodepenalty=\Lhighpen
\vspace{\Lemptyline}forall_nodes(v0,G) label[v0] = UNLABELED;\nwcodepenalty=\Llowpen
\vspace{\Lemptyline}forall_nodes(v0,G)\{\nwcodepenalty=\Llowpen
  if ( mate[v0] != nil ) continue;\nwcodepenalty=\Llowpen
  label[v0] = EVEN; T.clear(); T.append(v0); even_time[v0] = even_count++;\nwcodepenalty=\Llowpen
\vspace{\Lemptyline}  if( (w = find_aug_path(v0,v0)) != nil )\{\nwcodepenalty=\Llowpen
    node pw = parent[w];\nwcodepenalty=\Llowpen
    P.push(w); P.push(pw);\nwcodepenalty=\Llowpen
    find_path(pw,v0);\nwcodepenalty=\Llowpen
\vspace{\Lemptyline}    while(! P.empty())\{\nwcodepenalty=\Llowpen
      node a = P.pop();\nwcodepenalty=\Llowpen
      node b = P.pop();\nwcodepenalty=\Llowpen
      mate[a] = b;\nwcodepenalty=\Llowpen
      mate[b] = a;\nwcodepenalty=\Llowpen
    \} \nwcodepenalty=\Llowpen
\vspace{\Lemptyline}    forall(v,T) label[v] = UNLABELED;\nwcodepenalty=\Llowpen
    dbase.split(T);\nwcodepenalty=\Llowpen
    size_of_M++;\nwcodepenalty=\Lhighpen
  \}\nwcodepenalty=\Lhighpen
\}\nwcodepenalty=\Llowpen
\vspace{\Lemptyline}\nwendcode{}\nwbegindocs{38}\ensuremath{\mathit{find\nspaceunderscore\_aug\nspaceunderscore\_path}(v,\mathit{v0})} first scans all edges incident to $v$ to find out whether one of them has a free endpoint; this is the second heuristic improvement suggested by Kececioglu and Pecqueur. If so, the free endpoint is returned and no recursive calls are made. Otherwise, either the search structure is grown or an blossom is shrunk on an forward edge. On a backward-edge we do nothing. 

\nwenddocs{}\nwbegincode{39}\moddef{helper functions}\plusendmoddef\nwstartdeflinemarkup\nwenddeflinemarkup\nwcodepenalty=\Lhighpen
\vspace{\Lemptyline}  node find_aug_path(node v, node v0)\{ // we are growing a tree with root v0\nwcodepenalty=\Lhighpen
   edge e; node w; \nwcodepenalty=\Lhighpen
   forall_inout_edges(e,v)\{\nwcodepenalty=\Llowpen
     w = G.opposite(e,v); \nwcodepenalty=\Llowpen
     if( v == w ) continue;\nwcodepenalty=\Llowpen
     if( mate[w] == nil && label[w] == UNLABELED )\{\nwcodepenalty=\Llowpen
       parent[w] = v;\nwcodepenalty=\Llowpen
       T.append(w);\nwcodepenalty=\Llowpen
       label[w] = ODD;\nwcodepenalty=\Llowpen
       return w;\nwcodepenalty=\Llowpen
     \}\nwcodepenalty=\Llowpen
   \}\nwcodepenalty=\Llowpen
   // no immediate break-through\nwcodepenalty=\Llowpen
   forall_inout_edges(e,v)\{\nwcodepenalty=\Llowpen
     w = G.opposite(e,v); node bw = dbase(w);\nwcodepenalty=\Llowpen
     if (label[bw] == ODD || w == v) continue;\nwcodepenalty=\Llowpen
     if (label[bw] == UNLABELED)\{\nwcodepenalty=\Llowpen
       label[w] = ODD; parent[w] = v; T.append(w);\nwcodepenalty=\Llowpen
       node mw = mate[w];\nwcodepenalty=\Llowpen
       label[mw] = EVEN; T.append(mw); even_time[mw] = even_count++;\nwcodepenalty=\Llowpen
       node s = find_aug_path(mw,v0);\nwcodepenalty=\Llowpen
       if(s != nil)\nwcodepenalty=\Llowpen
         return s;\nwcodepenalty=\Llowpen
     \}\nwcodepenalty=\Llowpen
     else\{\nwcodepenalty=\Llowpen
       node bv = dbase(v);\nwcodepenalty=\Llowpen
       node bw = dbase(w);\nwcodepenalty=\Llowpen
       list<node> tmp;   \nwcodepenalty=\Llowpen
       if (even_time[bv] < even_time[bw])\{ //blossom_step along forward edge\nwcodepenalty=\Llowpen
         // walk down from bw to bv and perform unions\nwcodepenalty=\Llowpen
         while(bw != bv)\{// doing the unions carefully, so that only one make_rep is needed\nwcodepenalty=\Llowpen
           node mate_bw = mate[bw]; \nwcodepenalty=\Llowpen
           dbase.union_blocks(bw,mate_bw);\nwcodepenalty=\Llowpen
           bw = dbase(parent[mate_bw]);\nwcodepenalty=\Llowpen
           dbase.union_blocks(mate_bw,bw); \nwcodepenalty=\Llowpen
           tmp.push_front(mate_bw);\nwcodepenalty=\Llowpen
           source_bridge[mate_bw] = w;\nwcodepenalty=\Llowpen
           target_bridge[mate_bw] = v;\nwcodepenalty=\Llowpen
         \}\nwcodepenalty=\Llowpen
         dbase.make_rep(bv);\nwcodepenalty=\Llowpen
         forall(w,tmp)\{\nwcodepenalty=\Llowpen
          node s = find_aug_path(w,v0);\nwcodepenalty=\Llowpen
          if(s != nil)\nwcodepenalty=\Llowpen
            return s;\nwcodepenalty=\Llowpen
         \}\nwcodepenalty=\Llowpen
       \}\nwcodepenalty=\Llowpen
     \}\nwcodepenalty=\Llowpen
   \}\nwcodepenalty=\Llowpen
   return nil;\nwcodepenalty=\Llowpen
  \}\nwcodepenalty=\Llowpen
\vspace{\Lemptyline}\vspace{\Lemptyline}void find_path(node x, node y)\{\nwcodepenalty=\Llowpen
/* traces the even length alternating path from x to y; if non-trivial it starts with\nwcodepenalty=\Llowpen
the matching edge incident to x; collects the non-matching edges on this\nwcodepenalty=\Llowpen
path as pairs of nodes */\nwcodepenalty=\Llowpen
  if ( x == y ) return;\nwcodepenalty=\Llowpen
\vspace{\Lemptyline}  if( label[x] == EVEN )\{\nwcodepenalty=\Llowpen
    node mate_x = mate[x], par_mate_x = parent[mate_x];\nwcodepenalty=\Llowpen
    P.append(mate_x); P.append(par_mate_x);\nwcodepenalty=\Llowpen
    find_path(par_mate_x,y);   \nwcodepenalty=\Llowpen
    return;\nwcodepenalty=\Llowpen
  \}\nwcodepenalty=\Llowpen
  else\{ // x is ODD\nwcodepenalty=\Llowpen
    find_path(source_bridge[x],mate[x]);  \nwcodepenalty=\Llowpen
    P.append(source_bridge[x]); P.append(target_bridge[x]);\nwcodepenalty=\Llowpen
    find_path(target_bridge[x],y);\nwcodepenalty=\Llowpen
    return;\nwcodepenalty=\Lhighpen
  \}\nwcodepenalty=\Lhighpen
\}\nwcodepenalty=\Llowpen
\vspace{\Lemptyline}\nwendcode{}\nwbegindocs{40}\nwdocspar

\subsection{Phase Two: Construction and Augmentation of a Maximal Set of Saps}\label{phase2}\label{Phase2}

Recall the strategy for bipartite graphs. We first construct a layered network. Layer $i$ contains all nodes that can be reached from a free node by an alternating path of length $i$; layer zero consists of the free nodes. We stop the construction once we reach a layer containing again a free node. The layered network can be constructed using breadth-first search.

Then we construct a maximal set of edge-disjoint augmenting paths. We explore the layered network from a free node using \emph{depth-first search}. When we reach a free node and hence have found an augmenting path, the path corresponds to the recursion stack and hence is readily found. We delete the path and all its incident edges from the graph simply by tracing back the recursion and declaring all nodes on the path finished. Also, when we retreat from a node, we delete the node from further consideration as we can be sure that no free node can be reached through the node. 

How can we adapt this strategy to general graphs? We again explore \emph{depth-first}. For a warm-up, assume that we have a single bridge $xy$ with $y$ being a descendant of $x$. Both nodes are even. Consider the call for $x$ and let $e$ be the first edge on the tree path from $x$ to $y$. We have already explored some other subtrees rooted at $x$. If one of them results in a breakthrough,
we delete the augmenting path from the graph and no call along $e$ is ever made.  If none of them results in a breakthrough, we make the call along the edge $e$. When we return to $x$ from this call, we have not found an augmenting path using edge $e$. Note that all the odd nodes on the path from $x$ to $y$ have degree two in the search structure at this point and none of the even nodes can reach a free node. The odd nodes might be able to reach free nodes. We now explore some more edges out of $x$. If one of them results in a breakthrough, we delete the augmenting path and all nodes on it. This also deletes the edge $xy$ and hence none of the odd nodes on the path from $x$ to $y$ will ever become even. If none of these other edges out of $x$ results in a breakthrough, we explore the edge $xy$ and close a blossom. All odd nodes on the path from $x$ to $y$ become even and we make recursive calls for them. We start with the odd node closest to $x$ and then work our way back to $y$. Note that this is what we could have done, had we explored $xy$ before $e$. It is important that we continue exploring depth-first; hence the node \emph{closest} to $x$ is explored first.

For the description above, it is important that $xy$ is explored as a forward edge, i.e., from $x$ to $y$. To this end, we record for each node the time when it becomes even, i.e., we maintain a counter $t$, which we increment whenever a node becomes even and which we use to define \ensuremath{\mathit{even\nspaceunderscore\_time}[v]} for even vertices $v$. Actually, we need even-times only for nodes that are added as even nodes to the search structure as only such nodes can become base of a blossom. Consider the exploration of an edge $vu$ with $v$ and $u$ even. The edge represents the edge $(\ensuremath{\mathit{dbase}(v)},\ensuremath{\mathit{dbase}(u)})$. It is a forward edge iff $\ensuremath{\mathit{even\nspaceunderscore\_time}[\mathit{dbase}(v)]} < \ensuremath{\mathit{even\nspaceunderscore\_time}[\mathit{dbase}(u)]}$.

In phase two, we first make all nodes unlabeled and store with each node its current base in \ensuremath{\mathit{rep}}. Then $\ensuremath{\mathit{rep}[v]}$ is the node of $H$ into which $v$ is contracted. We then construct a maximal set of augmenting paths in $H$ and collect them in \ensuremath{\mathit{PG}}. At the end of the phase, we augment all paths in \ensuremath{\mathit{PG}}.

\nwenddocs{}\nwbegincode{41}\moddef{helper functions}\plusendmoddef\nwstartdeflinemarkup\nwenddeflinemarkup\nwcodepenalty=\Lhighpen
void phase_2()\{\nwcodepenalty=\Lhighpen
  node v;\nwcodepenalty=\Lhighpen
  labelHG.init(G,UNLABELED);\nwcodepenalty=\Llowpen
  forall(v,T) rep[v] = dbase(v);\nwcodepenalty=\Llowpen
  list<list<edge> > PG;\nwcodepenalty=\Llowpen
\vspace{\Lemptyline}  \LA{}find a maximal set of augmenting paths in HG and collect them in PG\RA{}\nwcodepenalty=\Llowpen
\vspace{\Lemptyline}  list<edge> aphG;\nwcodepenalty=\Llowpen
  forall(aphG,PG) augmentG(aphG);  // augment all paths found\nwcodepenalty=\Lhighpen
\vspace{\Lemptyline}  forall(v,T) contracted_into[v].clear();  // clear H\nwcodepenalty=\Lhighpen
\}\nwcodepenalty=\Llowpen
\vspace{\Lemptyline}\nwendcode{}\nwbegindocs{42}The search for a maximal set of augmenting paths iterates over all nodes of $H$ and starts a search at every unvisited free node. A node is unvisited if its label is \ensuremath{\mathit{UNLABELED}}. The call \ensuremath{\mathit{find\nspaceunderscore\_apHG}(\mathit{vh},\nspacedot.\nspacedot.\nspacedot.)} searches for an augmenting path from \ensuremath{\mathit{vh}}. The details of this procedure are discussed below. An unsuccessful search returns \ensuremath{\mathit{nil}}, a successful search returns the last node \ensuremath{\mathit{free}} of the augmenting path. In the case of a successful search, we construct the augmenting path as a \ensuremath{\mathit{list}\Ltemplateless \mathit{edge}\Ltemplategreater \ \mathit{aghP}} of the non-matching edges of the path. We start with the parent-edge of the free node and then call \ensuremath{\mathit{find\nspaceunderscore\_path\nspaceunderscore\_in\nspaceunderscore\_HG}(\mathit{apG},\mathit{zh},\mathit{vh})} to trace the augmenting path back to \ensuremath{\mathit{vh}}, the node where we started the search for an augmenting path; here \ensuremath{\mathit{zh}} is the other endpoint of the parent-edge \ensuremath{e} of \ensuremath{\mathit{free}}, i.e., \ensuremath{\mathit{rep}[G\nspacedot.\mathit{target}(e)]} if
\ensuremath{\mathit{free}} is equal to \ensuremath{\mathit{rep}[G\nspacedot.\mathit{source}(e)]} and \ensuremath{\mathit{rep}[G\nspacedot.\mathit{source}(e)]} otherwise. All augmenting paths are collected in \ensuremath{\mathit{PG}}, a list of list of edges.

\nwenddocs{}\nwbegincode{43}\moddef{find a maximal set of augmenting paths in HG and collect them in PG}\endmoddef\nwstartdeflinemarkup\nwenddeflinemarkup\nwcodepenalty=\Lhighpen
node vh;\nwcodepenalty=\Lhighpen
forall(vh,T)\{\nwcodepenalty=\Lhighpen
  if (vh != rep[vh] ) continue; // vh does not represent a node of H\nwcodepenalty=\Llowpen
  if(labelHG[vh] == UNLABELED && mateHG[vh] == nil)\{\nwcodepenalty=\Llowpen
    labelHG[vh] = EVEN;\nwcodepenalty=\Llowpen
    even_timeHG[vh] = tG++;\nwcodepenalty=\Llowpen
    node free = find_apHG(vh);\nwcodepenalty=\Llowpen
    if(free != nil)\{\nwcodepenalty=\Llowpen
      list<edge> apG;\nwcodepenalty=\Llowpen
      edge e = parentHG[free];\nwcodepenalty=\Llowpen
      apG.append(e);\nwcodepenalty=\Llowpen
      find_path_in_HG(apG, rep[rep[G.source(e)] == free ? G.target(e) : G.source(e)], vh);\nwcodepenalty=\Llowpen
      PG.append(apG);\nwcodepenalty=\Llowpen
    \}\nwcodepenalty=\Lhighpen
  \}\nwcodepenalty=\Lhighpen
\}\nwcodepenalty=\Llowpen
\nwendcode{}\nwbegindocs{44}The search for an augmenting path is essentially as described at the beginning of this section with some minor differences. It is realized by the recursive procedure \ensuremath{\mathit{find\nspaceunderscore\_apHG}(\mathit{vh})}; \ensuremath{\mathit{vh}} is the current node. It is a node of $G$ representing a node of $H$, i.e., $\ensuremath{\mathit{vh}} = \ensuremath{\mathit{rep}[\mathit{vh}]}$. Iterating over all edges incident to \ensuremath{\mathit{vh}} means to iterate over all edges \ensuremath{\mathit{eh}} incident to nodes $v$ of $G$ contracted into \ensuremath{\mathit{vh}}. If \ensuremath{\mathit{eh}} does not represent an edge  of $H$, we skip over it. Otherwise, we either perform a grow step or do nothing or shrink a blossom. Grow steps distinguish between the exploration of a non-matching edge ending in a free vertex or ending in a matching edge. In the former case, we return the free vertex, in the latter case, we grow the tree by two nodes and call \ensuremath{\mathit{find\nspaceunderscore\_apHG}} recursively.

Consider next a blossom step triggered by an edge (\ensuremath{\mathit{vh}},\ensuremath{\mathit{uh}}). Since blossom steps are to be performed on forward edges, we proceed only if \ensuremath{\mathit{even\nspaceunderscore\_time}[\mathit{baseH}(\mathit{vh})]<\mathit{even\nspaceunderscore\_time}[\mathit{baseH}(\mathit{uh})]}. We trace back the path from \ensuremath{\mathit{baseH}(\mathit{uh})} to \ensuremath{\mathit{baseH}(\mathit{vh})}, perform the appropriate union-operations, and collect the odd nodes of the path in \ensuremath{\mathit{tmp}}. The edge \ensuremath{\mathit{eh}} becomes the bridge for all nodes in \ensuremath{\mathit{tmp}}; the direction is from \ensuremath{\mathit{uh}} to \ensuremath{\mathit{vh}}. We finally make recursive calls for the nodes on these path, making sure that we start with the node closest to \ensuremath{\mathit{bh}}.

\nwenddocs{}\nwbegincode{45}\moddef{helper functions}\plusendmoddef\nwstartdeflinemarkup\nwenddeflinemarkup\nwcodepenalty=\Lhighpen
\vspace{\Lemptyline}node find_apHG(node vh)\{\nwcodepenalty=\Lhighpen
  edge eh; node v;\nwcodepenalty=\Lhighpen
  forall(v,contracted_into[vh])\{ // v is a node of G, vh = rep[vh] is a node of G representing a node of H\nwcodepenalty=\Llowpen
    forall_inout_edges(eh,v)\{\nwcodepenalty=\Llowpen
      if (!is_edge_of_H[eh]) continue;\nwcodepenalty=\Llowpen
      node uh = rep[G.opposite(v,eh)]; \nwcodepenalty=\Llowpen
      if(mateHG[vh] == uh) continue;\nwcodepenalty=\Llowpen
      if(labelHG[uh] == UNLABELED)\{ // grow step\nwcodepenalty=\Llowpen
        node mateHG_uh = mateHG[uh];\nwcodepenalty=\Llowpen
        if(mateHG_uh == nil)\{ // uh is free and we found an augmenting path\nwcodepenalty=\Llowpen
          labelHG[uh] = ODD;     \nwcodepenalty=\Llowpen
          parentHG[uh] = eh; \nwcodepenalty=\Llowpen
          return uh;\nwcodepenalty=\Llowpen
        \}\nwcodepenalty=\Llowpen
        else \{ // extend the path by two edges\nwcodepenalty=\Llowpen
          labelHG[uh] = ODD; labelHG[mateHG_uh] = EVEN; \nwcodepenalty=\Llowpen
          parentHG[uh] = eh;\nwcodepenalty=\Llowpen
          even_timeHG[mateHG_uh] = tG++; \nwcodepenalty=\Llowpen
          node s = find_apHG(mateHG_uh); \nwcodepenalty=\Llowpen
          if (s != nil)\nwcodepenalty=\Llowpen
            return s;\nwcodepenalty=\Llowpen
        \}\nwcodepenalty=\Llowpen
      \}\nwcodepenalty=\Llowpen
      else\{\nwcodepenalty=\Llowpen
        node bh = dbase(vh);\nwcodepenalty=\Llowpen
        node zh = dbase(uh);\nwcodepenalty=\Llowpen
        if(even_timeHG[bh] < even_timeHG[zh])\{ // blossom step along forward edge\nwcodepenalty=\Llowpen
          list<node> tmp;\nwcodepenalty=\Llowpen
          list<node> endpoints_of_M;\nwcodepenalty=\Llowpen
          while (zh != bh)\{\nwcodepenalty=\Llowpen
            endpoints_of_M.append(zh);\nwcodepenalty=\Llowpen
            zh = mateHG[zh];\nwcodepenalty=\Llowpen
            endpoints_of_M.append(zh);\nwcodepenalty=\Llowpen
            tmp.push_front(zh); // zh is odd and we add it to the front of tmp\nwcodepenalty=\Llowpen
            zh = dbase(rep[rep[G.source(parentHG[zh])] == zh ?\nwcodepenalty=\Llowpen
                              G.target(parentHG[zh]) : G.source(parentHG[zh])]);\nwcodepenalty=\Llowpen
          \}\nwcodepenalty=\Llowpen
\vspace{\Lemptyline}          forall(zh,endpoints_of_M) dbase.union_blocks(zh,bh);\nwcodepenalty=\Llowpen
          dbase.make_rep(bh);\nwcodepenalty=\Llowpen
\vspace{\Lemptyline}          forall(zh,tmp)\{\nwcodepenalty=\Llowpen
            //even_timeHG[zh] = tG++; // not needed\nwcodepenalty=\Llowpen
            bridgeHG[zh] = eh;   // assert(eh != nil); \nwcodepenalty=\Llowpen
            dirHG[zh] = (G.target(eh) == v ? 1 : -1);  // dirHG = 1 iff eh = (_,v)\nwcodepenalty=\Llowpen
          \}\nwcodepenalty=\Llowpen
\vspace{\Lemptyline}          forall(zh,tmp)\{ //the new even node closest to bh comes first\nwcodepenalty=\Llowpen
            node s = find_apHG(zh);\nwcodepenalty=\Llowpen
            if(s != nil) \nwcodepenalty=\Llowpen
              return s;\nwcodepenalty=\Llowpen
          \}\nwcodepenalty=\Llowpen
        \}\nwcodepenalty=\Llowpen
      \}\nwcodepenalty=\Llowpen
    \}\nwcodepenalty=\Llowpen
  \}\nwcodepenalty=\Lhighpen
  return nil;\nwcodepenalty=\Lhighpen
\}\nwcodepenalty=\Llowpen
\vspace{\Lemptyline}\nwendcode{}\nwbegindocs{46}We come to the actual construction of augmenting paths. \ensuremath{\mathit{find\nspaceunderscore\_path\nspaceunderscore\_in\nspaceunderscore\_H}(p,\mathit{vh},\mathit{uh})} constructs for two vertices $vh$ and $uh$ of $H$, where $vh$ is a descendant of $uh$, the even length alternating path connecting them. The non-matching edges of the path are returned in $p$ as a list of edges.
The strategy is simple. If $\ensuremath{\mathit{vh}} = \ensuremath{\mathit{uh}}$, there is nothing to do. If \ensuremath{\mathit{labelH}[\mathit{vh}]} is \ensuremath{\mathit{EVEN}}, we add the parent-edge of the mate of $v$ to the output and then walk from the other end of the parent edge to $uh$. If the label is \ensuremath{\mathit{ODD}}, i.e, \ensuremath{\mathit{vh}} became even in a blossom step, the path goes through the bridge of $vh$. Let $b = \ensuremath{\mathit{bridgeH}[\mathit{vh}]}$. The $vh$-side of the bridge is the source of $b$ if $\ensuremath{\mathit{dirH}[\mathit{vh}]} = 1$ and is the target of $b$ otherwise. So we walk from the $vh$-side of the bridge to the mate of $vh$ (recall that we construct an even length path), from the other side of the bridge to $uh$, and we add the bridge to the path. Note that we are only interested in the path as a set of edges and not as a sequence of edges. Hence there is no need to reverse the first subpath.

\nwenddocs{}\nwbegincode{47}\moddef{helper functions}\plusendmoddef\nwstartdeflinemarkup\nwenddeflinemarkup\nwcodepenalty=\Lhighpen
\vspace{\Lemptyline}void find_path_in_HG(list<edge>& p, node vh, node uh)\{\nwcodepenalty=\Lhighpen
 if(vh == uh) return;\nwcodepenalty=\Lhighpen
 if(labelHG[vh] == EVEN)\{\nwcodepenalty=\Llowpen
   node mvh = mateHG[vh];\nwcodepenalty=\Llowpen
   edge e = parentHG[mvh];\nwcodepenalty=\Llowpen
   p.append(e);\nwcodepenalty=\Llowpen
   find_path_in_HG(p, rep[rep[G.source(e)] == mvh ? G.target(e) : G.source(e)], uh);\nwcodepenalty=\Llowpen
   return;\nwcodepenalty=\Llowpen
 \}\nwcodepenalty=\Llowpen
 else\{\nwcodepenalty=\Llowpen
   edge bridgeHG_vh = bridgeHG[vh];\nwcodepenalty=\Llowpen
   find_path_in_HG(p, rep[dirHG[vh] == 1 ?\nwcodepenalty=\Llowpen
                      G.source(bridgeHG_vh) : G.target(bridgeHG_vh)], rep[mateHG[vh]]); \nwcodepenalty=\Llowpen
   p.append(bridgeHG[vh]);\nwcodepenalty=\Llowpen
   find_path_in_HG(p, rep[dirHG[vh] == 1 ?\nwcodepenalty=\Llowpen
                      G.target(bridgeHG_vh) : G.source(bridgeHG_vh)], uh);\nwcodepenalty=\Llowpen
   return;\nwcodepenalty=\Lhighpen
   \}\nwcodepenalty=\Lhighpen
\};\nwcodepenalty=\Llowpen
\vspace{\Lemptyline}\nwendcode{}\nwbegindocs{48}\nwdocspar

The function \ensuremath{\mathit{augment}} receives an augmenting path in $H$ as a list of its non-matching edges. It first fills in the parts inside the \ensuremath{\mathit{dbase}}-blossoms and then augments the path.

Let \ensuremath{\mathit{apH}} be an augmenting path in $H$. We construct the induced augmenting path \ensuremath{\mathit{ap}} in $G$ as a list of pairs of nodes, one for each non-matching edge of the path. Let $e = (u,v)$ be an edge of \ensuremath{\mathit{apH}}. We add the pair $(u,v)$ to \ensuremath{\mathit{ap}} and then call \ensuremath{\mathit{find\nspaceunderscore\_path\nspaceunderscore\_in\nspaceunderscore\_G}(\mathit{ap},u,\mathit{rep}[u])} and \ensuremath{\mathit{find\nspaceunderscore\_path\nspaceunderscore\_in\nspaceunderscore\_G}(\mathit{ap},v,\mathit{rep}[v]} to fill in the parts inside the \ensuremath{\mathit{dbase}}-blossoms. Once we have constructed \ensuremath{\mathit{ap}}, we augment it by simply mating each pair of nodes in \ensuremath{\mathit{ap}}. 

\ensuremath{\mathit{find\nspaceunderscore\_path\nspaceunderscore\_in\nspaceunderscore\_G}} does the obvious.

\nwenddocs{}\nwbegincode{49}\moddef{helper functions}\plusendmoddef\nwstartdeflinemarkup\nwenddeflinemarkup\nwcodepenalty=\Lhighpen
\vspace{\Lemptyline}void augmentG(const list<edge>& aph)\{ \nwcodepenalty=\Lhighpen
  list<node> ap; // augmenting path in G as a list of nodes, two nodes for each non-matching edge\nwcodepenalty=\Lhighpen
  edge e;\nwcodepenalty=\Llowpen
  forall(e,aph)\{\nwcodepenalty=\Llowpen
    node u = G.source(e); node v = G.target(e);\nwcodepenalty=\Llowpen
    ap.append(u); ap.append(v);\nwcodepenalty=\Llowpen
    find_path_in_G(ap, u, rep[u]); \nwcodepenalty=\Llowpen
    find_path_in_G(ap, v, rep[v]); \nwcodepenalty=\Llowpen
    \}\nwcodepenalty=\Llowpen
  while(!ap.empty())\{\nwcodepenalty=\Llowpen
    node u = ap.pop(); node v = ap.pop();\nwcodepenalty=\Llowpen
    mate[u] = v; mate[v] = u;\nwcodepenalty=\Llowpen
  \}\nwcodepenalty=\Llowpen
  size_of_M++;\nwcodepenalty=\Llowpen
\}\nwcodepenalty=\Llowpen
\vspace{\Lemptyline}void find_path_in_G(list<node>& p, node v, node u)\{ \nwcodepenalty=\Llowpen
/* finding the even path from v to u in the original graph using blossom bridges;\nwcodepenalty=\Llowpen
only returns the non-matching edges, for each edge the two endpoints */\nwcodepenalty=\Llowpen
  if(v == u) return;\nwcodepenalty=\Llowpen
  if(label[v] == EVEN)\{\nwcodepenalty=\Llowpen
    p.append(mate[v]); p.append(parent[mate[v]]); \nwcodepenalty=\Llowpen
    find_path_in_G(p, parent[mate[v]], u); \nwcodepenalty=\Llowpen
    return;\nwcodepenalty=\Llowpen
  \}\nwcodepenalty=\Llowpen
  else\{\nwcodepenalty=\Llowpen
    find_path_in_G(p, source_bridge[v], mate[v]); \nwcodepenalty=\Llowpen
    p.append(source_bridge[v]); p.append(target_bridge[v]);\nwcodepenalty=\Llowpen
    find_path_in_G(p, target_bridge[v], u);\nwcodepenalty=\Llowpen
    return;\nwcodepenalty=\Lhighpen
  \}\nwcodepenalty=\Lhighpen
\}\nwcodepenalty=\Llowpen
\nwendcode{}\nwbegindocs{50}\nwdocspar

Finally, we have to explain how we define the weight function. The current matching is encoded in \ensuremath{\mathit{mate}}. We iterate over all edges and set the weight of an edge to 2 if its endpoints are mates. We also unmate the endpoints so that only one edge in each bundle of parallel edges is given weight 2. Once we have assigned weights to all the edges, we again mate the endpoints of edges of weight 2.

\nwenddocs{}\nwbegincode{51}\moddef{setting up w}\endmoddef\nwstartdeflinemarkup\nwenddeflinemarkup\nwcodepenalty=\Lhighpen
\vspace{\Lemptyline}node z;\nwcodepenalty=\Lhighpen
forall(z,T)\nwcodepenalty=\Lhighpen
forall_out_edges(e,z)\{\nwcodepenalty=\Llowpen
  node u = G.source(e); node v = G.target(e);\nwcodepenalty=\Llowpen
  w[e] = 0;\nwcodepenalty=\Llowpen
  if (v == mate[u])\{\nwcodepenalty=\Llowpen
    w[e] = 2; \nwcodepenalty=\Llowpen
    mate[u] = mate[v] = nil;   // parallel edges; only one of them is in the matching.\nwcodepenalty=\Llowpen
  \}\nwcodepenalty=\Llowpen
\}\nwcodepenalty=\Llowpen
\vspace{\Lemptyline}forall(z,T)\nwcodepenalty=\Llowpen
forall_out_edges(e,z)\{\nwcodepenalty=\Llowpen
  node u = G.source(e); node v = G.target(e);\nwcodepenalty=\Llowpen
  if (w[e] == 2)\{\nwcodepenalty=\Llowpen
    mate[u] = v;\nwcodepenalty=\Llowpen
    mate[v] = u;\nwcodepenalty=\Lhighpen
  \}\nwcodepenalty=\Lhighpen
\}\nwcodepenalty=\Llowpen
\vspace{\Lemptyline}\nwendcode{}\nwbegindocs{52}\nwdocspar

\subsection{Further Optimizations}\label{Further Optimizations}

The LEDA implementation of~\cite{GT91} uses two powerful optimizations. Whenever, it grows a tree rooted as some free node, all nodes of the tree are collected in a set $T$. If the tree growing process finds an augmenting path, only the collected nodes are reinitialized for the next iteration. In this way, one pays only for what was touched in the tree growing process. If the tree growing process does not find an augmenting path, the nodes in the tree are not reinitialized, but keep their labels. Only $T$ is set to the empty set. In this way, the nodes in $T$ will never be added to a future tree as all unexplored edges are incident to odd nodes and odd nodes are not added to trees. We use this optimization also in the current program.  

We come to the second optimization. Suppose only odd nodes of a tree have unexplored incident edges. Then this tree will not grow further and we may delete it from further consideration. How can we determine that we explored all incident edges of all even nodes of a tree? We could let all even nodes point to their root and keep with the root a counter of the number of even nodes in the tree that still have unexplored incident edges. When we make a node even, we increase the counter, when we have explored all edges incident to an even node, we decrease the counter. If a counter reaches zero, the tree is stable and can be ignored in the future.

Alternatively. Consider the priority queue at the end of the first phase. It contains a number of edges. Let $e = xy$ be any edge in the priority queue. Then all nodes on the canonical paths of $x$ and $y$ can still be on an augmenting path. Also all unlabeled nodes can still be on augmenting paths. So we retire all nodes that are labeled and do not lie on one of these paths. 

We have a set $R$ of remaining nodes. We tentatively declare all nodes in $R$ inactive. For every edge $xy$ in the priority queue we trace back the canonical paths and declare their nodes active. Of course, we stop tracing once we encounter a node already declared active. We scan over all nodes in $R$ and delete all inactive nodes. We have not implemented the second optimization.









\section{Experiments and Evaluation}\label{experiments}

We compare three different implementations of matching algorithms, the LEDA-program based on the algorithm of Gabow and Tarjan~/\cite{Gabow:edmonds,Tarjan:book,LEDAbook}, the LEDA-program with the heuristic improvements suggested by Kececioglu/Pecqueur~\cite{Kececioglu:matching}, and our implementation of Gabow's general matching algorithm. The worst-case running time of the first two programs is $O(nm\alpha(n))$ and the worst-case running of the latter program is $O(n^{1/2}m\alpha(n))$. 
We performed three kinds of experiments and suggest some further experiments. 
\begin{description}
\item[Running Times:] We timed the implementation and two versions of LEDA's general matching algorithm (without and with the optimization suggested by Kececioglu and Pecqueur~\cite{Kececioglu:matching}) on various classes of sparse graphs with $m = O(n)$ edges: Graphs that force the algorithms into their worst case ($O(n^{3/2})$ for Gabow's algorithm and $O(n^2)$ for the other algorithms), graphs that can be handled in linear time by Gabow's algorithms and require time $O(n^{3/2})$ for the other algorithms, and random graphs that are handled in linear time by all algorithms. We also investigated the effect of switching to LEDA's matching algorithm once the distance to the maximum matching is about the number of iterations already performed. 
\item[Instruction Counts:] The measured running times do not conform with the theoretical asymptotics, but are strongly influenced by the cost of memory accesses. For an algorithm with asymptotic running time $T(n) = O(n^{3/2})$, the ratio $T(n)/T(n/2)$ could be approximately $2^{3/2} \approx 2.83$. However, due to the memory hierarchy, we see the factor $2.83$ only for very large $n \approx 10^5$. For smaller $n$, we see considerably larger factors. However, for instruction counts determined with the profiler \ensuremath{\mathit{gprof}}, we measure the right asymptotics.
\item[Breakdowns of Running Times:] Where does the implementation of Gabow's algorithm spend its time?
\item[An Unfair Comparison:] We compare the running times of our implementation of Gabow's algorithm and the implementation of the Micali-Vazirani algorithm by Huang and Stein~\cite{Huang-Stein} on random graphs. The latter implementation is in Python and requires 15 to 20 times more time. This factor cannot be taken as a sign of superiority of either algorithm. 
\end{description}

In Section~\ref{worst case graph}, we define a worst case family of graphs. The running time of a graph algorithm may depend on the presentation of the graph, i.e., the numbering of the nodes and the ordering of the adjacency list. We first define a worst case representation and then a worst case graph for which most representations are worst-case. We then report on running time experiments that differentiate between the different programs. In Section~\ref{random graphs}, we report our experiments on random graphs. The remaining section deal only with our implementation of Gabow's algorithm. We break down the running time over the various components of the program in Section~\ref{breakdown} and report on instruction counts and memory usage and explain the difference between measured asymptotics and theoretical asymptotics in Section~\ref{instruction counts}. In Section~\ref{unfair}, we time our program against the Python implementation of the Micali-Vazirani algorithm by Huang and Stein~\cite{Huang-Stein}. Finally, in Section~\ref{suggestions}, we suggest further experiments.

\subsection{A Worst Case Family of Graphs}\label{worst case graph}

We design a class of graphs
that force our programs into their worst case behavior. More precisely, we will
define for each $n$ and $m$ a graph $W_{n,m}$ with $O(n)$ nodes and $O(m)$
edges such that the running time of our programs on $W_{n,m}$ is 
$\Theta(nm)$ and $\Theta(\sqrt{n}m)$, respectively. The graph consists of 
a complete graph and a set of chains emanating from the complete graph and therefore we will first
study the behavior of our programs on complete graphs and
chains. The claim about the running time on $W_{n,m}$ will then follow easily.

The running time of graph and network algorithms depends on 
the representation of the input graph; reordering adjacency lists and renumbering
nodes may have a dramatic effect on the running time. We therefore have to be a
bit more precise about what we mean by \emph{the running time of our programs on $W_{n,m}$ is 
$\Theta(nm)$ and $\Theta(\sqrt{n}m)$, respectively}. Two interpretations are possible:
For every $n$ and $m$ 
\begin{itemize}
\item there is a representation of $W_{n,m}$, on which our programs run for 
$\Theta(nm)$ and $\Theta(\sqrt{n}m)$ steps, respectively, or
\item averaged over all representation of $W_{n,m}$, our algorithms run for 
$\Theta(nm)$ and $\Theta(\sqrt{n}m)$ steps, respectively.
\end{itemize}
We call a family of instances a \emph{robust} worst-case
example if the second interpretation applies. 
We will first construct a non-robust
worst-case example and then make it robust.

\subsubsection{A Worst-Case Representation}

We begin our study with the complete graph $G_C$ 
on about $\sqrt{m}$ nodes; we assume that the number of nodes is even.  We observe first that it is trivial to find a perfect matching. The
greedy heuristic suffices. It seems that we are not really heading towards a worst-case example, but
wait. Assume that we have determined a perfect matching in $G_C$. An important
observation about $G_C$ is that every edge is contained
in a short alternating path. More precisely, let 
$z$ be an arbitrary node of $G_C$. Then every edge is contained in an alternating path of length at
most four starting with the matching edge incident to $z$. This is easy to see. We may assume that the nodes
are numbered such that node $z = 0$ and nodes $2i$ and $2i+1$ are matched for $i = 0,1,\ldots$. The edge
  $(0,1)$ is an alternating path by itself, the matching edge $(2i,2i+1)$ with $i \ge 1$ lies on the paths
  $0,1,2i,2i+1$ and $0,1,2i+1,2i$. These paths also cover all the edges incident to $1$. Finally, the non-matching edge $(i,j)$ with $i > 1$ lies on one of the paths $0,1,i+1,i,j$ or $0,1,i-1,i,j$ depending on whether $i$ is even or odd. For $j = 0$, these path are non-simple, but this is of no importance.



  We come to $W_{n,m}$. It consists of a set of chains of even length and a complete graph $G_C$. One end of each chain, say $x$, is connected to $z$. Assume now that the greedy matching leaves the endpoints of each chain exposed, we start the search from $x$ with the edge connecting it to $z$, and the chain has length at least 8. Then a DFS starting at $x$ will explore all of $G_C$ before it starts traversing the path. A breadth-first strategy will also explore all of $G_C$.

  Let us first add $\Theta(n)$ chains of length 8 to $W_{n,m}$ and assume that each chain is explored as described in the previous paragraph. Then for both versions of the LEDA-implementation each augmentation has cost $\Theta(m)$ and the running time will be $\Theta(nm)$. Gabow's algorithm only needs $O(1)$ iterations and hence the running time will be $O(m)$.

Let us next add, in addition, one chain each of length $2k$, where $5 \le k \le \sqrt{n}$. Both LEDA algorithms will perform an additional $O(\sqrt{n})$ augmentations with no effect on the asymptotic running time. However, Gabow's algorithm will now need $\Theta(\sqrt{n})$ iterations and the
  running time will become $\Theta(\sqrt{n}m)$.

  The following program realizes these ideas. The construction of the complete graph and the final assembly of the complete graph and the chains is straight-forward. The construction of the chains requires some care. Let $k$ be an integer. We have nodes $0$ to $2k-1$. We first add the edge $(0,z)$. This will make sure that the search starting at $0$ explores $G_C$. Then we add the edges $(i,i+1)$ for $1 \le i \le 2k-2$ and finally the edge $(0,1)$. This ordering of the edges ensures that the greedy matching will leave the nodes $0$ and $2k-1$ exposed. We also add edges $(i,z)$ for all $i \ge 1$. This is not necessary in this section, but will be important in Section~\ref{robust example}. Tables~\ref{short chains} and~\ref{long chains} show running times on the two kind of instances.

I want two kind of tables: $W_{n,n}$ for increasing $n$ and $W_{n,m}$ for fixed $n$ and increasing $m$.

\nwenddocs{}\nwbegincode{53}\moddef{mc\_generator.cpp}\endmoddef\nwstartdeflinemarkup\nwenddeflinemarkup\nwcodepenalty=\Lhighpen
\vspace{\Lemptyline}node complete_g(graph& G, int m)  // sqrt(2m) nodes and about m edges\nwcodepenalty=\Lhighpen
\{ int n = 2* (int) sqrt(m/2.0); \nwcodepenalty=\Lhighpen
  array<node> a(n); int i, j;\nwcodepenalty=\Llowpen
  for (i = 0; i < n; i++) \nwcodepenalty=\Llowpen
    a[i] = G.new_node(); \nwcodepenalty=\Llowpen
  for (i = 0; i < n; i++) \nwcodepenalty=\Llowpen
    for (j = i+1; j < n; j++) G.new_edge(a[i],a[j]);   \nwcodepenalty=\Llowpen
\vspace{\Lemptyline}  return a[0];\nwcodepenalty=\Llowpen
\}\nwcodepenalty=\Llowpen
\vspace{\Lemptyline}void chain(graph& G, int k, node z)\{ // k new nodes and 2k - 1 edges \nwcodepenalty=\Llowpen
  array<node> a(k); int i;\nwcodepenalty=\Llowpen
\vspace{\Lemptyline}  for (i = 0; i < k; i++) a[i] = G.new_node();\nwcodepenalty=\Llowpen
\vspace{\Lemptyline}  G.new_edge(a[0],z);\nwcodepenalty=\Llowpen
\vspace{\Lemptyline}  for (i = 1; i < k-1; i++) \nwcodepenalty=\Llowpen
  \{ G.new_edge(a[i],a[i+1]);\nwcodepenalty=\Llowpen
    G.new_edge(a[i],z);      \nwcodepenalty=\Llowpen
  \}\nwcodepenalty=\Llowpen
  G.new_edge(a[0],a[1]);\nwcodepenalty=\Llowpen
\}\nwcodepenalty=\Llowpen
\vspace{\Lemptyline}void mc_worst_case_gen(graph& G, int n, int m, int mode = 1)\nwcodepenalty=\Llowpen
// sqrt(2m) + n + (mode == 1) n nodes and m + 2n + (mode == 1) 2n edges.\nwcodepenalty=\Llowpen
\{ node z = complete_g(G,m);\nwcodepenalty=\Llowpen
  int k = 4;\nwcodepenalty=\Llowpen
  for (int j = 0; j < n/8; j++) chain(G,2*k,z); \nwcodepenalty=\Llowpen
  if (mode == 1) \{\nwcodepenalty=\Llowpen
    for (k = 5; k < sqrt(n); k++)\nwcodepenalty=\Llowpen
    chain(G,2*k,z);\nwcodepenalty=\Lhighpen
  \}\nwcodepenalty=\Lhighpen
\}\nwcodepenalty=\Llowpen
\nwendcode{}\nwbegindocs{54}\nwdocspar

\begin{table}
  \begin{center}
\begin{tabular}{|r|r|r|r|r|r|r|r|} \hline
    $n$ & LEDA & $\sfrac{T(n)}{T(n/2)}$ & LEDA-KP& $\sfrac{T(n)}{T(n/2)}$ & Gabow & $\sfrac{T(n)}{T(n/2)}$ & \# iterations \\ \hline
10000 &  0.65  & -- &  1.85 & --& 0.39 & --  & 33 \\ \hline
    20000 & 2.92  & 4.49  &  8.23 & 4.44 & 1.39 & 3.56  & 47 \\ \hline
    40000 & 19.62 & 6.71 & 64.43 & 7.82 & 5.35 & 3.84  & 66 \\ \hline
80000 & 126.37  & 6.52 &   380.22 & 5.90 & 19.82 & 3.70 & 94 \\ \hline
    160000 & 588.78 & 4.65 & 1787.64 & 4.70    &  66.33 & 3.34 & 133 \\ \hline 
    320000 & 2486.23 &4.22  & 7644.66 & 4.27 & 202.98  &3.06 & 188 \\ \hline
  640000 & --   & -- & -- & -- & 607.62  & 2.99& 266 \\ \hline
  \end{tabular}
  \end{center}
\caption{\textbf{Worst Case Representation:} The running times of matching algorithms on graphs
generated by \ensuremath{\mathit{mc\nspaceunderscore\_worst\nspaceunderscore\_case\nspaceunderscore\_gen}}$(n,4n,1)$, i.e., we have $\Theta(n)$ short chains and, in addition, one long chain of length $2k$ for $k \in [5,\sqrt{n}]$. The running time of the LEDA algorithms on these instances is $O(n^2)$ and the running time of Gabow's algorithm is $O(n^{3/2})$. The first column shows $n$ and the next columns show the running time of the LEDA algorithm (Gabow/Tarjan) without and with the KP-heuristics, and of our implementation of Gabow's algorithm. For each of the three algorithms, we also show $T(n)/T(n/2)$. For Gabow's algorithm, we also show the number of iterations. The number of iterations grows as $\sqrt{n}$. Quadrupling $n$ doubles the number of iterations.\\ \protect
The factor $T(n)/T(n/2)$ should be $4$ for be for the first two algorithms and approximately 2.82 for Gabow's algorithm. We attempt an explanation in Section~\ref{instruction counts}, why we see larger factors. \label{long chains}
}
\end{table}

\begin{table}
  \begin{center}
\begin{tabular}{|r|r|r|r|r|r|r|r|} \hline
    $n$ & LEDA & $\sfrac{T(n)}{T(n/2)}$ & LEDA-KP& $\sfrac{T(n)}{T(n/2)}$ & Gabow & $\sfrac{T(n)}{T(n/2)}$ & \# iterations \\ \hline
    10000 &  0.59  & --    &  1.77 & --  & 0.01& --  & 2 \\ \hline
    20000 & 2.72  & 4.61   &  7.77 & 4,38 & 0.03& 3.00  & 2 \\ \hline
    40000 & 17.63 & 6.48   & 57,86 & 7.44 & 0.07& 2.33  & 2 \\ \hline
    80000 & 116.66  & 6.61 & 392.75 & 6.78 & 0.15& 2.14 & 2 \\ \hline
    160000 & 582.33 & 4.99 & 1790,87 & 4.56   &  0.33 & 2.20 & 2 \\ \hline
    320000 & 2485.37 &4.26 & 7523.94 & 4.20 & 0.66  &2.00 & 2 \\ \hline
    640000 & -- & -- & -- & -- & 1.31& 1.98& 2 \\ \hline
  1280000 &   &    &    &   & 2.60 & 1.98    & 2 \\ \hline
  2560000 &   &    &    &   & 5.39 & 2,07    & 2 \\ \hline
  \end{tabular}
  \end{center}
\caption{\textbf{Short Chains Only:} The running times of matching algorithms on graphs
generated by \ensuremath{\mathit{mc\nspaceunderscore\_worst\nspaceunderscore\_case\nspaceunderscore\_gen}}$(n,4n,0)$, i.e., we have $\Theta(n)$ short chains and no long chains. The running time of the LEDA algorithms on these instances is $O(n^2)$ and the running time of Gabow's algorithm is $O(n)$. The first column shows $n$ and the next columns show the running time of the LEDA algorithm without and with the KP-heuristics, and of our implementation of Gabow's algorithm. For each of the three algorithms, we also show $T(n)/T(n/2)$. For Gabow's algorithm, we also show the number of iterations. The number of iterations is constant, as short augmenting paths suffice. \\ \protect
We were surprised by the fact the added heuristics in LEDA-KP increase running times by about a factor of three. We have no explanation for this effect. 
The factor $T(n)/T(n/2)$ should be $4$ for the first two algorithms and approximately 2.82 for Gabow's algorithm. We explain in Section~\ref{instruction counts}, why we see larger factors. \label{short chains}
}
\end{table}

\subsubsection{A Robust Worst-Case Graph} The results of the preceding section crucially relied on the fact that we fed our algorithms a carefully constructed representation of $W_{n,m}$. It guarantees that the greedy matching constructs a perfect matching of the complete graphs and leaves the endpoints of all chains unmatched, and that each search for an augmenting path explores the entire complete graph. 
What changes when we run our algorithms on a random representation of
$W_{n,m}$, i.e., we use

\begin{Lcode}
{\Tt{}mc{\_}worst{\_}case{\_}gen(G,n,m);\nwendquote} \\ \nwcodepenalty=\Lhighpen
{\Tt{}\nwendquote} \\[\Lcodelineskip] \nwcodepenalty=\Lhighpen
{\Tt{}permute(G);\nwendquote}\nwcodepenalty=\Llowpen
\end{Lcode}
\nwenddocs{}\nwbegindocs{55}to generate $G$. Here \ensuremath{\mathit{permute}(G)} is a function that randomly permutes
the representation of $G$, i.e., the list of nodes, the list of edges, and all
adjacency lists. The adjacency lists are, however, not permuted independently but all in the same way. 

\ignore{

\nwenddocs{}\nwbegincode{56}\moddef{random\_representation.cpp}\endmoddef\nwstartdeflinemarkup\nwenddeflinemarkup\nwcodepenalty=\Lhighpen
\vspace{\Lemptyline}#include <LEDA/core/array.h>\nwcodepenalty=\Lhighpen
#include <LEDA/graph/graph.h>\nwcodepenalty=\Lhighpen
\vspace{\Lemptyline}\nwendcode{}\nwbegindocs{57}}

\nwenddocs{}\nwbegincode{58}\moddef{random\_representation.cpp}\plusendmoddef\nwstartdeflinemarkup\nwenddeflinemarkup\nwcodepenalty=\Lhighpen
\vspace{\Lemptyline}void permute(graph& G)\nwcodepenalty=\Lhighpen
\{ \nwcodepenalty=\Lhighpen
  list<edge> E = G.all_edges();\nwcodepenalty=\Llowpen
  E.permute();\nwcodepenalty=\Llowpen
  G.sort_edges(E); \nwcodepenalty=\Llowpen
\vspace{\Lemptyline}  list<node> V = G.all_nodes();\nwcodepenalty=\Llowpen
  V.permute();\nwcodepenalty=\Lhighpen
  G.sort_nodes(V);\nwcodepenalty=\Lhighpen
\vspace{\Lemptyline}\}\nwcodepenalty=\Llowpen
\vspace{\Lemptyline}\nwendcode{}\nwbegindocs{59}Table~\ref{random representation} shows
the running times of our matching algorithms on random representations of
our worst case graphs $W_{n,m}$.

\begin{table}[t]
  \begin{center}
\begin{tabular}{|r|r|r|r|r|r|r|r|} \hline
    $n$ & LEDA & $\sfrac{T(n)}{T(n/2)}$ & LEDA-KP& $\sfrac{T(n)}{T(n/2)}$ & Gabow & $\sfrac{T(n)}{T(n/2)}$ & \# iterations \\ \hline
    10000 &  0.29  & --       &  0.48 & --    & 0.28 & --  & 20.4 \\ \hline
    20000 & 1.05  & 3.62      &  5.45 &  11.35& 1.06 & 3.78  & 28.7 \\ \hline
    40000 & 11.43 & 10.88     & 18.83 &  3.45 & 4.56&  4.30  & 40.0 \\ \hline
    80000 & 94.43  & 8.25     & 260.54& 13.83  & 15.76& 3.45 & 56.4 \\ \hline
   160000 & 380.11 & 4.05    & 1103.28& 4.23   & 52.42 & 3.32 &  81.5 \\ \hline
  \end{tabular}
\end{center}
\caption{\textbf{Worst Case Graph:} The running times in seconds of matching algorithms on a random
representation of the graphs
generated by \ensuremath{\mathit{mc\nspaceunderscore\_worst\nspaceunderscore\_case\nspaceunderscore\_gen}}$(n,4n,1)$. All times are the averages of ten runs. 
The first column shows $n$. The next columns show
the running times of the LEDA algorithm without and with the KP-heuristics and of our implementation of Gabow's algorithm. The ratio $T(n)/T(n/2)$ is also shown. For Gabow's algorithm we also show the number of iterations. \\ \protect
The running times of the LEDA algorithm vary widely. For example, for $n = 80000$, the best time is $18.31$ and the worst time is $160.68$. For LEDA-KC, the best time is $58.55$ and the worst is $445.73$, and for Gabow, the best time is $12.83$ and the worst is $18.52$.}
\label{random representation}\label{robust example}
\end{table}

\begin{figure}[t]
\begin{center}
\includegraphics[width=0.7\textwidth]{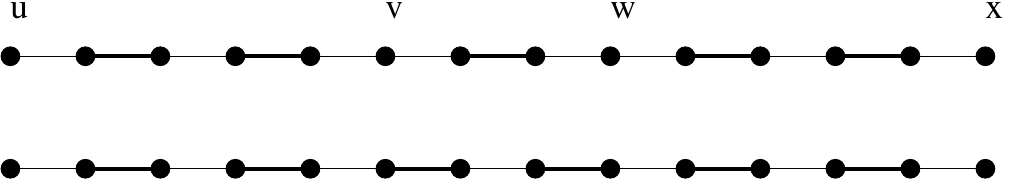}
\end{center}
\caption{In the top chain there are three augmenting paths, namely with
endpoints $u$ and $v$, $v$ and $w$, and $w$ and $x$, respectively. Using the
path from $v$ to $w$ for augmentation, merges the three paths into one. If one
endpoint of the augmenting path used is an endpoint of the containing chain, only two
paths are merged.}
\label{merging of augmenting paths}
\end{figure}

We attempt an explanation for the family $W_{n.n}$. For this graph class, about half of the edges are in the complete graph and most of the nodes are contained in chains. The greedy matching will either construct a perfect matching on $G_c$ or an almost perfect matching with one free vertex. The latter happens if it matches $z$ with a vertex on one of the chains. The greedy matching will also choose some chain edges and in this way break the chains into smaller chains separated by free vertices as shown in Figure~\ref{merging of augmenting paths}. 

We come to the augmentations. Each augmentation is of one of two
kinds: It either involves the edge $(x,z)$ and then turns the near perfect
matching on $G_C$ into a perfect matching, or it will merge (in generally
three) subchains into one, see Figure~\ref{merging of augmenting paths}. Once
the first kind of augmentation has happened and once augmenting paths have
length more than five, we will still have the
effect that all of $G_C$ will be explored in the search. Here we use the fact
that we connected all chain nodes to the special node $z$ of $G_C$. We therefore expect the same asymptotics as for the worst-case representation, although with a smaller constant.

\subsection{Random Graphs}\label{random graphs}

We tested the programs on random graphs with various edge densities. It is well known that sparse graphs with about $3n$ to $4n$ edges are hardest for matching algorithms and this is confirmed by our experiments. We therefore only report the running times for random graphs with $4n$ edges. For other small edge densities the running times are similar and for larger edge densities the running times are much smaller.

Table~\ref{tbl:random graphs} shows the running times. The running time of the LEDA program grows quadratically. As the graphs have a complete or almost almost complete matching the program performs a linear number of iterations. Each iteration seems to explore a large fraction of the graph. The heuristics by Kececioglu and Pecqueur~\cite{Kececioglu:matching} reduce the running times dramatically. In particular, the test whether the tip of the current alternating path has a direct connection to a free node reduces the length of augmenting paths dramatically. Gabow's algorithm seems to require a logarithmic number of iterations. Note that doubling $n$ seems to increase the number of iterations by one. It is known that for sufficiently dense random graphs with $m \ge 40n$, non-maximum matching always have an augmenting path of logarithmic length~\cite{MatchingSparseRandomGraphs}.

\begin{table}[h]
  \begin{center}
    \begin{tabular}{|r|r|r|r|r|r|r|r|} \hline
    $n$ & LEDA & $\sfrac{T(n)}{T(n/2)}$ & LEDA-KP& $\sfrac{T(n)}{T(n/2)}$ & Gabow & $\sfrac{T(n)}{T(n/2)}$ & \# iterations \\ \hline
       80000 & 0.74  & --   & 0.14 & --    & 0.39  & --    & 8.1 \\ \hline
      160000 & 2.80  & 3.78 & 0.42  & 3.00 & 1.31  & 3.35  & 9.2 \\ \hline
      320000 & 11.41 & 4.07 & 2.05  & 4.88 & 2.75  & 2.10  & 9.9\\ \hline
      640000 & 44.52 & 3.90 & 4.45  & 2.17 & 5.28  & 1.92  & 10.6 \\ \hline
     1280000 &326.03 & 7.32 & 12.34  & 2.77& 14.30 & 2.70 & 11.3 \\ \hline \hline
   10240000 & --    & --   & 221.66  &  19.96     & 233.67& 16.34   & 13.4 \\ \hline
    \end{tabular}
  \end{center}
  \caption{\label{tbl:random graphs}\textbf{Random Graphs:} The running time on random graphs with $n$ nodes and $4n$ edges. All times are the average of 10 runs and the same seed for the random generator was used for all three algorithms. The running time of the LEDA program grows quadratically. The other two programs fair much better and LEDA-KP outperforms Gabow's algorithm for graphs with up to $1.28 \cdot 10^6$ nodes. For $10.24 \cdot 10^6$ nodes, Gabow wins. The number of iterations seems to grow logarithmically for Gabow's algorithm. See the text for further explanations.   }
\end{table}

\subsection{Breakdown of Running Times}\label{breakdown}

We give the breakdown of the running time for solving the hard instance with $n = 2\cdot10^5$ and $m = 4n$.
Almost half of time is spent in \ensuremath{\mathit{phase\nspaceunderscore\_1}}, less than 10\% is spent in \ensuremath{\mathit{phase\nspaceunderscore\_2}}, and about 30\% of the time is spent in finishing off the computation by finding single augmenting paths. In \ensuremath{\mathit{phase\nspaceunderscore\_1}}, the most time consuming functions are \ensuremath{\mathit{scan\nspaceunderscore\_edge}} (23\%), \ensuremath{\mathit{shrink\nspaceunderscore\_path}} (12\%), and the various priority queue operations (10\%).

\subsection{Running Times, Instruction Counts, and Memory Hierarchy}\label{instruction counts}

The measured running times grow faster than the theoretical running times. This is readily explained. The analysis counts the number of executed instructions. In particular, all memory accesses take the same time, but real computers have a memory hierarchy consisting of several levels of cache and main memory.

The instruction counts can be determined with the help of a debugger. We used \emph{gprof} to determine the number of accesses to edge records in our implementation of Gabow's algorithm on the graphs with long and short chains, see Table~\ref{Instruction Counts}. For long chains, the running time grows as $n^{3/2}$, for short chains, the running time grows as $n$. In both cases, the instruction counts show the same growth rate. Note that gprof samples the instructions only every 1/200 sec and so the counts are only approximate. 

\begin{table}[t]
  \begin{center}
    \begin{tabular}{|r|r|r|r|r|}\hline
   $n$ & \multicolumn{2}{c|}{long chains} & \multicolumn{2}{c|}{short chains} \\ \hline
&  $I(n)$ & $I(n)/I(n/2)$ &  $I(n)$ & $I(n)/I(n/2)$ \\ \hline
10000 &   93570584 & --  &   2873994 & --\\ \hline
20000 &  268663789 & 2.87 &  5762337 & 2.01\\ \hline
40000 &  752844009 & 2.80 & 11471631 & 1.99\\ \hline
80000 & 2152713427 & 2.85 & 23014737 & 2.01\\ \hline
    \end{tabular}
    \end{center}
\caption{\textbf{Instruction Counts:} \label{Instruction Counts} The number of accesses to edge records in our implementation of Gabow's algorithm on graphs with long chains and only short chains. In the first case, the counts grow as $n^{3/2}$, and in the second case, the counts grow linearly. Note that $2^{3/2} \approx 2.82$.    }
\end{table}

We conducted our experiments on an INTEL x86\_64 processor with three levels of cache. The L1 cache has size 192 kB, L2 cache has size 5 MB, L3 cache has size 12 MB, and main memory has size 32 GB.

On a graph with $n$ nodes and $4n$ edges the LEDA and the LEDA-KP program use about $600n$ Bytes of memory and the Gabow program needs about $1200n$ Bytes of memory. For $n = 10^4$, this amounts to 6 MB and 12 MB, respectively. The ratio $r$ of access times between L3 and L2 is about 10 and between main memory and L3 it is about 2.

Let us restrict attention to adjacent layers of memory, say L3 and main memory, and let us consider a program
with linear space requirement. Let us also assume that a fraction $p$ of the memory accesses are to the faster memory. If we double input size only a fraction $p/2$ will be to the faster memory and hence memory access will slow down by a factor
\[    \frac{\sfrac{p}{2}\cdot 1 + (1 - \sfrac{p}{2}) r}{p\cdot 1 + (1 - p)\cdot r} = \frac{1}{2}\frac{p + (2 - p)r}{p + (1 - p)r} = \frac{1}{2}(1  +\frac{r}{p + (1-p)r}).  \]
For example, for $r = 2$, the factor grows from 1 to $\sfrac{3}{2}$ for $p$ growing from zero to one, and for $r = 10$, the factor grows from 1 to $\sfrac{11}{2}$ for $p$ growing from zero to one. We see this factor in addition to the theoretical growth of the running time due to larger input size. Since $r$ is larger for the the L2/L3 boundary than for the L3/Main-Memory boundary, the additional factor is larger for small $n$.

\subsection{An Unfair Comparison}\label{unfair}

We also timed the Python-implementation of the Micali-Vazirani algorithm by Huang-Stein~\cite{Huang-Stein} on random graphs with $m = 4n$ edges, see Table~\ref{MV}. The \CC\ implementation of Gabow's algorithm is about 15 to 20 times faster on the same task. This is well in line with the reported speed differences of Python and \CC\ and does not hint to a superiority of one algorithm over the other.

\begin{table}
  \begin{center}
  \begin{tabular}{|r|r|r|r|r|} \hline
    $n$  &  $T(n)$ & $\sfrac{T(n)}{T(n/2)}$ & Gabow $T_G(n)$  &  $\sfrac{T(n)}{T_G(n)}$ \\ \hline
    80000 & 8.71 &  & 0.39  &  22.33\\ \hline
    160000 & 21.42 & 2.45 & 1.31 & 16.35 \\ \hline
    320000 & 52.79 &  2.46  & 2.75 & 19.19 \\ \hline
    640000 &  94.11 & 1.78  & 5.28 & 17.82\\ \hline
  \end{tabular}
  \end{center}
\caption{\textbf{Huang-Stein Python Implementation of Micali-Vazirani:} The running times on random graphs with $n$ nodes and $4n$ edges. All times are the average of 10 runs. The running times for Gabow's algorithm are as in Table~\ref{tbl:random graphs} }\label{MV}
  \end{table}

\subsection{Suggestions for Further Experiments}\label{suggestions}

\paragraph{Vertical versus Horizontal Memory Allocation:} Our implementation uses a large number of node- and edge-arrays. They are allocated consecutively in memory. As a consequence, when accessing several arrays relating to a particular node or edge, different areas of memory are accessed. A case in point is the function \ensuremath{\mathit{scan\nspaceunderscore\_edge}}.

\begin{Lcode}
{\Tt{}void\LcodeS scan{\_}edge(const\LcodeS edge{\&}\LcodeS e,\LcodeS const\LcodeS node{\&}\LcodeS z){\nwlbrace}\nwendquote} \\ \nwcodepenalty=\Lhighpen
{\Tt{}\LcodeS \LcodeS node\LcodeS u\LcodeS =\LcodeS G.opposite(e,z);\nwendquote} \\ \nwcodepenalty=\Lhighpen
{\Tt{}\LcodeS \LcodeS if\LcodeS (mate[u]\LcodeS ==\LcodeS z\LcodeS ||\LcodeS label[base(u)]\LcodeS ==\LcodeS ODD)\LcodeS return;\LcodeS \nwendquote} \\ \nwcodepenalty=\Llowpen
{\Tt{}\LcodeS \LcodeS int\LcodeS p\LcodeS =\LcodeS d(z)\LcodeS +\LcodeS d(u);\LcodeS \nwendquote} \\ \nwcodepenalty=\Llowpen
{\Tt{}\LcodeS \LcodeS if\LcodeS (label[u]\LcodeS ==\LcodeS UNLABELED)\nwendquote} \\ \nwcodepenalty=\Llowpen
{\Tt{}\LcodeS \LcodeS \LcodeS \LcodeS PQ.insert(e,Delta\LcodeS +\LcodeS p);\nwendquote} \\ \nwcodepenalty=\Llowpen
{\Tt{}\LcodeS \LcodeS else\nwendquote} \\ \nwcodepenalty=\Llowpen
{\Tt{}\LcodeS \LcodeS \LcodeS \LcodeS PQ.insert(e,Delta\LcodeS +\LcodeS p/2);\nwendquote} \\ \nwcodepenalty=\Lhighpen
{\Tt{}\LcodeS \LcodeS return;\nwendquote} \\ \nwcodepenalty=\Lhighpen
{\Tt{}\LcodeS \LcodeS {\nwrbrace}\nwendquote}\nwcodepenalty=\Llowpen
\end{Lcode}
\nwenddocs{}\nwbegindocs{60}It accesses \ensuremath{\mathit{mate}[u]}, \ensuremath{\mathit{bd}[u]}, \ensuremath{\mathit{bdelta}[u]} (both is \ensuremath{d(u)}), and some more information related to $u$ in \ensuremath{\mathit{base}(u)}. It might be more efficient to associate a record with $u$ that contains all array-values related to $u$. We have not implemented this approach.

\paragraph{Static Graphs:} LEDA offers several implementations of graphs, in particular fully dynamic graphs that support node and vertex additions and deletions, and static graphs that cannot be changed. Static graphs have a much smaller memory footprint than dynamic graphs and are sufficient for Gabow's algorithm. We have used dynamic graphs for lack of knowledge. A change to static graphs would reduce running times.

\section{Conclusions} We have given an \CC\ implementation of Gabow's general matching algorithm and reported on running time experiments. For worst-case graphs, the improvement in running time is dramatic. A comparison with a \CC\  implementation of the Micali-Vazirani algorithm would be highly interesting. We hope that somebody picks up our suggestions for further experiments.

\paragraph*{Acknowledgment:} We want to thank Stefan N\"aher for help with LEDA and Hal Gabow for useful comments. 

\bibliographystyle{alpha}
\bibliography{ref}

\newcommand{\etalchar}[1]{$^{#1}$}
\begin{thebibliography}{vdBCK{\etalchar{+}}23}

\bibitem[BMSH05]{MatchingSparseRandomGraphs}
H.~Bast, K.~Mehlhorn, G.~Sch\"{a}fer, and H.Tamaki.
\newblock Matching Algorithms are Fast in Sparse Random
  Graphs
\newblock {\em Theory of Computing Systems}, 31(1):3--14, 2005.
\newblock preliminary version in STACS 2004, LNCS 2996, 81 -- 92.

\bibitem[CKL{\etalchar{+}}22]{AlmostLinearTimeMaxFlow}
Li~Chen, Rasmus Kyng, Yang~P. Liu, Richard Peng, Maximilian~Probst Gutenberg,
  and Sushant Sachdeva.
\newblock Maximum flow and minimum-cost flow in almost-linear time.
\newblock In {\em 63rd {IEEE} Annual Symposium on Foundations of Computer
  Science, {FOCS} 2022, Denver, CO, USA, October 31 - November 3, 2022}, pages
  612--623. {IEEE}, 2022.

\bibitem[Edm65]{Edmonds:matching}
J.~Edmonds.
\newblock Maximum matching and a polyhedron with 0,1 - vertices.
\newblock {\em Journal of Research of the National Bureau of Standards},
  69B:125--130, 1965.

\bibitem[Gab76]{Gabow:edmonds}
H.~N. Gabow.
\newblock An efficient implementation of {E}dmonds' algorithm for maximum
  matching on graphs.
\newblock {\em Journal of the ACM}, 23:221--234, 1976.

\bibitem[Gab17]{Gabow:GeneralMatching}
Harald Gabow.
\newblock The weighted matching approach to maximum cardinality matching.
\newblock {\em Fundamenta Informaticae}, 154:109--130, 2017.

\bibitem[GK04]{Goldberg-Karzanov}
A.V. Goldberg and A.V. Karzanov.
\newblock Maximum skew-symmetric flows and matchings.
\newblock {\em Math. Programming, Series A}, 100:537 -- 568, 2004.

\bibitem[GT85]{Gabow-Tarjan:union-find}
H.~N. Gabow and R.~E. Tarjan.
\newblock A linear-time algorithm for a special case of disjoint set union.
\newblock {\em Journal of Computer and System Sciences}, 30(2):209--221, 1985.

\bibitem[GT91]{GT91}
H.~N. Gabow and R.~E. Tarjan.
\newblock Faster scaling algorithms for general graph-matching problems.
\newblock {\em J.~ACM}, 38(4):815--853, 1991.

\bibitem[HK73]{HK75}
J.~E. Hopcroft and R.~M. Karp.
\newblock An $n^{5/2}$ algorithm for maximum matchings in bipartite graphs.
\newblock {\em SIAM Journal of Computing}, 2(4):225--231, 1973.

\bibitem[HS17]{Huang-Stein}
Michael Huang and Clifford Stein.
\newblock Extending search phases in the micali-vazirani algorithm.
\newblock In {\em 16th Symposium on Experimental Algorithmics (SEA)}, LIPIcs,
  pages 10:1--10:19, 2017.

\bibitem[KL93]{Knuth-Levy}
D.~Knuth and S.~Levy.
\newblock {\em The {CWEB} System of Structured Documentation, Version 3.0}.
\newblock Addison-Wesley, 1993.

\bibitem[KP98]{Kececioglu:matching}
J.D. Kececioglu and J.~Pecqueur.
\newblock Computing maximum-cardinality matchings in sparse general graphs.
\newblock In {\em Proceedings of the 2nd Workshop on Algorithm Engineering
  ({WAE}'98)}, pages 121--132. Max-Planck-Institut f\"ur Informatik, 1998.

\bibitem[LED]{LEDAsystem}
{LEDA} ({L}ibrary of {E}fficient {D}ata {T}ypes and {A}lgorithms).
\newblock \myurl{www.algorithmic-solutions.com}.

\bibitem[MN99]{LEDAbook}
K.~Mehlhorn and S.~N\"aher.
\newblock {\em The {L}{E}{D}{A} {P}latform for {C}ombinatorial and {G}eometric
  {C}omputing}.
\newblock Cambridge University Press, 1999.

\bibitem[MR91]{Mattingly-Ritchey}
R.~Bruce Mattingly and Nathan~P. Ritchey.
\newblock Implementing on {$O\sqrt{N}M)$} cardinality matching algorithm.
\newblock In David~S. Johnson and Catherine~C. McGeoch, editors, {\em Network
  Flows And Matching, Proceedings of a {DIMACS} Workshop, New Brunswick, New
  Jersey, USA, October 14-16, 1991}, volume~12 of {\em {DIMACS} Series in
  Discrete Mathematics and Theoretical Computer Science}, pages 539--556.
  {DIMACS/AMS}, 1991.

\bibitem[MV80]{MV80}
S.~Micali and V.~Vazirani.
\newblock An ${O}(\sqrt{ |V|} \cdot |{E}|)$ algorithm for finding maximum
  matching in general graphs.
\newblock In {\em Proc.\ 21st {IEEE} Symposium on Foundations of Computer
  Science (FOCS)}, pages 17--27, 1980.

\bibitem[Tar83]{Tarjan:book}
R.~E. Tarjan.
\newblock {\em Data Structures and Network Algorithms}.
\newblock SIAM, 1983.

\bibitem[Vaz]{Vazirani23}
Vijay~V. Vazirani.
\newblock A theory of alternating paths and blossoms, from the perspective of
  minimum length.
\newblock available
  at~\url{https://www.ics.uci.edu/~vazirani/Matching_paper.pdf}.

\bibitem[Vaz94]{Vazirani94}
V.~V. Vazirani.
\newblock A theory of alternating paths and blossoms for proving correctness of
  the ${O}(\sqrt{V}{E})$ general graph maximum matching algorithm.
\newblock {\em Combinatorica}, 14(1):71--109, 1994.

\bibitem[Vaz12]{Vazirani12}
Vijay~V. Vazirani.
\newblock An improved definition of blossoms and a simpler proof of the {MV}
  matching algorithm.
\newblock {\em CoRR}, abs/1210.4594, 2012.

\bibitem[Vaz20]{Vazirani20}
Vijay~V. Vazirani.
\newblock A proof of the {MV} matching algorithm.
\newblock {\em CoRR}, abs/2012.03582, 2020.

\bibitem[vdBCK{\etalchar{+}}23]{AlmostLinearTimeMinCostFlow}
Jan van~den Brand, Li~Chen, Rasmus Kyng, Yang~P. Liu, Richard Peng,
  Maximilian~Probst Gutenberg, Sushant Sachdeva, and Aaron Sidford.
\newblock A deterministic almost-linear time algorithm for minimum-cost flow.
\newblock {\em CoRR}, abs/2309.16629, 2023.

\bibitem[vdBLN{\etalchar{+}}20]{NearlyLinearTimeBipartiteMatching}
Jan van~den Brand, Yin~Tat Lee, Danupon Nanongkai, Richard Peng, Thatchaphol
  Saranurak, Aaron Sidford, Zhao Song, and Di~Wang.
\newblock Bipartite matching in nearly-linear time on moderately dense graphs.
\newblock In Sandy Irani, editor, {\em 61st {IEEE} Annual Symposium on
  Foundations of Computer Science, {FOCS} 2020, Durham, NC, USA, November
  16-19, 2020}, pages 919--930. {IEEE}, 2020.

\end{thebibliography}

\appendix

\section{LEDA's Matching Algorithm}

The implementation is discussed in detail in~\cite[Section 7.7]{LEDAbook}.

\nwenddocs{}\nwbegincode{61}\moddef{LEDA}\endmoddef\nwstartdeflinemarkup\nwenddeflinemarkup\nwcodepenalty=\Lhighpen
\vspace{\Lemptyline}/*******************************************************************************\nwcodepenalty=\Lhighpen
+\nwcodepenalty=\Lhighpen
+  LEDA 7.0  \nwcodepenalty=\Llowpen
+\nwcodepenalty=\Llowpen
+\nwcodepenalty=\Llowpen
+  _mc_matching.c\nwcodepenalty=\Llowpen
+\nwcodepenalty=\Llowpen
+\nwcodepenalty=\Llowpen
+  Copyright (c) 1995-2023\nwcodepenalty=\Llowpen
+  by Algorithmic Solutions Software GmbH\nwcodepenalty=\Llowpen
+  All rights reserved.\nwcodepenalty=\Llowpen
+ \nwcodepenalty=\Llowpen
*******************************************************************************/\nwcodepenalty=\Llowpen
\vspace{\Lemptyline}\vspace{\Lemptyline}\vspace{\Lemptyline}#include <LEDA/graph/graph_alg.h>\nwcodepenalty=\Llowpen
#include <LEDA/graph/node_partition.h>\nwcodepenalty=\Llowpen
#include <LEDA/graph/node_slist.h>\nwcodepenalty=\Llowpen
#include <LEDA/system/assert.h>\nwcodepenalty=\Llowpen
#include <LEDA/graph/mc_matching.h>\nwcodepenalty=\Llowpen
\vspace{\Lemptyline}#include <LEDA/core/array.h>\nwcodepenalty=\Llowpen
#include <LEDA/core/string.h>\nwcodepenalty=\Llowpen
\vspace{\Lemptyline}LEDA_BEGIN_NAMESPACE\nwcodepenalty=\Llowpen
\vspace{\Lemptyline}static bool return_false(string s)\nwcodepenalty=\Llowpen
\{ cerr << "CHECK_MAX_CARD_MATCHING: " << s << "\\n"; return false; \}\nwcodepenalty=\Llowpen
\vspace{\Lemptyline}bool CHECK_MAX_CARD_MATCHING(const graph& G, const list<edge>& M,\nwcodepenalty=\Llowpen
                                    const node_array<int>& OSC)\nwcodepenalty=\Llowpen
\{ int n = leda_max(2,G.number_of_nodes());\nwcodepenalty=\Llowpen
  int K = 1;\nwcodepenalty=\Llowpen
  array<int> count(n);\nwcodepenalty=\Llowpen
  int i;\nwcodepenalty=\Llowpen
  for (i = 0; i < n; i++) count[i] = 0;\nwcodepenalty=\Llowpen
  node v; edge e;\nwcodepenalty=\Llowpen
\vspace{\Lemptyline}  forall_nodes(v,G) \nwcodepenalty=\Llowpen
  \{ if ( OSC[v] < 0 || OSC[v] >= n ) \nwcodepenalty=\Llowpen
     return return_false("negative label or label larger than n - 1");\nwcodepenalty=\Llowpen
    count[OSC[v]]++;\nwcodepenalty=\Llowpen
    if (OSC[v] > K) K = OSC[v];\nwcodepenalty=\Llowpen
  \}\nwcodepenalty=\Llowpen
\vspace{\Lemptyline}  int S = count[1];\nwcodepenalty=\Llowpen
  for (i = 2; i <= K; i++) S += count[i]/2;\nwcodepenalty=\Llowpen
  if ( S != M.length() )\nwcodepenalty=\Llowpen
    return_false("OSC does not prove optimality");\nwcodepenalty=\Llowpen
\vspace{\Lemptyline}  forall_edges(e,G)\nwcodepenalty=\Llowpen
  \{ node v = G.source(e); node w = G.target(e);\nwcodepenalty=\Llowpen
    if ( v == w || OSC[v] == 1 || OSC[w] == 1 ||\nwcodepenalty=\Llowpen
            ( OSC[v] == OSC[w] && OSC[v] >= 2) ) continue;\nwcodepenalty=\Llowpen
    return return_false("OSC is not a cover");\nwcodepenalty=\Llowpen
  \}\nwcodepenalty=\Llowpen
  return true;\nwcodepenalty=\Llowpen
\}\nwcodepenalty=\Llowpen
\vspace{\Lemptyline}\vspace{\Lemptyline}enum LABEL \{ODD, EVEN, UNLABELED\};\nwcodepenalty=\Llowpen
\vspace{\Lemptyline}\vspace{\Lemptyline}static void shrink_path(node b, node v, node w, \nwcodepenalty=\Llowpen
          node_partition& base, node_array<node>& mate, \nwcodepenalty=\Llowpen
          node_array<node>& pred, node_array<node>& source_bridge, \nwcodepenalty=\Llowpen
          node_array<node>& target_bridge, node_slist& Q)\nwcodepenalty=\Llowpen
\{ node x = base(v);\nwcodepenalty=\Llowpen
  while (x != b)\nwcodepenalty=\Llowpen
  \{ base.union_blocks(x,b);\nwcodepenalty=\Llowpen
    x = mate[x];\nwcodepenalty=\Llowpen
    base.union_blocks(x,b);\nwcodepenalty=\Llowpen
    base.make_rep(b);\nwcodepenalty=\Llowpen
    Q.append(x);\nwcodepenalty=\Llowpen
    source_bridge[x] = v;  \nwcodepenalty=\Llowpen
    target_bridge[x] = w;\nwcodepenalty=\Llowpen
    x = base(pred[x]);\nwcodepenalty=\Llowpen
  \}\nwcodepenalty=\Llowpen
\}\nwcodepenalty=\Llowpen
\vspace{\Lemptyline}\vspace{\Lemptyline}static void find_path(list<node>& P, node x, node y,\nwcodepenalty=\Llowpen
                 node_array<int>&  label, node_array<node>& pred,\nwcodepenalty=\Llowpen
                 node_array<node>& mate, \nwcodepenalty=\Llowpen
                 node_array<node>& source_bridge,\nwcodepenalty=\Llowpen
                 node_array<node>& target_bridge)\nwcodepenalty=\Llowpen
\{ if ( x == y ) \nwcodepenalty=\Llowpen
  \{\nwcodepenalty=\Llowpen
    P.append(x);  \nwcodepenalty=\Llowpen
    return;\nwcodepenalty=\Llowpen
  \}\nwcodepenalty=\Llowpen
\vspace{\Lemptyline}  if ( label[x] == EVEN ) \nwcodepenalty=\Llowpen
  \{ \nwcodepenalty=\Llowpen
    P.append(x);\nwcodepenalty=\Llowpen
    P.append(mate[x]);\nwcodepenalty=\Llowpen
    find_path(P,pred[mate[x]],y,label,pred,mate,\nwcodepenalty=\Llowpen
                   source_bridge,target_bridge);   \nwcodepenalty=\Llowpen
    return;\nwcodepenalty=\Llowpen
  \}\nwcodepenalty=\Llowpen
  else // x is ODD\nwcodepenalty=\Llowpen
  \{  \nwcodepenalty=\Llowpen
    P.append(x);\nwcodepenalty=\Llowpen
\vspace{\Lemptyline}    list<node> P2;\nwcodepenalty=\Llowpen
    find_path(P2,source_bridge[x],mate[x],label,pred,mate,\nwcodepenalty=\Llowpen
                             source_bridge,target_bridge);  \nwcodepenalty=\Llowpen
    P2.reverse_items();\nwcodepenalty=\Llowpen
    P.conc(P2);\nwcodepenalty=\Llowpen
\vspace{\Lemptyline}    find_path(P,target_bridge[x],y,label,pred,mate,\nwcodepenalty=\Llowpen
                      source_bridge,target_bridge);\nwcodepenalty=\Llowpen
\vspace{\Lemptyline}    return;\nwcodepenalty=\Llowpen
  \}\nwcodepenalty=\Llowpen
\}\nwcodepenalty=\Llowpen
\vspace{\Lemptyline}\vspace{\Lemptyline}list<edge> MAX_CARD_MATCHING(const graph& G, node_array<int>& OSC, int heur)\nwcodepenalty=\Llowpen
\{ \nwcodepenalty=\Llowpen
\vspace{\Lemptyline}    node_array<node> mate(G,nil);\nwcodepenalty=\Llowpen
    node_partition base(G);    // now base(v) = v for all nodes v\nwcodepenalty=\Llowpen
\vspace{\Lemptyline}\{ node v;\nwcodepenalty=\Llowpen
  forall_nodes(v,G) assert(base(v) == v);\nwcodepenalty=\Llowpen
\}\nwcodepenalty=\Llowpen
\vspace{\Lemptyline}    node_array<int>  label(G,EVEN);     \nwcodepenalty=\Llowpen
    node_array<node> pred(G,nil);\nwcodepenalty=\Llowpen
\vspace{\Lemptyline}\vspace{\Lemptyline}    double strue = 0;\nwcodepenalty=\Llowpen
    node_array<double>  path1(G,0);\nwcodepenalty=\Llowpen
    node_array<double>  path2(G,0);\nwcodepenalty=\Llowpen
\vspace{\Lemptyline}\vspace{\Lemptyline}    node_array<node> source_bridge(G,nil);\nwcodepenalty=\Llowpen
    node_array<node> target_bridge(G,nil);\nwcodepenalty=\Llowpen
\vspace{\Lemptyline}    switch (heur) \{\nwcodepenalty=\Llowpen
\vspace{\Lemptyline}    case 0: break;\nwcodepenalty=\Llowpen
\vspace{\Lemptyline}    case 1: \{ edge e;\nwcodepenalty=\Llowpen
              forall_edges(e,G)\nwcodepenalty=\Llowpen
              \{ node v = G.source(e); node w = G.target(e);\nwcodepenalty=\Llowpen
                if ( v != w && mate[v] == nil && mate[w] == nil)\nwcodepenalty=\Llowpen
                \{ mate[v] = w; label[v] = UNLABELED;\nwcodepenalty=\Llowpen
                  mate[w] = v; label[w] = UNLABELED;\nwcodepenalty=\Llowpen
                \}\nwcodepenalty=\Llowpen
              \}\nwcodepenalty=\Llowpen
              break;\nwcodepenalty=\Llowpen
            \}\nwcodepenalty=\Llowpen
    \}\nwcodepenalty=\Llowpen
\vspace{\Lemptyline}\vspace{\Lemptyline}  node v; edge e;\nwcodepenalty=\Llowpen
\vspace{\Lemptyline}  forall_nodes(v,G)\nwcodepenalty=\Llowpen
  \{ if ( mate[v] != nil ) continue;\nwcodepenalty=\Llowpen
\vspace{\Lemptyline}    node_slist Q(G); Q.append(v);\nwcodepenalty=\Llowpen
    list<node> T; T.append(v);\nwcodepenalty=\Llowpen
    bool breakthrough = false;\nwcodepenalty=\Llowpen
\vspace{\Lemptyline}    while (!breakthrough && !Q.empty()) // grow tree rooted at v\nwcodepenalty=\Llowpen
    \{\nwcodepenalty=\Llowpen
      node v = Q.pop();\nwcodepenalty=\Llowpen
\vspace{\Lemptyline}      forall_inout_edges(e,v)\nwcodepenalty=\Llowpen
      \{ node w = G.opposite(v,e); \nwcodepenalty=\Llowpen
\vspace{\Lemptyline}        if ( base(v) == base(w) || label[base(w)] == ODD ) \nwcodepenalty=\Llowpen
           continue;   // do nothing\nwcodepenalty=\Llowpen
\vspace{\Lemptyline}        if ( label[w] == UNLABELED )\nwcodepenalty=\Llowpen
          \{ \nwcodepenalty=\Llowpen
            label[w] = ODD;            T.append(w);\nwcodepenalty=\Llowpen
            pred[w] = v;\nwcodepenalty=\Llowpen
            label[mate[w]] = EVEN;     T.append(mate[w]);\nwcodepenalty=\Llowpen
            Q.append(mate[w]);\nwcodepenalty=\Llowpen
 \}\nwcodepenalty=\Llowpen
        else  // base(w) is EVEN\nwcodepenalty=\Llowpen
          \{ \nwcodepenalty=\Llowpen
            node hv = base(v);\nwcodepenalty=\Llowpen
            node hw = base(w);\nwcodepenalty=\Llowpen
\vspace{\Lemptyline}            strue++; \nwcodepenalty=\Llowpen
            path1[hv] = path2[hw] = strue;\nwcodepenalty=\Llowpen
\vspace{\Lemptyline}            while ((path1[hw] != strue && path2[hv] != strue) &&\nwcodepenalty=\Llowpen
                   (mate[hv] != nil || mate[hw] != nil) )\nwcodepenalty=\Llowpen
            \{ \nwcodepenalty=\Llowpen
              if (mate[hv] != nil)\nwcodepenalty=\Llowpen
              \{ hv = base(pred[mate[hv]]);\nwcodepenalty=\Llowpen
                path1[hv] = strue;\nwcodepenalty=\Llowpen
              \}\nwcodepenalty=\Llowpen
\vspace{\Lemptyline}              if (mate[hw] != nil)\nwcodepenalty=\Llowpen
              \{ hw = base(pred[mate[hw]]);\nwcodepenalty=\Llowpen
                path2[hw] = strue;\nwcodepenalty=\Llowpen
              \}\nwcodepenalty=\Llowpen
            \}\nwcodepenalty=\Llowpen
\vspace{\Lemptyline}            if (path1[hw] == strue || path2[hv] == strue) \nwcodepenalty=\Llowpen
              \{ \nwcodepenalty=\Llowpen
                node b = (path1[hw] == strue) ? hw : hv;    // Base\nwcodepenalty=\Llowpen
\vspace{\Lemptyline}                shrink_path(b,v,w,base,mate,pred,source_bridge,target_bridge,Q);\nwcodepenalty=\Llowpen
                shrink_path(b,w,v,base,mate,pred,source_bridge,target_bridge,Q);\nwcodepenalty=\Llowpen
 \} \nwcodepenalty=\Llowpen
            else  \nwcodepenalty=\Llowpen
              \{ \nwcodepenalty=\Llowpen
                list<node> P;\nwcodepenalty=\Llowpen
\vspace{\Lemptyline}                find_path(P,v,hv,label,pred,mate,source_bridge,target_bridge);\nwcodepenalty=\Llowpen
                P.push(w);\nwcodepenalty=\Llowpen
\vspace{\Lemptyline}                while(! P.empty())\nwcodepenalty=\Llowpen
                \{ node a = P.pop();\nwcodepenalty=\Llowpen
                  node b = P.pop();\nwcodepenalty=\Llowpen
                  mate[a] = b;\nwcodepenalty=\Llowpen
                  mate[b] = a;\nwcodepenalty=\Llowpen
                \}\nwcodepenalty=\Llowpen
\vspace{\Lemptyline}                T.append(w);\nwcodepenalty=\Llowpen
                forall(v,T) label[v] = UNLABELED;  \nwcodepenalty=\Llowpen
                base.split(T);\nwcodepenalty=\Llowpen
\vspace{\Lemptyline}                breakthrough = true;\nwcodepenalty=\Llowpen
                break;        \nwcodepenalty=\Llowpen
 \} \nwcodepenalty=\Llowpen
 \} \nwcodepenalty=\Llowpen
      \} \nwcodepenalty=\Llowpen
\vspace{\Lemptyline}    \}\nwcodepenalty=\Llowpen
  \}\nwcodepenalty=\Llowpen
\vspace{\Lemptyline}\vspace{\Lemptyline}  list<edge> M;\nwcodepenalty=\Llowpen
\vspace{\Lemptyline}   forall_edges(e,G) \nwcodepenalty=\Llowpen
   \{ node v = source(e);\nwcodepenalty=\Llowpen
     node w = target(e);\nwcodepenalty=\Llowpen
     if ( v != w  &&  mate[v] == w ) \nwcodepenalty=\Llowpen
     \{ M.append(e);\nwcodepenalty=\Llowpen
       mate[v] = v;\nwcodepenalty=\Llowpen
       mate[w] = w;\nwcodepenalty=\Llowpen
      \}\nwcodepenalty=\Llowpen
    \}\nwcodepenalty=\Llowpen
\vspace{\Lemptyline}\vspace{\Lemptyline}  forall_nodes(v,G) OSC[v] = -1;\nwcodepenalty=\Llowpen
\vspace{\Lemptyline}  int number_of_unlabeled = 0;\nwcodepenalty=\Llowpen
  node arb_u_node = 0;\nwcodepenalty=\Llowpen
\vspace{\Lemptyline}  forall_nodes(v,G) \nwcodepenalty=\Llowpen
   if ( label[v] == UNLABELED ) \nwcodepenalty=\Llowpen
   \{ number_of_unlabeled++;\nwcodepenalty=\Llowpen
     arb_u_node = v;\nwcodepenalty=\Llowpen
   \}\nwcodepenalty=\Llowpen
\vspace{\Lemptyline}  int L = 0;\nwcodepenalty=\Llowpen
  if ( number_of_unlabeled > 0 )\nwcodepenalty=\Llowpen
  \{ OSC[arb_u_node] = 1;\nwcodepenalty=\Llowpen
    if (number_of_unlabeled > 2) L = 2;\nwcodepenalty=\Llowpen
    forall_nodes(v,G) \nwcodepenalty=\Llowpen
      if ( label[v] == UNLABELED && v != arb_u_node ) OSC[v] = L;\nwcodepenalty=\Llowpen
  \}\nwcodepenalty=\Llowpen
\vspace{\Lemptyline}  int K = ( L == 0? 2 : 3);\nwcodepenalty=\Llowpen
\vspace{\Lemptyline}  forall_nodes(v,G)\nwcodepenalty=\Llowpen
   if ( base(v) != v && OSC[base(v)] == -1 ) OSC[base(v)] = K++;\nwcodepenalty=\Llowpen
\vspace{\Lemptyline}  forall_nodes(v,G)\nwcodepenalty=\Llowpen
  \{ if ( base(v) == v && OSC[v] == -1 )\nwcodepenalty=\Llowpen
    \{ if ( label[v] == EVEN ) OSC[v] = 0;\nwcodepenalty=\Llowpen
      if ( label[v] == ODD  ) OSC[v] = 1;\nwcodepenalty=\Llowpen
    \}\nwcodepenalty=\Llowpen
    if ( base(v) != v ) OSC[v] = OSC[base(v)];\nwcodepenalty=\Llowpen
  \}\nwcodepenalty=\Llowpen
\vspace{\Lemptyline} return M;\nwcodepenalty=\Lhighpen
\}\nwcodepenalty=\Lhighpen
\vspace{\Lemptyline}\vspace{\Lemptyline}LEDA_END_NAMESPACE\nwcodepenalty=\Llowpen
\vspace{\Lemptyline}\vspace{\Lemptyline}\vspace{\Lemptyline}\nwendcode{}\nwbegindocs{62}\nwdocspar
\section{LEDA's Matching Algorithm plus the Kececioglu-Pecqueur Heuristics}

And now with the heuristics suggested by Kececioglu and Pecqueur~\cite{Kececioglu:matching}.

\nwenddocs{}\nwbegincode{63}\moddef{LEDA-KP}\endmoddef\nwstartdeflinemarkup\nwenddeflinemarkup\nwcodepenalty=\Lhighpen
\vspace{\Lemptyline}class G_card_matching_KP\{\nwcodepenalty=\Lhighpen
private:\nwcodepenalty=\Lhighpen
  const graph& G;\nwcodepenalty=\Llowpen
  list<edge> M; // matching in G\nwcodepenalty=\Llowpen
  node_array<node> mate; // mates in M\nwcodepenalty=\Llowpen
  node_partition base; // blossoms in G\nwcodepenalty=\Llowpen
  node_array<node> parent; // parent[v] is the parent of v in the alternating tree\nwcodepenalty=\Llowpen
  node_array<node> source_bridge;  // bridges close blossoms; a blossom consists of two paths\nwcodepenalty=\Llowpen
  node_array<node> target_bridge;  // x--z and y--z plus the edge xy; z is the base of the blossom.\nwcodepenalty=\Llowpen
  // the nodes on the path from x to z store x as source_bridge and y as target_bridge.\nwcodepenalty=\Llowpen
  node_array<LABEL> label;\nwcodepenalty=\Llowpen
  node_array<int> even_time;\nwcodepenalty=\Llowpen
  int even_count;\nwcodepenalty=\Llowpen
  int size_of_M;\nwcodepenalty=\Llowpen
  node_array<int> num;\nwcodepenalty=\Llowpen
  list<node> T;\nwcodepenalty=\Llowpen
  list<node> P;    // initialized as empty list by default constructor\nwcodepenalty=\Llowpen
\vspace{\Lemptyline}  node find_aug_path(node v, node v0)\{ // we are growing a tree with root v0\nwcodepenalty=\Llowpen
   edge e; node w; \nwcodepenalty=\Llowpen
   forall_inout_edges(e,v)\{\nwcodepenalty=\Llowpen
     w = G.opposite(e,v); \nwcodepenalty=\Llowpen
     if( v == w ) continue;\nwcodepenalty=\Llowpen
     if( mate[w] == nil && label[w] == UNLABELED )\{\nwcodepenalty=\Llowpen
       parent[w] = v;\nwcodepenalty=\Llowpen
       T.append(w);\nwcodepenalty=\Llowpen
       label[w] = ODD;\nwcodepenalty=\Llowpen
       return w;\nwcodepenalty=\Llowpen
     \}\nwcodepenalty=\Llowpen
   \}\nwcodepenalty=\Llowpen
   // no immediate break-through\nwcodepenalty=\Llowpen
   forall_inout_edges(e,v)\{\nwcodepenalty=\Llowpen
     w = G.opposite(e,v);\nwcodepenalty=\Llowpen
     if (label[base(w)] == ODD || w == v) continue;\nwcodepenalty=\Llowpen
     if (label[base(w)] == UNLABELED)\{\nwcodepenalty=\Llowpen
       label[w] = ODD; parent[w] = v; T.append(w);\nwcodepenalty=\Llowpen
       node mw = mate[w];\nwcodepenalty=\Llowpen
       label[mw] = EVEN; T.append(mw); even_time[mw] = even_count++;\nwcodepenalty=\Llowpen
       node s = find_aug_path(mw,v0);\nwcodepenalty=\Llowpen
       if(s != nil)\nwcodepenalty=\Llowpen
         return s;\nwcodepenalty=\Llowpen
     \}\nwcodepenalty=\Llowpen
     else\{\nwcodepenalty=\Llowpen
       node bv = base(v);\nwcodepenalty=\Llowpen
       node bw = base(w);\nwcodepenalty=\Llowpen
       list<node> tmp;   \nwcodepenalty=\Llowpen
       if (even_time[bv] < even_time[bw])\{ //blossom_step along forward edge\nwcodepenalty=\Llowpen
         // walk down from bw to bv and perform unions\nwcodepenalty=\Llowpen
         while(bw != bv)\{// doing the unions carefully, so that only one make_rep is needed\nwcodepenalty=\Llowpen
           node mate_bw = mate[bw]; \nwcodepenalty=\Llowpen
           base.union_blocks(bw,mate_bw);\nwcodepenalty=\Llowpen
           bw = base(parent[mate_bw]);\nwcodepenalty=\Llowpen
           base.union_blocks(mate_bw,bw); \nwcodepenalty=\Llowpen
           tmp.push_front(mate_bw);\nwcodepenalty=\Llowpen
           source_bridge[mate_bw] = w;\nwcodepenalty=\Llowpen
           target_bridge[mate_bw] = v;\nwcodepenalty=\Llowpen
         \}\nwcodepenalty=\Llowpen
         base.make_rep(bv);\nwcodepenalty=\Llowpen
         forall(w,tmp)\{\nwcodepenalty=\Llowpen
          node s = find_aug_path(w,v0);\nwcodepenalty=\Llowpen
          if(s != nil)\nwcodepenalty=\Llowpen
            return s;\nwcodepenalty=\Llowpen
         \}\nwcodepenalty=\Llowpen
       \}\nwcodepenalty=\Llowpen
     \}\nwcodepenalty=\Llowpen
   \}\nwcodepenalty=\Llowpen
   return nil;\nwcodepenalty=\Llowpen
  \}\nwcodepenalty=\Llowpen
\vspace{\Lemptyline}\vspace{\Lemptyline}\vspace{\Lemptyline}  void find_path(node x, node y)\{\nwcodepenalty=\Llowpen
/* traces the even length alternating path from x to y; if non-trivial it starts with\nwcodepenalty=\Llowpen
the matching edge incident to x; collects the non-matching edges on this\nwcodepenalty=\Llowpen
path as pairs of nodes */\nwcodepenalty=\Llowpen
    if ( x == y ) return;\nwcodepenalty=\Llowpen
\vspace{\Lemptyline}    if ( label[x] == EVEN )\{\nwcodepenalty=\Llowpen
      P.append(mate[x]); P.append(parent[mate[x]]);\nwcodepenalty=\Llowpen
      find_path(parent[mate[x]],y);   \nwcodepenalty=\Llowpen
      return;\nwcodepenalty=\Llowpen
    \}\nwcodepenalty=\Llowpen
    else\{ // x is ODD\nwcodepenalty=\Llowpen
      find_path(source_bridge[x],mate[x]);  \nwcodepenalty=\Llowpen
      P.append(source_bridge[x]); P.append(target_bridge[x]);\nwcodepenalty=\Llowpen
      find_path(target_bridge[x],y);\nwcodepenalty=\Llowpen
      return;\nwcodepenalty=\Llowpen
    \}\nwcodepenalty=\Llowpen
  \}\nwcodepenalty=\Llowpen
\vspace{\Lemptyline}\vspace{\Lemptyline}\vspace{\Lemptyline}\vspace{\Lemptyline}public:\nwcodepenalty=\Llowpen
  G_card_matching_KP(const graph& g): G(g), mate(node_array<node>(G,nil)), base(node_partition(G)),\nwcodepenalty=\Llowpen
                                 parent(node_array<node>(G)),\nwcodepenalty=\Llowpen
                                 source_bridge(node_array<node>(G)), target_bridge(node_array<node>(G)),\nwcodepenalty=\Llowpen
                                 label(node_array<LABEL>(G,UNLABELED)),even_time(node_array<int>(G,0)),\nwcodepenalty=\Llowpen
                                 even_count(0), size_of_M(0), num(node_array<int>(G))\{\nwcodepenalty=\Llowpen
    // most arrays do not need initialization\nwcodepenalty=\Llowpen
  \}\nwcodepenalty=\Llowpen
\vspace{\Lemptyline}  int init()\{\nwcodepenalty=\Llowpen
    edge e;\nwcodepenalty=\Llowpen
    forall_edges(e,G)\{\nwcodepenalty=\Llowpen
      node u = G.source(e); node v = G.target(e);\nwcodepenalty=\Llowpen
      if (u != v && mate[u] == nil && mate[v] == nil) \{\nwcodepenalty=\Llowpen
        mate[u] = v;\nwcodepenalty=\Llowpen
        mate[v] = u;\nwcodepenalty=\Llowpen
        size_of_M++;\nwcodepenalty=\Llowpen
      \}\nwcodepenalty=\Llowpen
    \}\nwcodepenalty=\Llowpen
    return size_of_M;\nwcodepenalty=\Llowpen
  \};\nwcodepenalty=\Llowpen
\vspace{\Lemptyline}\vspace{\Lemptyline}  list<edge> solve(node_array<int>& OSC,int heur = 1)\{\nwcodepenalty=\Llowpen
    node v, v0, w; edge e;\nwcodepenalty=\Llowpen
    int count = 1;\nwcodepenalty=\Llowpen
    forall_nodes(v,G) num[v] = count++;\nwcodepenalty=\Llowpen
    if (heur == 1) init();\nwcodepenalty=\Llowpen
\vspace{\Lemptyline}    forall_nodes(v0,G)\{\nwcodepenalty=\Llowpen
      if ( mate[v0] != nil ) continue;\nwcodepenalty=\Llowpen
\vspace{\Lemptyline}      label[v0] = EVEN; T.clear(); T.append(v0); even_time[v0] = even_count++;\nwcodepenalty=\Llowpen
\vspace{\Lemptyline}      if( (w = find_aug_path(v0,v0)) != nil )\{\nwcodepenalty=\Llowpen
          node pw = parent[w];  \nwcodepenalty=\Llowpen
          P.push(w); P.push(pw);\nwcodepenalty=\Llowpen
          find_path(pw,v0);\nwcodepenalty=\Llowpen
\vspace{\Lemptyline}          while(! P.empty())\{\nwcodepenalty=\Llowpen
            node a = P.pop();\nwcodepenalty=\Llowpen
            node b = P.pop();\nwcodepenalty=\Llowpen
            mate[a] = b;\nwcodepenalty=\Llowpen
            mate[b] = a;\nwcodepenalty=\Llowpen
          \} \nwcodepenalty=\Llowpen
\vspace{\Lemptyline}          forall(v,T) label[v] = UNLABELED;  \nwcodepenalty=\Llowpen
          base.split(T);\nwcodepenalty=\Llowpen
          size_of_M++;\nwcodepenalty=\Llowpen
      \}\nwcodepenalty=\Llowpen
    \}\nwcodepenalty=\Llowpen
\vspace{\Lemptyline}    list<edge> M;\nwcodepenalty=\Llowpen
\vspace{\Lemptyline}    forall_edges(e,G) \{\nwcodepenalty=\Llowpen
      node v = source(e);\nwcodepenalty=\Llowpen
      node w = target(e);\nwcodepenalty=\Llowpen
      if ( v != w  &&  mate[v] == w ) \{\nwcodepenalty=\Llowpen
        M.append(e);\nwcodepenalty=\Llowpen
        mate[v] = v;\nwcodepenalty=\Llowpen
        mate[w] = w;\nwcodepenalty=\Llowpen
      \}\nwcodepenalty=\Llowpen
    \}\nwcodepenalty=\Llowpen
\vspace{\Lemptyline}\vspace{\Lemptyline}    forall_nodes(v,G) OSC[v] = -1;\nwcodepenalty=\Llowpen
\vspace{\Lemptyline}    int number_of_unlabeled = 0;\nwcodepenalty=\Llowpen
    node arb_u_node = 0;\nwcodepenalty=\Llowpen
\vspace{\Lemptyline}    forall_nodes(v,G) \nwcodepenalty=\Llowpen
      if ( label[v] == UNLABELED ) \{\nwcodepenalty=\Llowpen
        number_of_unlabeled++;\nwcodepenalty=\Llowpen
        arb_u_node = v;\nwcodepenalty=\Llowpen
      \}\nwcodepenalty=\Llowpen
\vspace{\Lemptyline}    int L = 0;\nwcodepenalty=\Llowpen
    if ( number_of_unlabeled > 0 )\{\nwcodepenalty=\Llowpen
      OSC[arb_u_node] = 1;\nwcodepenalty=\Llowpen
      if (number_of_unlabeled > 2) L = 2;\nwcodepenalty=\Llowpen
      forall_nodes(v,G) \nwcodepenalty=\Llowpen
        if ( label[v] == UNLABELED && v != arb_u_node ) OSC[v] = L;\nwcodepenalty=\Llowpen
    \}\nwcodepenalty=\Llowpen
\vspace{\Lemptyline}    int K = ( L == 0? 2 : 3);\nwcodepenalty=\Llowpen
\vspace{\Lemptyline}    forall_nodes(v,G)\nwcodepenalty=\Llowpen
      if ( base(v) != v && OSC[base(v)] == -1 ) OSC[base(v)] = K++;\nwcodepenalty=\Llowpen
\vspace{\Lemptyline}    forall_nodes(v,G)\{\nwcodepenalty=\Llowpen
      if ( base(v) == v && OSC[v] == -1 )\{\nwcodepenalty=\Llowpen
        if ( label[v] == EVEN ) OSC[v] = 0;\nwcodepenalty=\Llowpen
        if ( label[v] == ODD  ) OSC[v] = 1;\nwcodepenalty=\Llowpen
      \}\nwcodepenalty=\Llowpen
      if ( base(v) != v ) OSC[v] = OSC[base(v)];\nwcodepenalty=\Llowpen
    \}\nwcodepenalty=\Llowpen
\vspace{\Lemptyline}    return M;\nwcodepenalty=\Lhighpen
  \}\nwcodepenalty=\Lhighpen
\vspace{\Lemptyline}\};\nwcodepenalty=\Llowpen
\vspace{\Lemptyline}\nwendcode{}\nwbegindocs{64}\nwdocspar
\section{The Program for the Running Time Experiments}

\nwenddocs{}\nwbegincode{65}\moddef{RunningTimes}\endmoddef\nwstartdeflinemarkup\nwenddeflinemarkup\nwcodepenalty=\Lhighpen
\vspace{\Lemptyline}\vspace{\Lemptyline}#line 271 "Experiments.lw"\nwcodepenalty=\Lhighpen
#include <LEDA/graph/graph.h>\nwcodepenalty=\Lhighpen
#include <LEDA/graph/node_partition.h>\nwcodepenalty=\Llowpen
#include <LEDA/graph/node_slist.h>\nwcodepenalty=\Llowpen
#include <LEDA/graph/mc_matching.h>\nwcodepenalty=\Llowpen
#include <LEDA/system/assert.h>\nwcodepenalty=\Llowpen
#include <LEDA/core/random.h>\nwcodepenalty=\Llowpen
\vspace{\Lemptyline}#include <LEDA/core/array.h>\nwcodepenalty=\Llowpen
#include <LEDA/core/string.h>\nwcodepenalty=\Llowpen
#include <LEDA/core/list.h>\nwcodepenalty=\Llowpen
#include <LEDA/core/stack.h>\nwcodepenalty=\Llowpen
\vspace{\Lemptyline}#include <LEDA/graph/graph_gen.h>\nwcodepenalty=\Llowpen
#include <LEDA/core/misc.h>\nwcodepenalty=\Llowpen
#include <LEDA/graph/mcb_matching.h>\nwcodepenalty=\Llowpen
#include <LEDA/core/array.h>\nwcodepenalty=\Llowpen
#include <LEDA/graph/graph.h>\nwcodepenalty=\Llowpen
#include <assert.h>\nwcodepenalty=\Llowpen
#include <math.h>\nwcodepenalty=\Llowpen
#define BOOK\nwcodepenalty=\Llowpen
#include <LEDA/core/IO_interface.h>\nwcodepenalty=\Llowpen
\vspace{\Lemptyline}using namespace leda;\nwcodepenalty=\Llowpen
\vspace{\Lemptyline}//enum LABEL \{ODD, EVEN, UNLABELED\};\nwcodepenalty=\Llowpen
\vspace{\Lemptyline}\vspace{\Lemptyline}#include "Gabow.h"\nwcodepenalty=\Llowpen
// #include "new_mc_matching.cpp"\nwcodepenalty=\Llowpen
\vspace{\Lemptyline}node complete_g(graph& G, int m)  // sqrt(2m) nodes and about m edges\nwcodepenalty=\Llowpen
\{ int n = 2* (int) sqrt(m/2.0); \nwcodepenalty=\Llowpen
  array<node> a(n); int i, j;\nwcodepenalty=\Llowpen
  for (i = 0; i < n; i++) \nwcodepenalty=\Llowpen
    a[i] = G.new_node(); \nwcodepenalty=\Llowpen
  for (i = 0; i < n; i++) \nwcodepenalty=\Llowpen
    for (j = i+1; j < n; j++) G.new_edge(a[i],a[j]);   \nwcodepenalty=\Llowpen
\vspace{\Lemptyline}  return a[0];\nwcodepenalty=\Llowpen
\}\nwcodepenalty=\Llowpen
\vspace{\Lemptyline}void chain(graph& G, int k, node z)\{ // k new nodes and 2k - 1 edges \nwcodepenalty=\Llowpen
  array<node> a(k); int i;\nwcodepenalty=\Llowpen
\vspace{\Lemptyline}  for (i = 0; i < k; i++) a[i] = G.new_node();\nwcodepenalty=\Llowpen
\vspace{\Lemptyline}  G.new_edge(a[0],z);\nwcodepenalty=\Llowpen
\vspace{\Lemptyline}  for (i = 1; i < k-1; i++) \nwcodepenalty=\Llowpen
  \{ G.new_edge(a[i],a[i+1]);\nwcodepenalty=\Llowpen
    G.new_edge(a[i],z);      \nwcodepenalty=\Llowpen
  \}\nwcodepenalty=\Llowpen
  G.new_edge(a[0],a[1]);\nwcodepenalty=\Llowpen
\}\nwcodepenalty=\Llowpen
\vspace{\Lemptyline}#define LONG  // LONG // SHORT\nwcodepenalty=\Llowpen
#define RANDOM\nwcodepenalty=\Llowpen
#define KP\nwcodepenalty=\Llowpen
#define PERMUTE \nwcodepenalty=\Llowpen
\vspace{\Lemptyline}\vspace{\Lemptyline}\vspace{\Lemptyline}#ifdef LONG\nwcodepenalty=\Llowpen
int mode = 1;\nwcodepenalty=\Llowpen
#endif\nwcodepenalty=\Llowpen
#ifdef SHORT\nwcodepenalty=\Llowpen
int mode = 0;\nwcodepenalty=\Llowpen
#endif\nwcodepenalty=\Llowpen
\vspace{\Lemptyline}\vspace{\Lemptyline}void mc_worst_case_gen(graph& G, int n, int m, int mode)\nwcodepenalty=\Llowpen
// sqrt(2m) + n + (mode == 1) n nodes and m + 2n + (mode == 1) 2n edges.\nwcodepenalty=\Llowpen
\{ node z = complete_g(G,m);\nwcodepenalty=\Llowpen
  int k = 4;\nwcodepenalty=\Llowpen
  for (int j = 0; j < n/8; j++) chain(G,2*k,z); \nwcodepenalty=\Llowpen
  if (mode == 1) \{\nwcodepenalty=\Llowpen
    for (k = 5; k < sqrt(n); k++)\nwcodepenalty=\Llowpen
    chain(G,2*k,z);\nwcodepenalty=\Llowpen
  \}\nwcodepenalty=\Llowpen
\}\nwcodepenalty=\Llowpen
\vspace{\Lemptyline}\vspace{\Lemptyline}void permute(graph& G)\nwcodepenalty=\Llowpen
\{ \nwcodepenalty=\Llowpen
  list<edge> E = G.all_edges();\nwcodepenalty=\Llowpen
  E.permute();\nwcodepenalty=\Llowpen
  G.sort_edges(E); \nwcodepenalty=\Llowpen
\vspace{\Lemptyline}  list<node> V = G.all_nodes();\nwcodepenalty=\Llowpen
  V.permute();\nwcodepenalty=\Llowpen
  G.sort_nodes(V);\nwcodepenalty=\Llowpen
\vspace{\Lemptyline}\}\nwcodepenalty=\Llowpen
\vspace{\Lemptyline}#include "class_new_mc_matching.cpp"\nwcodepenalty=\Llowpen
\vspace{\Lemptyline}\vspace{\Lemptyline}float matching_test(int n, int m, int seed = 290849)\nwcodepenalty=\Llowpen
\{  graph G;\nwcodepenalty=\Llowpen
  rand_int.set_seed(seed);\nwcodepenalty=\Llowpen
#ifdef RANDOM\nwcodepenalty=\Llowpen
  random_graph(G,n,m);\nwcodepenalty=\Llowpen
#else\nwcodepenalty=\Llowpen
  mc_worst_case_gen(G,n,m,mode);\nwcodepenalty=\Llowpen
#ifdef PERMUTE\nwcodepenalty=\Llowpen
  permute(G);\nwcodepenalty=\Llowpen
#endif\nwcodepenalty=\Llowpen
#endif\nwcodepenalty=\Llowpen
\vspace{\Lemptyline}   int number_of_iterations;\nwcodepenalty=\Llowpen
\vspace{\Lemptyline}\vspace{\Lemptyline}  list<edge> M;\nwcodepenalty=\Llowpen
  node_array<int> OSC(G);\nwcodepenalty=\Llowpen
  float T = used_time(); float UT;\nwcodepenalty=\Llowpen
#ifdef LEDA\nwcodepenalty=\Llowpen
  M = MAX_CARD_MATCHING(G,OSC,1);\nwcodepenalty=\Llowpen
#endif\nwcodepenalty=\Llowpen
#ifdef KP\nwcodepenalty=\Llowpen
  G_card_matching_KP P(G);\nwcodepenalty=\Llowpen
  M = P.solve(OSC);\nwcodepenalty=\Llowpen
#endif\nwcodepenalty=\Llowpen
#ifdef GABOW\nwcodepenalty=\Llowpen
  G_card_matching P(G);\nwcodepenalty=\Llowpen
  M = P.solve(OSC,number_of_iterations);\nwcodepenalty=\Llowpen
#endif\nwcodepenalty=\Llowpen
\vspace{\Lemptyline}  UT = used_time(T); std::cout << "\\ntime =  " << UT;\nwcodepenalty=\Llowpen
  CHECK_MAX_CARD_MATCHING(G,M,OSC);\nwcodepenalty=\Llowpen
  return UT;\nwcodepenalty=\Llowpen
\}\nwcodepenalty=\Llowpen
\vspace{\Lemptyline}\vspace{\Lemptyline}\vspace{\Lemptyline}int main(int argc, char** argv)\nwcodepenalty=\Llowpen
\{\nwcodepenalty=\Llowpen
  if (argc < 2) \{\nwcodepenalty=\Llowpen
    cerr << "usage: " << argv[0] << argv[1] << " n " << endl;\nwcodepenalty=\Llowpen
   return 1;\nwcodepenalty=\Llowpen
  \}\nwcodepenalty=\Llowpen
\vspace{\Lemptyline}  int n = atoi(argv[1]);\nwcodepenalty=\Llowpen
  int seed = atoi(argv[2]);\nwcodepenalty=\Llowpen
  float UT = matching_test(n,4*n,seed);\nwcodepenalty=\Llowpen
  //print_statistics();\nwcodepenalty=\Lhighpen
  return 0;\nwcodepenalty=\Lhighpen
\}\nwcodepenalty=\Llowpen
\vspace{\Lemptyline}\vspace{\Lemptyline}\nwendcode{}\nwbegindocs{66}\end{document}